\documentclass[a4paper,12pt]{article}
\pdfoutput=1
\usepackage{jheppub}
\usepackage{pgf,tikz,pgfplots}
\usepackage{tikz}
\usetikzlibrary{cd}
\pgfplotsset{compat=1.15}
\usepackage{mathrsfs}
\usetikzlibrary{arrows}
\usepackage{dsfont}
\usepackage{amsfonts}
\usepackage{amsmath,amssymb}
\usepackage{physics}
\usepackage{amsthm}
\usepackage{setspace}
\usepackage{relsize}
\usepackage{todonotes}
\usepackage{xpatch}
\usepackage[none]{hyphenat}
\xpatchcmd\section{\large}{\Large}{}{}
\xpatchcmd\subsection{\normalsize}{\large}{}{}
\xpatchcmd\subsubsection{\normalsize}{\normalsize}{}{}

\newcommand{\cm}{\operatorname{\text{\usefont{U}{BOONDOX-cal}{m}{n}m}}}

\newcommand{\CH}{\mathcal{H}}
\newcommand{\CD}{\mathcal{D}}
\newcommand{\CE}{\mathcal{E}}
\newcommand{\CQ}{\mathcal{Q}}
\newcommand{\CB}{\mathcal{B}}
\newcommand{\CC}{\mathcal{C}}
\newcommand{\CO}{\mathcal{O}}
\newcommand{\CT}{\mathcal{T}}

\newcommand{\CZ}{\mathcal{Z}}
\newcommand{\CM}{\mathcal{M}}

\newcommand{\SH}{\mathscr{H}}

\newcommand{\DZ}{\mathds{Z}}
\newcommand{\trl}{\mathbf{1}}

\newcommand{\be}{\begin{equation}}
\newcommand{\ee}{\end{equation}}
\newcommand{\bea}{\begin{eqnarray}}
\newcommand{\eea}{\end{eqnarray}}

\newcommand{\al}{\alpha}

\newtheorem{theorem}{Theorem}[section]

\title{Non-anomalous non-invertible symmetries in 1+1D from gapped boundaries of SymTFTs}
\author[1]{Pavel Putrov}
\author[1]{and Rajath Radhakrishnan}
\affiliation[1]{International Centre for Theoretical Physics,\\ Strada Costiera 11, Trieste 34151, Italy}

\abstract{We study the anomalies of non-invertible symmetries in 1+1D QFTs using gapped boundaries of its SymTFT. We establish the explicit relation between Lagrangian algebras which determine gapped boundaries of the SymTFT, and algebras which determine non-anomalous/gaugeable topological line operators in the 1+1D QFT. If the Lagrangian algebras in the SymTFT are known, this provides a method to compute algebras in all fusion categories that share the same SymTFT. We find necessary conditions that a line operator in the SymTFT must satisfy for the corresponding line operator in the 1+1D QFT to be non-anomalous. We use this constraint to show that a non-invertible symmetry admits a 1+1D trivially gapped phase if and only if the SymTFT admits a magnetic Lagrangian algebra. We define a process of transporting non-anomalous line operators between fusion categories which share the same SymTFT and apply this method to the three Haagerup fusion categories. }

\begin{document} 

\maketitle

\newpage

\section{Introduction}

A quantum field theory contains a rich collection of extended operators of various dimensions. Among them, topological operators are a particularly rich, and yet relatively simple set of operators which can be often analysed explicitly. These operators play a very important role in constraining the dynamics of a QFT as they implement generalized symmetries \cite{Gaiotto:2014kfa}. The topological nature of these operators puts strong constraints on them to the extent that a ``topological bootstrap" approach can be used to classify them. This reveals that such operators are captured by an algebraic structure called a higher-fusion category \cite{Kapustin:2010ta,Bhardwaj:2022yxj}. In a Topological Quantum Field Theory, which is a QFT in which all operators are topological, the topological operators have to obey additional constraints like modularity. This has led to the (partial) classification of TQFTs in various dimensions \cite{Lan:2018vjb,Lan:2018bui,Johnson-Freyd:2020usu,Kong:2020jne,Johnson-Freyd:2020twl,Johnson-Freyd:2021tbq,Ng:2023wsc}. 

While the topological operators that can appear in a $d$ dimensional QFT are not as constrained as those in a TQFT, it is remarkable that their properties can be captured by a $d+1$ dimensional TQFT called the SymTFT \cite{Freed:2012bs,Gaiotto:2020iye,Kong:2020cie,Apruzzi:2021nmk,Freed:2022qnc,Chatterjee:2022kxb,Kaidi:2022cpf,Kaidi:2023maf}. The SymTFT allows us to study topological properties of a QFT with symmetry $\CC$ and other QFTs obtained from gauging ``sub-symmetries" of $\CC$ in a unified setting. 

Given a 1+1D QFT with a finite invertible symmetry $G$ and anomaly given by $\omega$ in $H^3(G,U(1))$, a natural question to consider is whether $G$ contains non-anomalous subgroups. A subgroup $H\subset G$ is non-anomalous iff the anomaly $\omega$ trivializes on the subgroup. More precisely, $H$ is non-anomalous iff $\omega|_{H}$ is trivial in $H^3(H,U(1))$. An alternate approach to this problem is to use the SymTFT of the symmetry $G$ with anomaly $\omega$. This is the twisted Dijkgraaf-Witten theory DW$(G,\omega)$ determined by the gauge group $G$ and 3-cocycle $\omega$ \cite{Dijkgraaf:1989pz,Gaiotto:2020iye}. Non-anomalous subgroups of $G$ are in correspondence with gapped boundaries of this SymTFT. In 1+1D, gauging symmetries can produce dual QFTs. For example, gauging non-anomalous subgroups $K,H \subseteq G$ such that $K$ and $H$ are conjugate to each other are physically equivalent. This defines an equivalence relation on the gaugeable subgroups of $G$. It is then natural to ask:
\vspace{0.2cm}

\textit{Given a gapped boundary of the 2+1D SymTFT DW$(G,\omega)$ how do we determine the equivalence class of physically equivalent gaugings in 1+1D?}
\vspace{0.2cm}

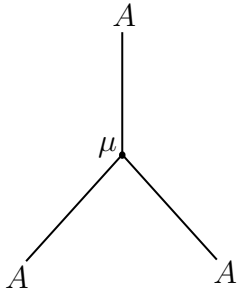
\begin{figure}[h!]
    \centering

\tikzset{every picture/.style={line width=0.75pt}} %set default line width to 0.75pt        

\begin{tikzpicture}[x=0.75pt,y=0.75pt,yscale=-0.8,xscale=0.8]

%Straight Lines [id:da08636887019594053] 
\draw    (331,157) -- (271,224) ;
%Straight Lines [id:da2382307740137487] 
\draw    (331,157) -- (390,223) ;
%Straight Lines [id:da6391111084686333] 
\draw    (331,80) -- (331,157) ;
%Shape: Ellipse [id:dp5043570118221843] 
\draw  [color={rgb, 255:red, 0; green, 0; blue, 0 }  ,draw opacity=1 ][fill={rgb, 255:red, 0; green, 0; blue, 0 }  ,fill opacity=1 ] (329.59,157.27) .. controls (329.59,156.31) and (330.22,155.54) .. (331,155.54) .. controls (331.78,155.54) and (332.41,156.31) .. (332.41,157.27) .. controls (332.41,158.22) and (331.78,159) .. (331,159) .. controls (330.22,159) and (329.59,158.22) .. (329.59,157.27) -- cycle ;

% Text Node
\draw (323,60.4) node [anchor=north west][inner sep=0.75pt]    {$A$};
% Text Node
\draw (256,225.4) node [anchor=north west][inner sep=0.75pt]    {$A$};
% Text Node
\draw (386,222.4) node [anchor=north west][inner sep=0.75pt]    {$A$};
% Text Node
\draw (314,144.4) node [anchor=north west][inner sep=0.75pt]    {$\mu $};

\end{tikzpicture}
    \caption{Gauging $A$ involves choosing a point operator $\mu$ at a trivalent junction of $A$ lines.}
    \label{fig:point operator on 3A}
\end{figure}

\begin{figure}[h!]
    \centering

\tikzset{every picture/.style={line width=0.75pt}} %set default line width to 0.75pt        

\begin{tikzpicture}[x=0.75pt,y=0.75pt,yscale=-0.8,xscale=0.8]

%Straight Lines [id:da08636887019594053] 
\draw    (299.62,169.45) -- (265.96,200.07) ;
%Straight Lines [id:da2382307740137487] 
\draw    (299.62,169.45) -- (332.72,199.61) ;
%Straight Lines [id:da6391111084686333] 
\draw    (299.62,134.26) -- (299.62,169.45) ;
%Shape: Ellipse [id:dp5043570118221843] 
\draw  [color={rgb, 255:red, 0; green, 0; blue, 0 }  ,draw opacity=1 ][fill={rgb, 255:red, 0; green, 0; blue, 0 }  ,fill opacity=1 ] (298.83,169.57) .. controls (298.83,169.14) and (299.19,168.78) .. (299.62,168.78) .. controls (300.06,168.78) and (300.41,169.14) .. (300.41,169.57) .. controls (300.41,170.01) and (300.06,170.36) .. (299.62,170.36) .. controls (299.19,170.36) and (298.83,170.01) .. (298.83,169.57) -- cycle ;
%Straight Lines [id:da683556190047111] 
\draw    (299.57,134.71) -- (333.34,104.17) ;
%Straight Lines [id:da2018999954013032] 
\draw    (299.57,134.71) -- (266.58,104.47) ;
%Shape: Ellipse [id:dp5025719369535697] 
\draw  [color={rgb, 255:red, 0; green, 0; blue, 0 }  ,draw opacity=1 ][fill={rgb, 255:red, 0; green, 0; blue, 0 }  ,fill opacity=1 ] (298.78,134.59) .. controls (298.78,134.15) and (299.13,133.8) .. (299.57,133.8) .. controls (300,133.8) and (300.35,134.15) .. (300.35,134.59) .. controls (300.35,135.03) and (300,135.38) .. (299.57,135.38) .. controls (299.13,135.38) and (298.78,135.03) .. (298.78,134.59) -- cycle ;
%Straight Lines [id:da83922543203353] 
\draw    (69.51,150.98) -- (36.34,119.92) ;
%Straight Lines [id:da46126968774494237] 
\draw    (69.87,151.71) -- (36.78,181.88) ;
%Straight Lines [id:da730047385250135] 
\draw    (107.64,151.47) -- (69.28,151.24) ;
%Shape: Ellipse [id:dp42021589691018824] 
\draw  [color={rgb, 255:red, 0; green, 0; blue, 0 }  ,draw opacity=1 ][fill={rgb, 255:red, 0; green, 0; blue, 0 }  ,fill opacity=1 ] (68.52,150.07) .. controls (67.78,150.06) and (67.18,150.65) .. (67.17,151.38) .. controls (67.17,152.11) and (67.76,152.7) .. (68.5,152.7) .. controls (69.24,152.71) and (69.84,152.12) .. (69.85,151.39) .. controls (69.85,150.66) and (69.26,150.07) .. (68.52,150.07) -- cycle ;
%Straight Lines [id:da39474945863730515] 
\draw    (108.51,151.88) -- (141.59,183.04) ;
%Straight Lines [id:da7628000069594155] 
\draw    (107.38,151.15) -- (140.56,121.08) ;
%Shape: Ellipse [id:dp63673083946734] 
\draw  [color={rgb, 255:red, 0; green, 0; blue, 0 }  ,draw opacity=1 ][fill={rgb, 255:red, 0; green, 0; blue, 0 }  ,fill opacity=1 ] (107.52,150.43) .. controls (108,150.43) and (108.38,150.76) .. (108.38,151.16) .. controls (108.38,151.56) and (107.99,151.88) .. (107.51,151.88) .. controls (107.03,151.87) and (106.65,151.55) .. (106.65,151.15) .. controls (106.65,150.75) and (107.04,150.43) .. (107.52,150.43) -- cycle ;
%Straight Lines [id:da8473000523340479] 
\draw    (520,89) -- (520,128) ;
%Straight Lines [id:da17965151251453637] 
\draw    (520,178) -- (520,214) ;
%Shape: Circle [id:dp39460695817083014] 
\draw   (495,153) .. controls (495,139.19) and (506.19,128) .. (520,128) .. controls (533.81,128) and (545,139.19) .. (545,153) .. controls (545,166.81) and (533.81,178) .. (520,178) .. controls (506.19,178) and (495,166.81) .. (495,153) -- cycle ;
%Shape: Ellipse [id:dp2821359086745586] 
\draw  [color={rgb, 255:red, 0; green, 0; blue, 0 }  ,draw opacity=1 ][fill={rgb, 255:red, 0; green, 0; blue, 0 }  ,fill opacity=1 ] (520.14,127.27) .. controls (520.62,127.27) and (521,127.6) .. (521,128) .. controls (521,128.4) and (520.61,128.72) .. (520.13,128.72) .. controls (519.66,128.72) and (519.27,128.39) .. (519.27,127.99) .. controls (519.28,127.59) and (519.67,127.27) .. (520.14,127.27) -- cycle ;
%Shape: Ellipse [id:dp518300656492785] 
\draw  [color={rgb, 255:red, 0; green, 0; blue, 0 }  ,draw opacity=1 ][fill={rgb, 255:red, 0; green, 0; blue, 0 }  ,fill opacity=1 ] (520.01,177.55) .. controls (520.49,177.56) and (520.87,177.88) .. (520.87,178.28) .. controls (520.87,178.68) and (520.48,179) .. (520,179) .. controls (519.52,179) and (519.14,178.67) .. (519.14,178.27) .. controls (519.14,177.87) and (519.53,177.55) .. (520.01,177.55) -- cycle ;
%Straight Lines [id:da6424300925621167] 
\draw    (631,90) -- (631,216) ;
%Shape: Ellipse [id:dp12562760173113685] 
\draw  [color={rgb, 255:red, 0; green, 0; blue, 0 }  ,draw opacity=1 ][fill={rgb, 255:red, 0; green, 0; blue, 0 }  ,fill opacity=1 ] (108.25,150.15) .. controls (107.51,150.14) and (106.91,150.73) .. (106.91,151.46) .. controls (106.9,152.19) and (107.5,152.78) .. (108.23,152.78) .. controls (108.97,152.79) and (109.57,152.2) .. (109.58,151.47) .. controls (109.58,150.75) and (108.99,150.15) .. (108.25,150.15) -- cycle ;
%Shape: Ellipse [id:dp4166418908830992] 
\draw  [color={rgb, 255:red, 0; green, 0; blue, 0 }  ,draw opacity=1 ][fill={rgb, 255:red, 0; green, 0; blue, 0 }  ,fill opacity=1 ] (299.7,133.28) .. controls (298.96,133.28) and (298.36,133.86) .. (298.35,134.59) .. controls (298.35,135.32) and (298.94,135.91) .. (299.68,135.92) .. controls (300.42,135.92) and (301.02,135.33) .. (301.03,134.61) .. controls (301.03,133.88) and (300.44,133.28) .. (299.7,133.28) -- cycle ;
%Shape: Ellipse [id:dp08422951293681002] 
\draw  [color={rgb, 255:red, 0; green, 0; blue, 0 }  ,draw opacity=1 ][fill={rgb, 255:red, 0; green, 0; blue, 0 }  ,fill opacity=1 ] (520.01,126.69) .. controls (519.27,126.69) and (518.67,127.27) .. (518.67,128) .. controls (518.66,128.73) and (519.26,129.32) .. (519.99,129.33) .. controls (520.73,129.33) and (521.33,128.74) .. (521.34,128.02) .. controls (521.34,127.29) and (520.75,126.69) .. (520.01,126.69) -- cycle ;
%Shape: Ellipse [id:dp3428545076589905] 
\draw  [color={rgb, 255:red, 0; green, 0; blue, 0 }  ,draw opacity=1 ][fill={rgb, 255:red, 0; green, 0; blue, 0 }  ,fill opacity=1 ] (299.51,168.25) .. controls (298.77,168.24) and (298.17,168.83) .. (298.16,169.56) .. controls (298.16,170.28) and (298.75,170.88) .. (299.49,170.88) .. controls (300.23,170.89) and (300.83,170.3) .. (300.83,169.57) .. controls (300.84,168.84) and (300.24,168.25) .. (299.51,168.25) -- cycle ;
%Shape: Ellipse [id:dp15175888976783303] 
\draw  [color={rgb, 255:red, 0; green, 0; blue, 0 }  ,draw opacity=1 ][fill={rgb, 255:red, 0; green, 0; blue, 0 }  ,fill opacity=1 ] (520.01,176.68) .. controls (519.27,176.68) and (518.67,177.26) .. (518.66,177.99) .. controls (518.66,178.72) and (519.25,179.31) .. (519.99,179.32) .. controls (520.73,179.32) and (521.33,178.74) .. (521.34,178.01) .. controls (521.34,177.28) and (520.75,176.69) .. (520.01,176.68) -- cycle ;

% Text Node
\draw (328.62,87.92) node [anchor=north west][inner sep=0.75pt]    {$A$};
% Text Node
\draw (254.25,199.59) node [anchor=north west][inner sep=0.75pt]    {$A$};
% Text Node
\draw (324.18,200.21) node [anchor=north west][inner sep=0.75pt]    {$A$};
% Text Node
\draw (254.06,86.38) node [anchor=north west][inner sep=0.75pt]    {$A$};
% Text Node
\draw (138.61,103.72) node [anchor=north west][inner sep=0.75pt]    {$A$};
% Text Node
\draw (137.07,177.78) node [anchor=north west][inner sep=0.75pt]    {$A$};
% Text Node
\draw (22.45,178.26) node [anchor=north west][inner sep=0.75pt]    {$A$};
% Text Node
\draw (22.73,102.79) node [anchor=north west][inner sep=0.75pt]    {$A$};
% Text Node
\draw (502.62,206.92) node [anchor=north west][inner sep=0.75pt]    {$A$};
% Text Node
\draw (501.62,86.92) node [anchor=north west][inner sep=0.75pt]    {$A$};
% Text Node
\draw (612.62,86.92) node [anchor=north west][inner sep=0.75pt]    {$A$};
% Text Node
\draw (180,142.4) node [anchor=north west][inner sep=0.75pt]    {$\longleftrightarrow $};
% Text Node
\draw (561,145.4) node [anchor=north west][inner sep=0.75pt]    {$\longleftrightarrow $};

\end{tikzpicture}
    \caption{Fig: Conditions on $\mu$ for the gauging to be consistent.}
    \label{fig:conditions on mu}
\end{figure}
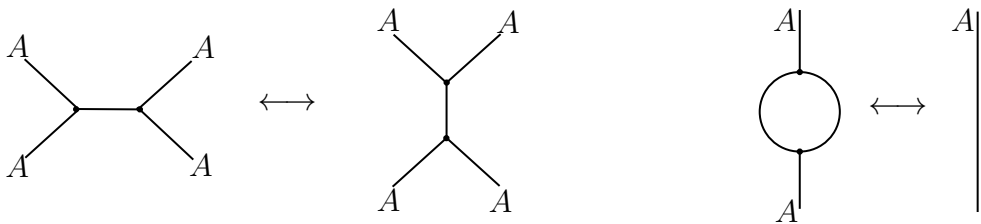

More generally, if a 1+1D QFT has symmetries described by a fusion category $\CC$, then the appropriate generalization of ``non-anomalous subgroup" is a line operator $A$ (generically non-simple) which admits the structure of an algebra \cite{Frohlich:2009gb,Bhardwaj:2017xup}. This constraint is an algebraic generalization of the anomalies for invertible symmetries. For a 1+1D QFT on a spacetime 2-manfiold $\Sigma$ gauging a line operator $A$ involves constructing a mesh of $A$ lines on $\Sigma$. Since this mesh involves trivalent junctions, we have to make a choice of a point operator at a trivalent junction of $A$ lines (see Fig. \ref{fig:point operator on 3A}).
We want to choose $\mu$ such that the resulting theory is independent of the choice of the mesh. This is guaranteed if the conditions in Fig. \ref{fig:conditions on mu} are satisfied.
If there is at least one $\mu$ which satisfies these constraints, we can regard $A$ as a non-anomalous line operator as it can be consistently gauged. If there are multiple $\mu$ which satisfy these constraints, then the choice of $\mu$ can be thought of as a choice of ``generalized discrete torsion" on $A$. In this case the object $A$ admits an algebra structure provided by $\mu:A\times A \rightarrow A$. Once again, gauging two algebras $A_1$ and $A_2$ might be physically equivalent. In fact, in some cases gauging $A$ might be a self-duality of the QFT. For example, in a 1+1D QFT with symmetries described by the Ising fusion category, gauging $1+\psi$, where $\psi$ is the non-trivial order two line is a self-duality \cite{Frohlich:2006ch,Chang:2018iay}. While gauging $1+\psi$ produces the same 1+1 QFT, it is not a trivial operation. Indeed, gauging $1+\psi$ in the Ising CFT is the famous Krammers-Wannier duality which non-trivially exchanges local operators with twisted-sector operators \cite{PhysRev.60.252,Frohlich:2006ch,Chang:2018iay}. Therefore, it is important to keep track of physically equivalent gaugings. In this general setting, it is natural to ask:
\vspace{0.2cm}

\textit{Given a gapped boundary $\CB$ of the SymTFT, $\CZ(\CC)$, of a fusion category $\CC$ how do we determine the corresponding equivalence class of physically equivalent gaugings in $\CC$?}
\vspace{0.2cm}

\begin{figure}[h!]
    \centering

\tikzset{every picture/.style={line width=0.75pt}} %set default line width to 0.75pt        

\begin{tikzpicture}[x=0.75pt,y=0.75pt,yscale=-1,xscale=1]
%uncomment if require: \path (0,330); %set diagram left start at 0, and has height of 330

%Shape: Parallelogram [id:dp9637167781997072] 
\draw  [color={rgb, 255:red, 0; green, 0; blue, 0 }  ,draw opacity=1 ][fill={rgb:red, 74; green, 74; blue, 74 }  ,fill opacity=0.38 ] (355.15,233.1) -- (355.17,115.24) -- (432,94) -- (431.98,211.86) -- cycle ;
%Straight Lines [id:da4704014722379637] 
\draw [color={rgb, 255:red, 139; green, 6; blue, 24 }  ,draw opacity=1 ]   (261.32,170.33) -- (392.06,168.67) ;
%Straight Lines [id:da3972251383566747] 
\draw    (392.06,168.67) -- (433,153) ;
%Straight Lines [id:da13554412005424887] 
\draw  [dash pattern={on 4.5pt off 4.5pt}]  (256.64,117.14) -- (358.17,116.25) ;
%Straight Lines [id:da8708932189608322] 
\draw  [dash pattern={on 4.5pt off 4.5pt}]  (316.14,94.24) -- (434.66,93.33) ;
%Straight Lines [id:da11782865805110054] 
\draw  [dash pattern={on 4.5pt off 4.5pt}]  (313.45,212.76) -- (431.98,211.78) ;
%Straight Lines [id:da4553859030359517] 
\draw  [dash pattern={on 4.5pt off 4.5pt}]  (248.62,233.07) -- (355.15,232.1) ;
%Shape: Ellipse [id:dp7814451638776253] 
\draw  [color={rgb, 255:red, 139; green, 6; blue, 24 }  ,draw opacity=1 ][fill={rgb, 255:red, 139; green, 6; blue, 24 }  ,fill opacity=1 ] (390.65,168.4) .. controls (390.65,167.44) and (391.28,166.67) .. (392.06,166.67) .. controls (392.83,166.67) and (393.46,167.44) .. (393.46,168.4) .. controls (393.46,169.36) and (392.83,170.13) .. (392.06,170.13) .. controls (391.28,170.13) and (390.65,169.36) .. (390.65,168.4) -- cycle ;

% Text Node
\draw (361.17,118.65) node [anchor=north west][inner sep=0.75pt]  [font=\small]  {$\CB_{\CC}$};
% Text Node
\draw (323.5,150.55) node [anchor=north west][inner sep=0.75pt]    {$x$};
% Text Node
\draw (395.06,135.07) node [anchor=north west][inner sep=0.75pt]    {$F( x)$};
% Text Node
\draw (256,130.4) node [anchor=north west][inner sep=0.75pt]    {$\CZ(\CC)$};

\end{tikzpicture}
    \caption{The bulk-to-boundary map $F$ determines the outcome of fusing a bulk line operator on the boundary $\CB_{\CC}$.}
    \label{fig:general F}
\end{figure}
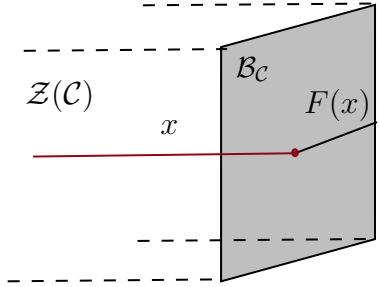

In this paper, we answer this question by studying the bulk-to-boundary map which tells us how to map line operators in $\CZ(\CC)$ to line operators in $\CC$, $F:\CZ(\CC)\to \CC$. Physically, $F(x)$ for some line operator $x\in \CC$ is the result of perpendicular fusion of $x$ on the gapped boundary $\CB_{\CC}$ of $\CZ(\CC)$ corresponding to the fusion category $\CC$ (see Fig. \ref{fig:general F}). We also study a boundary-to-bulk map $K: \CC\to \CZ(\CC)$ which can be used to determine the gapped boundary of $\CZ(\CC)$ corresponding to a non-anomalous line operator in $\CC$. 

A gapped boundary of $\CZ(\CC)$ is almost completely determined by the line operators which can end on it. In fact, the lines which can end on a gapped boundary form a Lagrangian algebra $L$ \cite{Kapustin:2010hk,kitaev2012models,kong2014anyon}.  Conversely, the Lagrangian algebra $L$ completely determines the gapped boundary. Ending the full Lagrangian algebra object $L$ on the gapped boundary $\CB_{\CC}$, produces  certain line operator $F(L) \in \CC$. We show that 
\begin{itemize}
    \item $F(L)$ is a non-anomalous line operator in $\CC$. That is, it can be gauged. (Theorem \ref{th: F(L)})
    \item The equivalence class of non-anomalous line operators whose gauging is physically equivalent is contained in $F(L)$. (Theorem \ref{th: F(L)})
    \item The multiplication on an algebra $A$ in $\CC$ can be determined from the multiplication on the corresponding Lagrangian algebra $L$ through an explicit formula. (Equation \eqref{eq:relating multiplications}) 
\end{itemize}
This has various applications:
\begin{itemize}
\item If all Lagrangian algebras of a SymTFT are known, then the bulk-to-boundary map $F$ determines all algebras in the fusion category describing the line operators on the chosen boundary. This is an alternative to using NIM-reps to classify algebra objects in a fusion category \cite{grossman2012quantum,grossman2016brauer,Komargodski:2020mxz,Diatlyk:2023fwf}. 

\item The map between Lagrangian algebras in the SymTFT $\CZ(\CC)$ and non-anomalous line operators in $\CC$ can be used to determine necessary conditions for a line operator in $\CC$ to be non-anomalous. We use this condition to show that $\CC$-symmetric trivially gapped phases exist if and only if $\CZ(\CC)$ contains a magnetic Lagrangian algebra with respect to $\CB_{\CC}$. This relation was first established in \cite{Zhang:2023wlu} through interval compactification of the SymTFT. 

\item We show that in many cases, the explicit structure of $F(L)$ also allows us to classify line operators in $\CZ(\CC)$ admitting the structure of a Lagrangian algebra in terms of algebras in $\CC$. We use this to define the notion of transporting non-anomalous line operators from one fusion category to another which share the same SymTFT. 

\item When $\CC$ is a modular tensor category, both a Lagrangian algebra $L$ in the SymTFT $\CZ(\CC)$ and $F(L)$ have a purely bulk interpretation. In this case, $L$ determines the action of a topological surface operator, say $S_L$, on the line operators of $\CC$. Then $F(L)$ is the equivalence class of algebras which can be higher-gauged to construct the surface $S_L$.  
\end{itemize}

The plan of the paper is as follows. In section \ref{sec:review} we will briefly review non-invertible symmetries in 1+1D and their gauging. In section \ref{sec:SymTFT and C-symmetric} we will introduce 2+1D SymTFTs $\CZ(\CC)$ and describe their gapped boundaries in terms of Lagrangian algebras. We will review $\CC$-symmetric TQFTs and explain the relation between 1D gapped boundaries of a $\CC$-symmetric TQFT and gapped interfaces between gapped boundaries of $\CZ(\CC)$. In section \ref{sec: gaugeable lines from SymTFT} we will explain the explicit map between non-anomalous line operators in $\CC$ and gapped-boundaries $\CZ(\CC)$. We begin by proving Theorem \ref{th: F(L)}. Then we provide an explicit formula relating the multiplication on a Lagrangian algebra of $\CZ(\CC)$ and the generalized discrete torsion of the corresponding non-anomalous line operator in $\CC$. We then study a boundary-to-bulk map $K$ and give a physical picture of how it can be used to determine the Lagrangian algebra corresponding to an algebra in the boundary. Finally, we will discuss the special case of invertible symmetries in detail and prove Theorem \ref{th: non-anomalous}. In section \ref{sec:Examples} we discuss various examples involving invertible and non-invertible symmetries. We end this section with an example illustrating how Theorem \ref{th: F(L)} can be used to find Lagrangian algebra objects in $\CZ(\CC)$ using non-anomalous line operators in $\CC$. Finally, in section \ref{sec:Applications} we discuss various applications of our results. We introduce the notion of ``transporting algebra objects" between fusion categories and illustrate it using the Haagerup fusion categories. We conclude with some interesting open problems.

\section{Review: Non-invertible symmetries in 1+1D and their gauging}

\label{sec:review}

In this section, we will briefly review non-invertible symmetries of 1+1D QFTs and their gauging. We refer the reader to the excellent exposition of this topic in the references \cite[Section 5]{Huang:2021zvu}, \cite{Choi:2023xjw} and \cite{Diatlyk:2023fwf} for more details. 

Symmetries of a 1+1D QFT $\CQ$ are implemented by topological line operators\footnote{We assume that $\CQ$ has a unique vacuum. In this case, there are no non-trivial topological local operators.}. The structure of a finite set of line operators is described by a fusion category $\CC$.\footnote{See the reviews \cite{Schafer-Nameki:2023jdn,Brennan:2023mmt,Shao:2023gho}.} The simple objects in the category $a,b,c,\dots$ label simple line operators of the QFT.\footnote{A line operator is simple if it cannot be decomposed into a sum of other line operators. Equivalently, the only point operator that a simple line operator hosts on it is the identity point operator and its complex multiples.} Their fusion rules are captured by the non-negative integers $N_{ab}^c$ as follows.
\be
a \times b =\sum_c N_{ab}^c ~c~.
\ee
The fusion of four line operators obeys the Pentagon equations whose solutions are given in terms of the $F$ matrices. 

A line operator $A$ in $\CC$ is \textit{non-anomalous} if it can be gauged to produce a new QFT $\CQ/A$. For this gauging to be consistent, the line operator $A$ must admit the structure of a symmetric separable Frobenius algebra \cite{Frohlich:2009gb,Bhardwaj:2017xup}. To explain this structure, suppose $A$ admits the decomposition
\be
A=\sum_a N_{A}^a ~a~,
\ee
into simple line operators $a\in \CC$, where $N_A^{a}$ are non-negative integers. The object $A$ admits the structure of an associative algebra if there exist complex numbers
\be
\mu_{(a,\alpha),(b,\beta)}^{(c,\gamma),\delta}~,
\ee
where $\alpha,\beta,\gamma,\delta$ denote indices running from $1,\dots,N_A^a~$;$~1,\dots, N_A^b~$;$~1,\dots,N_A^c$ and $1,\dots, N_{ab}^c$, respectively. These complex numbers should satisfy the following constraint involving the $F$-symbol:
\be
\label{eq: m associativity}
\mu_{ab}^c\mu_{de}^f= \mu_{ag}^f \mu_{be}^g F_{abe}^f~.
\ee
In writing the above constraint, we have assumed that both the fusion coefficients and the algebra $A$ are multiplicity-free. If there is multiplicity, extra indices need to be added. We will assume that $N_A^{\trl}=1$, where $\trl$ is the trivial line operator. In this case, $A$ is called a haploid algebra. A haploid algebra admits a unique symmetric separable algebra structure (see \cite[Footnote 20]{Choi:2023xjw} and references therein). Therefore, we will not discuss these extra conditions in this review. 

Note that a line operator $A$ might admit distinct algebra structures. In other words, there maybe more than one inequivalent solution to the constraints \eqref{eq: m associativity}. We will call this a choice of \textit{generalized discrete torsion} for gauging $A$. This is a generalization of the fact that when a non-anomalous line operator $A$ is of the form 
\be
A=\sum_{g \in G} g~,
\ee
for some group $G$, then the distinct multiplications on $A$ are classified by the discrete torsion $H^2(G,U(1))$.

Two distinct algebras $A_1$ and $A_2$ might correspond to physically equivalent gauging procedures leading to dual QFTs $\CQ/A_1$ and $\CQ/A_2$. This is captured by Morita equivalence of algebras.\footnote{See \cite{Diatlyk:2023fwf} for more details.} We will denote a Morita equivalence class with a representative algebra $A$ as $[A]$. Note that every Morita equivalent class has a haploid representative. Therefore, the condition $N_{A}^{\trl}=1$ can be imposed without loss of generality. In the following sections, we will use the terms ``non-anomalous line operator" and ``algebra object" interchangeably.

\section{SymTFTs and $\CC$-symmetric TQFTs}

\label{sec:SymTFT and C-symmetric}

\subsection{SymTFTs}

\begin{figure}[h!]
    \centering

\tikzset{every picture/.style={line width=0.75pt}} %set default line width to 0.75pt        

\begin{tikzpicture}[x=0.75pt,y=0.75pt,yscale=-1,xscale=1]
%uncomment if require: \path (0,204); %set diagram left start at 0, and has height of 204

%Shape: Parallelogram [id:dp4935765422656929] 
\draw  [color={rgb, 255:red, 0; green, 0; blue, 0 }  ,draw opacity=1 ][fill={rgb, 255:red, 74; green, 74; blue, 74 }  ,fill opacity=0.38 ] (355.15,170.1) -- (355.17,52.24) -- (432,31) -- (431.98,148.86) -- cycle ;
%Shape: Parallelogram [id:dp5398824968454786] 
\draw  [color={rgb, 255:red, 0; green, 0; blue, 0 }  ,draw opacity=1 ][fill={rgb, 255:red, 74; green, 74; blue, 74 }  ,fill opacity=0.38 ] (236.63,171) -- (236.65,53.14) -- (313.47,31.9) -- (313.45,149.76) -- cycle ;
%Straight Lines [id:da09307695617394862] 
\draw  [dash pattern={on 4.5pt off 4.5pt}]  (236.64,53.14) -- (355.17,52.25) ;
%Straight Lines [id:da17050390160019235] 
\draw  [dash pattern={on 4.5pt off 4.5pt}]  (313.47,31.9) -- (432,31) ;
%Straight Lines [id:da17181517829214366] 
\draw  [dash pattern={on 4.5pt off 4.5pt}]  (313.45,149.76) -- (431.98,148.78) ;
%Straight Lines [id:da3912848741681707] 
\draw  [dash pattern={on 4.5pt off 4.5pt}]  (236.62,171.07) -- (355.15,170.1) ;

% Text Node
\draw (242.52,59.01) node [anchor=north west][inner sep=0.75pt]  [font=\small]  {$\mathcal{B}_{\mathcal{Q}}$};
% Text Node
\draw (359.41,56.01) node [anchor=north west][inner sep=0.75pt]  [font=\small]  {$\mathcal{B}_{\CC}$};
% Text Node
\draw (315,90.4) node [anchor=north west][inner sep=0.75pt]    {$\CZ( \CC)$};

\end{tikzpicture}
    \caption{The SymTFT $\CZ(\CC)$ with two boundary conditions. The boundary $\CB_{\CC}$ is gapped while the $\CB_{\CQ}$ is generically gappless. The category of line operators on  $\CB_{\CC}$ is $\CC$.}
    \label{fig:SymTFT}
\end{figure}
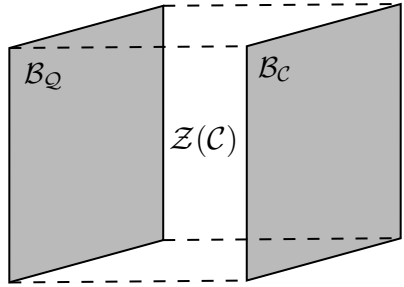

Let $\CC$ be a finite subcategory of symmetries of a 1+1D QFT $\CQ$. The SymTFT of $\CC$ is the Turaev-Viro-Barrett-Westbury 2+1D TQFT described by the Drinfeld centre $\CZ(\CC)$ of $\CC$ \cite{Lan:2018vjb,Lan:2018bui,Johnson-Freyd:2020usu,Kong:2020jne,Johnson-Freyd:2020twl,Johnson-Freyd:2021tbq,Ng:2023wsc,Turaev:1992hq,Barrett:1993ab}. The simple line operators in $\CZ(\CC)$ can be written in the form
\be
(a,e_a)~,
\ee
where $a$ is a (generically non-simple) line operator in $\CC$ and $e_a$ are isomorphisms called half-braidings
\be
e_a(b): a\times b \xrightarrow{\sim} b \times a  ~,\; \forall ~ b \in \CC~,
\ee
satisfying several consistency conditions (see, for example, \cite{muger2003subfactors,etingof2016tensor}). The SymTFT allows us to separate the data of the symmetry $\CC$ from the 1+1D QFT $\CQ$ on which it is acting (see Fig. \ref{fig:SymTFT}).

Given the full data of the fusion category $\CC$, the line operators and modular data of $\CZ(\CC)$ can be computed using the string-net model \cite{Levin:2004mi,Lin:2020bak}. We refer the readers to the review in \cite[Appendix A]{Cordova:2023bja} for more details. 

\subsubsection{Lagrangian algebras and gapped boundaries}

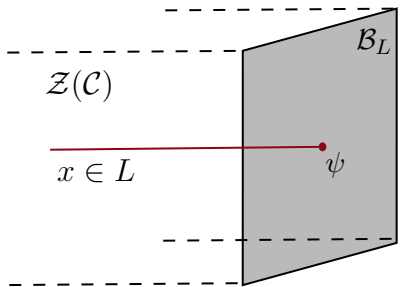
\begin{figure}[h!]
    \centering

\tikzset{every picture/.style={line width=0.75pt}} %set default line width to 0.75pt        

\begin{tikzpicture}[x=0.75pt,y=0.75pt,yscale=-1,xscale=1]
%uncomment if require: \path (0,330); %set diagram left start at 0, and has height of 330

%Shape: Parallelogram [id:dp84461910785525] 
\draw  [color={rgb, 255:red, 0; green, 0; blue, 0 }  ,draw opacity=1 ][fill={rgb, 255:red, 74; green, 74; blue, 74 }  ,fill opacity=0.38 ] (351.15,246.1) -- (351.17,128.24) -- (428,107) -- (427.98,224.86) -- cycle ;
%Straight Lines [id:da714765442747638] 
\draw [color={rgb, 255:red, 139; green, 6; blue, 24 }  ,draw opacity=1 ]   (255,178) -- (389.58,176.55) ;
%Shape: Ellipse [id:dp06189842406096169] 
\draw  [color={rgb, 255:red, 139; green, 6; blue, 24 }  ,draw opacity=1 ][fill={rgb, 255:red, 139; green, 6; blue, 24 }  ,fill opacity=1 ] (389.58,176.55) .. controls (389.58,175.59) and (390.21,174.82) .. (390.98,174.82) .. controls (391.76,174.82) and (392.39,175.59) .. (392.39,176.55) .. controls (392.39,177.51) and (391.76,178.28) .. (390.98,178.28) .. controls (390.21,178.28) and (389.58,177.51) .. (389.58,176.55) -- cycle ;
%Straight Lines [id:da9630485063154574] 
\draw  [dash pattern={on 4.5pt off 4.5pt}]  (233.64,129.14) -- (352.17,128.25) ;
%Straight Lines [id:da4046196775672194] 
\draw  [dash pattern={on 4.5pt off 4.5pt}]  (312.47,107.9) -- (431,107) ;
%Straight Lines [id:da09607122408834579] 
\draw  [dash pattern={on 4.5pt off 4.5pt}]  (311.45,223.76) -- (429.98,222.78) ;
%Straight Lines [id:da8486075282162301] 
\draw  [dash pattern={on 4.5pt off 4.5pt}]  (234.62,246.07) -- (353.15,245.1) ;

% Text Node
\draw (406.41,116.01) node [anchor=north west][inner sep=0.75pt]  [font=\small]  {$\mathcal{B}_{L}$};
% Text Node
\draw (251,136.4) node [anchor=north west][inner sep=0.75pt]    {$\CZ(\CC)$};
% Text Node
\draw (257,181.4) node [anchor=north west][inner sep=0.75pt]    {$x\in L$};
% Text Node
\draw (391,177.4) node [anchor=north west][inner sep=0.75pt]  [font=\small]  {$\psi $};

\end{tikzpicture}
    \caption{A simple line operator $x \in L$ can end on the gapped boundary $\CB_{L}$.}
    \label{fig:gapped boundary}
\end{figure}

A gapped boundary of the SymTFT $\CZ(\CC)$ is completely determined by a Lagrangian algebra $L$ \cite{Kapustin:2010hk,kitaev2012models,kong2014anyon}. $L$ is a line operator in $\CZ(\CC)$ and can be written as 
\be
L=\sum_x N_L^x ~ x~,
 \ee
where $x$ are simple line operators and $N_L^x$ are non-negative integers. If $N_L^x\neq 0$, then $x$ can end on the gapped boundary $\CB_{L}$ (see Fig. \ref{fig:gapped boundary}).

We will focus on simple gapped boundaries on which the only non-trivial point operator is the identity operator and its complex multiples.\footnote{A general gapped boundary can be written as a sum of simple gapped boundaries.} Therefore, the trivial line operator in the bulk should end on the gapped boundary in a unique way. In other words, we require that $L$ has exactly one copy of the trivial line,  $N_L^{\trl}=1$, where $\trl$ is the trivial line operator. The line operator $L$ admits the structure of a commutative and associative algebra. This algebra structure is determined by the following complex numbers
\be
m_{(x,i),(y,j)}^{(z,k),l}
\ee
where $i,j,k,l$ denote indices running from $1,\dots,N_L^x~$;$~1,\dots, N_L^y~$;$~1,\dots,N_L^z$ and $1,\dots, N_{xy}^z$, respectively. When the SymTFT $\CZ(\CC)$ has fusion coefficients valued in $N_{xy}^{z} \in \{0,1\}$, we will denote the multiplication coefficients in the algebra $L$ as $m_{(x,i),(y,j)}^{(z,k)}$. Moreover, if $N_L^x\in \{0,1\}$,  we will denote it as $m_{xy}^z$. 
Since $L$ is an associative algebra. The complex numbers $m_{xy}^z$ must satisfy the constraint \eqref{eq: m associativity}. Also, for $L$ to be a commutative algebra, these complex numbers must satisfy the following constraint involving the $R$-symbol (braiding):
\be
\label{eq:m commutativity}
m_{xy}^z= m_{yx}^z R_{xy}^z~.
\ee
In writing this equation, we have assumed that $\CC$ and $A$ are multiplicity-free. Moreover, for $L$ to be a Lagrangian algebra, the non-negative integers $N_L^a$ must satisfy
\be
\label{eq:Lagrangian algebra constraints}
N_L^x~ N_L^y \leq \sum_z N_{xy}^z~ N_L^z, ~~ \text{and} ~~ \text{dim}(L)=\sum_{x} N_L^x d_x = \sqrt{\text{dim}(\CZ(\CC))}= \text{dim}(\CC)~.
\ee
A line operator $L$ with multiplication specified by the set of complex numbers $m_{(x,i),(y,j)}^{(z,k),l}$ is a Lagrangian algebra if and only if the constraints \eqref{eq: m associativity}, \eqref{eq:m commutativity} and \eqref{eq:Lagrangian algebra constraints} are satisfied \cite[Corollary 3.8]{Cong:2017ffh}.\footnote{Strictly speaking, this result requires unitarity. We will assume that the fusion category is unitary throughout this paper.} Note that the conditions \eqref{eq:Lagrangian algebra constraints} depend only on the fusion rules of $\CZ(\CC)$ and a choice of the line operator $L$. However, they are not a set of sufficient conditions for $L$ to be Lagrangian \cite{davydov2016unphysical}. The constraints \eqref{eq:Lagrangian algebra constraints} can be used to enumerate a set of candidate Lagrangian algebras for which we can try to solve \eqref{eq: m associativity}, \eqref{eq:Lagrangian algebra constraints}. However, this requires the knowledge of the full $F$ and $R$ matrices of the SymTFT $\CZ(\CC)$. In sections \ref{sec:D8 example} and \ref{sec:Applications}, as a consequence of our main result, we will consider an alternate method to determine line operators in the bulk SymTFT which are guaranteed to admit a Lagrangian algebra structure using non-anomalous line operators on a gapped boundary.

\subsection{$\CC$-symmetric TQFTs}
\label{sec:Csym TQFTs}

In this section, we will briefly review the classification of $1+1$D $\CC$-symmetric TQFTs in terms of module categories over $\CC$. The topological local operators in a 1+1D TQFT form an algebra, $E$, where the multiplication on $E$ captures the O.P.E of the local operators. Moreover, the 2-point function of the local operators defines a commutative Frobenius algebra structure on $E$ \cite{Moore:2006dw}. Arbitrary correlation functions of local operators on any closed 2-manifold can be determined using $E$. The topological line operators in the TQFT are labeled by bimodules of the Frobenius algebra, BiMod$(E)$ (for example, see \cite[Section 2.4.1]{Carqueville:2016nqk}).\footnote{BiMod$(E)$ is a multifusion category \cite{etingof2005fusion} which captures the structure of both line operators and local operators of the TQFT} 

\begin{figure}[h!]
    \centering

\tikzset{every picture/.style={line width=0.75pt}} %set default line width to 0.75pt        

\begin{tikzpicture}[x=0.75pt,y=0.75pt,yscale=-1,xscale=1]

%Straight Lines [id:da4872597782772172] 
\draw [color={rgb, 255:red, 74; green, 144; blue, 226 }  ,draw opacity=1 ]   (319,57) -- (319,185) ;
%Straight Lines [id:da013728570424669462] 
\draw    (318.5,121) -- (262,181.86) ;
%Shape: Ellipse [id:dp21668372059990015] 
\draw  [color={rgb, 255:red, 74; green, 144; blue, 226 }  ,draw opacity=1 ][fill={rgb, 255:red, 74; green, 144; blue, 226 }  ,fill opacity=1 ] (317,121) .. controls (317,119.77) and (317.88,118.77) .. (318.96,118.77) .. controls (320.04,118.77) and (320.92,119.77) .. (320.92,121) .. controls (320.92,122.23) and (320.04,123.23) .. (318.96,123.23) .. controls (317.88,123.23) and (317,122.23) .. (317,121) -- cycle ;

% Text Node
\draw (274,139.71) node [anchor=north west][inner sep=0.75pt]    {$c$};
% Text Node
\draw (329,148.92) node [anchor=north west][inner sep=0.75pt]    {$\cm_i$};
% Text Node
\draw (330,67.46) node [anchor=north west][inner sep=0.75pt]    {$\cm_j$};
% Text Node
\draw (325,110.4) node [anchor=north west][inner sep=0.75pt]    {$\alpha $};

\end{tikzpicture}
    \caption{Action of $\CC$ on 1D gapped boundaries.}
    \label{fig:symmetry action on gapped boundaries}
\end{figure}
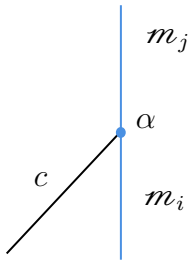

A $\CC$-symmetric TQFT is determined by a functor
\be
\CC \to \text{BiMod}(E)~.
\ee
which assigns line operators in $\CC$ to line operators in BiMod$(E)$ in a consistent way. In other words, we must be able to identify line operators in $\text{BiMod}(E)$ which form the chosen fusion category $\CC$. Let $\CM$ be the category of boundary conditions of a 1+1D TQFT with simple objects $\{\cm_1,\dots \cm_n\} \in \CM$ labeling the distinct boundary conditions. Then there is an action of the fusion category $\CC$ on $\CM$ given by 
\be
c \times \cm_i = \sum_{j} N_{ci}^j ~ \cm_j ~.
\ee
where $N_{ci}^j$ are non-negative integers. If $N_{ci}^j \neq 0$, we get Fig. \ref{fig:symmetry action on gapped boundaries} where $\alpha$ labels a basis for the local operators at the junction of $c,\cm_i$ and $\cm_j$.

It is known that the full defining data of a 1+1D $\CC$-symmetric TQFT can be obtained from how $\CC$ acts on the set of boundary conditions \cite{Moore:2006dw,Thorngren:2019iar,Komargodski:2020mxz,Huang:2021zvu}. In fact, $\CM$ is a module category over $\CC$. All physical data in the $\CC$-symmetric TQFT can be computed using the data of this module category. 
\be
\CC\text{-module category }\leftrightarrow \text{1+1D }\CC\text{-symmetric TQFT}~.  
\ee
A $\CC$-module category is in turn completely determined by an algebra $A$ in $\CC$ \cite{ostrik2003module} (see also \cite{Choi:2023xjw,Diatlyk:2023fwf}). 
\be
\text{algebra $A$ in $\CC$ } \longleftrightarrow \CC\text{-module category } \CM\longleftrightarrow 1+1\text{D } \CC\text{-symmetric TQFT}~.
\ee
In other words, every gaugeable line operator $A$ in $\CC$ determines a $\CC$-module category which in turn determines a 1+1D $\CC$-symmetric TQFT. Given an algebra $A$, the module category $\CM$ is the category of modules of $A$ in $\CC$. Note that the module category only depends on the Morita equivalence class $[A]$ of the algebra $A$. Conversely, given a module category $\CM$, $[A]$ can be determined as follows. Choose some $\cm \in \CM$, then a line operator $a$  can form a junction on the gapped boundary $\cm$ if and only if $a\times \cm$ contains $\cm$. Let us define the non-simple line operator
\be
\label{eq:Am definition}
A_{\cm}:=\sum_{a\in \CC} N_{\cm}^a ~ a~,
\ee
where $N_{\cm}^a$ is the dimension of Hilbert space of operators at the junction of $a$ with the gapped boundary $\cm$. The object $A_{\cm}$ is called the Internal Hom, and it admits the structure of a haploid symmetric separable Frobenius algebra \cite{ostrik2003module}. In other words, $A_{\cm}$ is non-anomalous and can be gauged. If we choose a different boundary condition in $\CM$, we get a Morita equivalent haploid symmetric separable Frobenius algebra.

\subsubsection{$\CC$-symmetric TQFTs from SymTFTs}

In this section, we will review the construction of 1+1D $\CC$-symmetric TQFTs from the SymTFT $\CZ(\CC)$. Most of this section is based on the papers \cite{Zhang:2023wlu,Bhardwaj:2023idu,Bhardwaj:2023fca,Bhardwaj:2023bbf}, except for the last part where we explain how to obtain the gapped boundaries of the 1+1D TQFT from the SymTFT. 

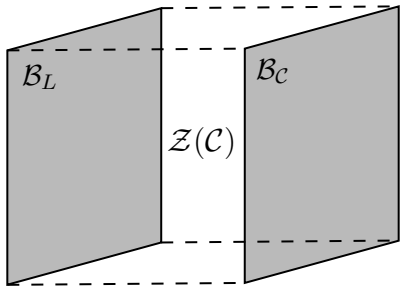
\begin{figure}[h!]
    \centering

\tikzset{every picture/.style={line width=0.75pt}} %set default line width to 0.75pt        

\begin{tikzpicture}[x=0.75pt,y=0.75pt,yscale=-1,xscale=1]
%uncomment if require: \path (0,204); %set diagram left start at 0, and has height of 204

%Shape: Parallelogram [id:dp4935765422656929] 
\draw  [color={rgb, 255:red, 0; green, 0; blue, 0 }  ,draw opacity=1 ][fill={rgb, 255:red, 74; green, 74; blue, 74 }  ,fill opacity=0.38 ] (355.15,170.1) -- (355.17,52.24) -- (432,31) -- (431.98,148.86) -- cycle ;
%Shape: Parallelogram [id:dp5398824968454786] 
\draw  [color={rgb, 255:red, 0; green, 0; blue, 0 }  ,draw opacity=1 ][fill={rgb, 255:red, 74; green, 74; blue, 74 }  ,fill opacity=0.38 ] (236.63,171) -- (236.65,53.14) -- (313.47,31.9) -- (313.45,149.76) -- cycle ;
%Straight Lines [id:da09307695617394862] 
\draw  [dash pattern={on 4.5pt off 4.5pt}]  (236.64,53.14) -- (355.17,52.25) ;
%Straight Lines [id:da17050390160019235] 
\draw  [dash pattern={on 4.5pt off 4.5pt}]  (313.47,31.9) -- (432,31) ;
%Straight Lines [id:da17181517829214366] 
\draw  [dash pattern={on 4.5pt off 4.5pt}]  (313.45,149.76) -- (431.98,148.78) ;
%Straight Lines [id:da3912848741681707] 
\draw  [dash pattern={on 4.5pt off 4.5pt}]  (236.62,171.07) -- (355.15,170.1) ;

% Text Node
\draw (242.52,59.01) node [anchor=north west][inner sep=0.75pt]  [font=\small]  {$\mathcal{B}_{L}$};
% Text Node
\draw (359.41,56.01) node [anchor=north west][inner sep=0.75pt]  [font=\small]  {$\mathcal{B}_{\CC}$};
% Text Node
\draw (315,90.4) node [anchor=north west][inner sep=0.75pt]    {$\CZ( \CC)$};

\end{tikzpicture}
    \caption{SymTFT with two gapped boundaries $\CB_{L}$ and $\CB_{\CC}$ can be compactified to get a $\CC$-symmetric TQFT.}
    \label{fig:sandwich picture}
\end{figure}

Consider the sandwich picture in Fig. \ref{fig:sandwich picture}. Interval compactification of this picture results in a 1+1D TQFT $\CT_L$ with $\CC$-symmetry.\footnote{Similar procedures can be used to obtain TQFTs in higher-dimensions \cite{Cordova:2023bja,Antinucci:2023ezl}. For example, interval compactification of the Crane-Yetter TQFT with two fixed gapped boundaries  results in a 2+1D TQFT. This played a crucial role in proving that all Spin TQFTs admit a modular extension \cite{johnson2024minimal}. See also \cite{Nardoni:2024sos}.} When $\CB_{L}=\CB_{\CC}$, we get the regular $\CC$-symmetric TQFT defined in \cite{Huang:2021zvu} in which the symmetry $\CC$ is completely spontaneously broken. For other choices of $\CB_{L}$ we may get TQFTs where $\CC$ is partially or fully preserved. 

Various crucial properties of the TQFT can be deduced from this picture \cite{Zhang:2023wlu,Bhardwaj:2023idu,Bhardwaj:2023fca}. For example, the number of local operators in the resulting TQFT can be counted by looking at the line operators in $\CZ(\CC)$ which can end on both $\CB_{L}$ and $\CB_{\CC}$. More precisely, the number of local operators in the 1+1D TQFT is given by
\be
\label{eq:number of local operators}
\sum_{x\in \CZ(\CC)} N_{\CC}^x ~ N^x_{L}~.
\ee
where $N^{x}_{\CC}$ counts the number of copies of $x$ in the canonical Lagrangian algebra of $\CZ(\CC)$ corresponding to the gapped boundary $\CB_{\CC}$. 
Moreover, further details of $\CT_L$ including how the symmetry $\CC$ is spontaneously broken, the action of the symmetry on the vacua and order parameters can be determined using this picture \cite{Bhardwaj:2023idu,Bhardwaj:2023fca}. 

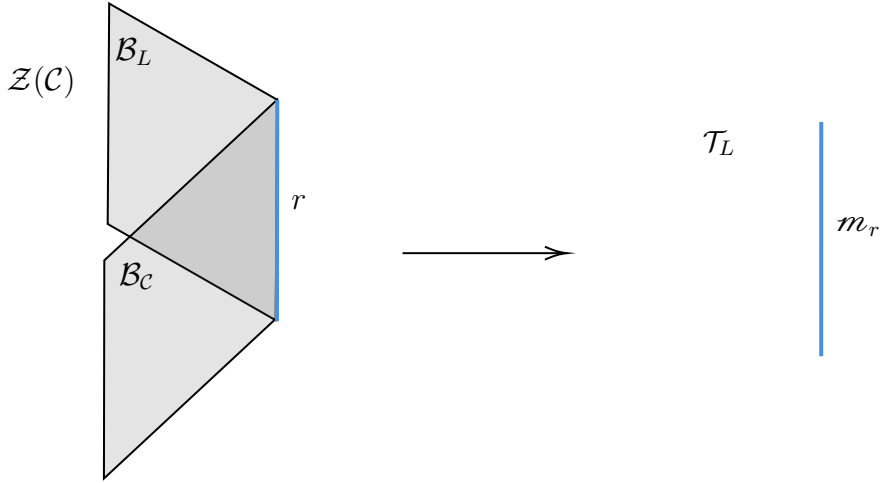
\begin{figure}[h!]
    \centering

\tikzset{every picture/.style={line width=0.75pt}} %set default line width to 0.75pt        

\begin{tikzpicture}[x=0.75pt,y=0.75pt,yscale=-1,xscale=1]

%Shape: Parallelogram [id:dp7009737618627823] 
\draw  [fill={rgb, 255:red, 74; green, 74; blue, 74 }  ,fill opacity=0.15 ] (227.4,72.83) -- (226.65,183.73) -- (142.81,135.32) -- (143.56,24.42) -- cycle ;
%Shape: Parallelogram [id:dp9238300653667753] 
\draw  [fill={rgb, 255:red, 74; green, 74; blue, 74 }  ,fill opacity=0.15 ] (227.4,72.83) -- (227.21,182.56) -- (140.94,263.36) -- (141.13,153.63) -- cycle ;
%Straight Lines [id:da3619900216413029] 
\draw [color={rgb, 255:red, 74; green, 144; blue, 226 }  ,draw opacity=1 ][line width=1.5]    (227.3,184.22) -- (227.47,72.89) ;
%Straight Lines [id:da38555187791037104] 
\draw    (290,150) -- (370,150) ;
\draw [shift={(372,150)}, rotate = 180] [color={rgb, 255:red, 0; green, 0; blue, 0 }  ][line width=0.75]    (10.93,-3.29) .. controls (6.95,-1.4) and (3.31,-0.3) .. (0,0) .. controls (3.31,0.3) and (6.95,1.4) .. (10.93,3.29)   ;
%Straight Lines [id:da4739357041102966] 
\draw [color={rgb, 255:red, 74; green, 144; blue, 226 }  ,draw opacity=1 ][line width=1.5]    (498.98,201.86) -- (499,84) ;

% Text Node
\draw (147,153.4) node [anchor=north west][inner sep=0.75pt]    {$\mathcal{B}_{\CC}$};
% Text Node
\draw (145,40.4) node [anchor=north west][inner sep=0.75pt]    {$\mathcal{B}_{L}$};
% Text Node
\draw (233,119.4) node [anchor=north west][inner sep=0.75pt]    {$r$};
% Text Node
\draw (438.41,87.01) node [anchor=north west][inner sep=0.75pt]  [font=\small]  {$\CT_L$};
% Text Node
\draw (505,130.4) node [anchor=north west][inner sep=0.75pt]    {$\cm_r$};
% Text Node
\draw (91,54.4) node [anchor=north west][inner sep=0.75pt]    {$\CZ( \CC)$};

\end{tikzpicture}
    \caption{A 1D gapped interface between the 2D gapped boundaries $\CB_{L}$ and $\CB_{\CC}$ turns into a 1D gapped boundary of the 1+1D $\CC$-symmetric TQFT $\CT_L$.}
    \label{fig:taco}
\end{figure}

The gapped boundaries of $\CT_L$ can also be determined from the sandwich picture. To that end, we consider the configuration in Fig. \ref{fig:taco} where $r$ is a simple gapped interface between the gapped boundaries $\CB_L$ and $\CB_{\CC}$. Compactifying this diagram gives $\CT_L$ where the gapped interface $r$ becomes a 1D gapped boundary $\cm_r$ of $\CT_L$. 

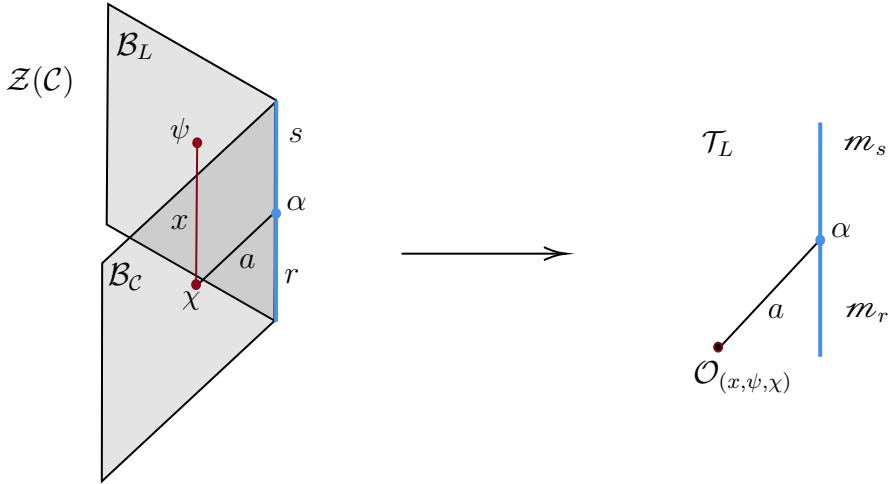
\begin{figure}
    \centering

\tikzset{every picture/.style={line width=0.75pt}} %set default line width to 0.75pt        

\begin{tikzpicture}[x=0.75pt,y=0.75pt,yscale=-1,xscale=1]

%Shape: Parallelogram [id:dp7009737618627823] 
\draw  [fill={rgb, 255:red, 74; green, 74; blue, 74 }  ,fill opacity=0.15 ] (227.4,72.83) -- (226.65,183.73) -- (142.81,135.32) -- (143.56,24.42) -- cycle ;
%Shape: Parallelogram [id:dp9238300653667753] 
\draw  [fill={rgb, 255:red, 74; green, 74; blue, 74 }  ,fill opacity=0.15 ] (226.84,74.01) -- (226.65,183.73) -- (140.38,264.53) -- (140.57,154.81) -- cycle ;
%Straight Lines [id:da18933110176244428] 
\draw [color={rgb, 255:red, 139; green, 6; blue, 24 }  ,draw opacity=1 ]   (188,92) -- (187.43,167.78) ;
%Straight Lines [id:da3619900216413029] 
\draw [color={rgb, 255:red, 74; green, 144; blue, 226 }  ,draw opacity=1 ][line width=1.5]    (227.3,184.22) -- (227.47,72.89) ;
%Straight Lines [id:da38555187791037104] 
\draw    (290,150) -- (370,150) ;
\draw [shift={(372,150)}, rotate = 180] [color={rgb, 255:red, 0; green, 0; blue, 0 }  ][line width=0.75]    (10.93,-3.29) .. controls (6.95,-1.4) and (3.31,-0.3) .. (0,0) .. controls (3.31,0.3) and (6.95,1.4) .. (10.93,3.29)   ;
%Straight Lines [id:da4739357041102966] 
\draw [color={rgb, 255:red, 74; green, 144; blue, 226 }  ,draw opacity=1 ][line width=1.5]    (498.98,201.86) -- (499,84) ;
%Shape: Ellipse [id:dp8436270035301053] 
\draw  [color={rgb, 255:red, 139; green, 6; blue, 24 }  ,draw opacity=1 ][fill={rgb, 255:red, 139; green, 6; blue, 24 }  ,fill opacity=1 ] (185.47,165.56) .. controls (185.47,164.32) and (186.35,163.33) .. (187.43,163.33) .. controls (188.51,163.33) and (189.39,164.32) .. (189.39,165.56) .. controls (189.39,166.79) and (188.51,167.78) .. (187.43,167.78) .. controls (186.35,167.78) and (185.47,166.79) .. (185.47,165.56) -- cycle ;
%Straight Lines [id:da7085338924099465] 
\draw    (189.39,164.56) -- (227.39,128.56) ;
%Straight Lines [id:da22880396435582095] 
\draw    (449,197) -- (498.99,142.93) ;
%Shape: Ellipse [id:dp46337964064590864] 
\draw  [color={rgb, 255:red, 139; green, 6; blue, 24 }  ,draw opacity=1 ][fill={rgb, 255:red, 139; green, 6; blue, 24 }  ,fill opacity=1 ] (186.04,94.23) .. controls (186.04,93) and (186.92,92) .. (188,92) .. controls (189.08,92) and (189.96,93) .. (189.96,94.23) .. controls (189.96,95.46) and (189.08,96.46) .. (188,96.46) .. controls (186.92,96.46) and (186.04,95.46) .. (186.04,94.23) -- cycle ;
%Shape: Ellipse [id:dp9146632749803072] 
\draw  [color={rgb, 255:red, 139; green, 6; blue, 24 }  ,draw opacity=1 ][fill={rgb, 255:red, 0; green, 0; blue, 0 }  ,fill opacity=1 ] (446.08,197) .. controls (446.08,195.77) and (446.96,194.77) .. (448.04,194.77) .. controls (449.12,194.77) and (450,195.77) .. (450,197) .. controls (450,198.23) and (449.12,199.23) .. (448.04,199.23) .. controls (446.96,199.23) and (446.08,198.23) .. (446.08,197) -- cycle ;
%Shape: Ellipse [id:dp4338505794070868] 
\draw  [color={rgb, 255:red, 74; green, 144; blue, 226 }  ,draw opacity=1 ][fill={rgb, 255:red, 74; green, 144; blue, 226 }  ,fill opacity=1 ] (225.43,129.78) .. controls (225.43,128.55) and (226.31,127.56) .. (227.39,127.56) .. controls (228.47,127.56) and (229.35,128.55) .. (229.35,129.78) .. controls (229.35,131.02) and (228.47,132.01) .. (227.39,132.01) .. controls (226.31,132.01) and (225.43,131.02) .. (225.43,129.78) -- cycle ;
%Shape: Ellipse [id:dp6683381197796407] 
\draw  [color={rgb, 255:red, 74; green, 144; blue, 226 }  ,draw opacity=1 ][fill={rgb, 255:red, 74; green, 144; blue, 226 }  ,fill opacity=1 ] (497.03,143.16) .. controls (497.03,141.93) and (497.91,140.93) .. (498.99,140.93) .. controls (500.07,140.93) and (500.95,141.93) .. (500.95,143.16) .. controls (500.95,144.39) and (500.07,145.39) .. (498.99,145.39) .. controls (497.91,145.39) and (497.03,144.39) .. (497.03,143.16) -- cycle ;

% Text Node
\draw (142,153.4) node [anchor=north west][inner sep=0.75pt]    {$\mathcal{B}_{\CC}$};
% Text Node
\draw (146,37.4) node [anchor=north west][inner sep=0.75pt]    {$\mathcal{B}_{L}$};
% Text Node
\draw (230,156.4) node [anchor=north west][inner sep=0.75pt]    {$r$};
% Text Node
\draw (438.41,87.01) node [anchor=north west][inner sep=0.75pt]  [font=\small]  {$\CT_L$};
% Text Node
\draw (510,173.4) node [anchor=north west][inner sep=0.75pt]    {$\cm_r$};
% Text Node
\draw (91,54.4) node [anchor=north west][inner sep=0.75pt]    {$\CZ( \CC)$};
% Text Node
\draw (173.39,128.96) node [anchor=north west][inner sep=0.75pt]    {$x$};
% Text Node
\draw (232,85.4) node [anchor=north west][inner sep=0.75pt]    {$s$};
% Text Node
\draw (509,89.4) node [anchor=north west][inner sep=0.75pt]    {$\cm_s$};
% Text Node
\draw (207.39,148.96) node [anchor=north west][inner sep=0.75pt]    {$a$};
% Text Node
\draw (471.39,173.96) node [anchor=north west][inner sep=0.75pt]    {$a$};
% Text Node
\draw (173,80.4) node [anchor=north west][inner sep=0.75pt]  [font=\small]  {$\psi $};
% Text Node
\draw (179,167.4) node [anchor=north west][inner sep=0.75pt]  [font=\small]  {$\chi $};
% Text Node
\draw (434,202.4) node [anchor=north west][inner sep=0.75pt]    {$\mathcal{O}_{( x,\psi ,\chi )}$};
% Text Node
\draw (231,119.4) node [anchor=north west][inner sep=0.75pt]    {$\alpha $};
% Text Node
\draw (503,133.4) node [anchor=north west][inner sep=0.75pt]    {$\alpha $};

\end{tikzpicture}
    \caption{The action of $a\in \CC$ on the 1D gapped boundaries of $\CT_L$ can be determined by the action of $a\in \CC$ on gapped interfaces in the SymTFT. Upon collapsing the wedge on the left, the line operator $x$ becomes a twisted-sector local operator $\CO_{(x,\psi,\chi)}.$}
    \label{fig:action on gapped boundaries from SymTFT}
\end{figure}

In section \ref{sec:Csym TQFTs}, we described how the line operators in $\CC$ act on the 1D gapped boundaries of $\CC$. This can be determined from the SymTFT as follows. Consider a bulk line operator $x$ in the Lagrangian algebra $L$. The line $x$ can end trivially on the gapped boundary $\CB_{L}$. Generically, $x$ cannot end on the gapped boundary $\CB_{\CC}$ but forms a junction with a line operator $a\in \CC$. The action of $a$ on the gapped boundary $\cm_r$ of the 1+1D TQFT $\CT_L$ is determined by the action of $a$ on the corresponding gapped interface $r$ in the SymTFT as in Fig. \ref{fig:action on gapped boundaries from SymTFT}. 

\section{Gaugeable line operators in 1+1D and SymTFTs}

\label{sec: gaugeable lines from SymTFT}

Consider the symmetries of a 1+1D QFT described by a fusion category $\CC$. Given a line operator $A$ in the category $\CC$, how do we determine if it is anomalous? One approach is to determine whether $A$ admits the structure of an algebra. This involves solving the constraints \eqref{eq: m associativity} which requires the $F$ matrices of the fusion category $\CC$.  

In this section, we will consider an alternate approach by studying gapped boundaries of the SymTFT $\CZ(\CC)$. An important defining property of the SymTFT is that physically distinct gaugings in $\CC$ are in one-to-one correspondence with gapped boundaries of $\CZ(\CC)$ \cite[Proposition 4.8]{davydov2013witt}.
\be
\text{$[A] \in \CC$} \longleftrightarrow L \in \CZ(\CC)~.
\ee
In the rest of this section, we will study this relation explicitly. 

\subsection{Non-anomalous line operators from SymTFT}

In this section, we will explore the one-to-one correspondence between gapped boundaries of $\CZ(\CC)$ and non-anomalous line operators in $\CC$ from a physical perspective. We will describe various properties of this correspondence which can be used to answer the following question: Given a Lagrangian algebra $L$ of $\CZ(\CC)$, which Morita equivalence class of gaugeable algebra $[A]$ in $\CC$ does it correspond to? To answer this question, 
let us consider the perpendicular fusion of a line operator $l$ on a gapped boundary $\CB$
\be
\label{eq:bulk to boundary}
F_{\CB} (x) = \sum_{a} N_{F_{\CB}(x)}^a ~a ~,
\ee
where the line operators $a$ live on the gapped boundary $\CB$ and $N_{l}^a$ are non-negative integers. In general, $F_{\CB} (x)$ is part of the data of a bulk-to-boundary map which relates the line operators and fusion spaces of the bulk TQFT $\CZ(\CC)$ to those of the category of line operators on the gapped boundary $\CB$. Consider the canonical gapped boundary $\CB_{\CC}$ of $\CZ(\CC)$ on which the line operators form the fusion category $\CC$. In this case, we will denote $F_{\CB_{\CC}}$ as $F$ to simplify notation.\footnote{This notation is not to be confused with the associator of $\CC$ given by the $F$ matrices.} The action of $F$ on the simple line operators of $\CZ(\CC)$ has a simple description. Recall that simple line operators in $\CZ(\CC)$ can be written as
\be
(a,e_a)~,~ a\in \CC
\ee
where $e_a$ are half-braidings. In this notation, $F$ is given by\footnote{This is called a forgetful functor from $\CZ(\CC)$\ to $\CC$ since it forgets about the half-braidings $e_a$. More generally, for any gapped boundary $\CB$, $F_{\CB}$ is a tensor functor \cite[Section V. B]{Cong:2017hcl}.} 
\be
F((a,e_a))=a~.
\ee
Since $F$ determines the relation between bulk and boundary line operators, it is perhaps unsurprising that it can be used to determine the relation between Lagrangian algebras in $\CZ(\CC)$ and non-anomalous line operators in $\CC$. Indeed, consider a Lagrangian algebra $L$ in $\CZ(\CC)$. Consider the perpendicular fusion of $L$ with the gapped boundary $\CB_{\CC}$, given by  
\be
F(L) = \sum_{a} N_{F(L)}^a ~a ~, ~ ~ a \in \CC~.
\ee
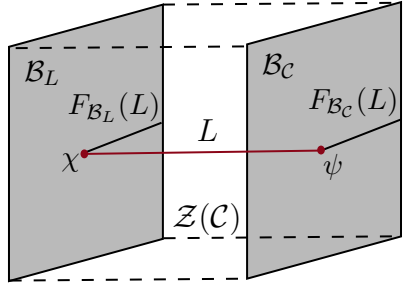
\begin{figure}[h!]
    \centering

\tikzset{every picture/.style={line width=0.75pt}} %set default line width to 0.75pt        

\begin{tikzpicture}[x=0.75pt,y=0.75pt,yscale=-1,xscale=1]
%uncomment if require: \path (0,330); %set diagram left start at 0, and has height of 330

%Shape: Parallelogram [id:dp8336414950009718] 
\draw  [color={rgb, 255:red, 0; green, 0; blue, 0 }  ,draw opacity=1 ][fill={rgb, 255:red, 74; green, 74; blue, 74 }  ,fill opacity=0.38 ] (369.15,265.1) -- (369.17,147.24) -- (446,126) -- (445.98,243.86) -- cycle ;
%Shape: Parallelogram [id:dp6828681308090923] 
\draw  [color={rgb, 255:red, 0; green, 0; blue, 0 }  ,draw opacity=1 ][fill={rgb, 255:red, 74; green, 74; blue, 74 }  ,fill opacity=0.38 ] (250.29,266.33) -- (250.31,148.47) -- (327.14,127.24) -- (327.12,245.09) -- cycle ;
%Straight Lines [id:da7423092163243079] 
\draw [color={rgb, 255:red, 139; green, 6; blue, 24 }  ,draw opacity=1 ]   (289.32,202.33) -- (406.06,200.67) ;
%Straight Lines [id:da6764191749897529] 
\draw    (406.06,200.67) -- (446,185) ;
%Straight Lines [id:da9250587780299152] 
\draw    (286.51,202.33) -- (326.45,186.66) ;
%Straight Lines [id:da6015727681759209] 
\draw  [dash pattern={on 4.5pt off 4.5pt}]  (253.64,149.14) -- (372.17,148.25) ;
%Straight Lines [id:da14891354129884204] 
\draw  [dash pattern={on 4.5pt off 4.5pt}]  (327.14,127.24) -- (445.66,126.33) ;
%Straight Lines [id:da708264958420786] 
\draw  [dash pattern={on 4.5pt off 4.5pt}]  (327.45,244.76) -- (445.98,243.78) ;
%Straight Lines [id:da6666188120269951] 
\draw  [dash pattern={on 4.5pt off 4.5pt}]  (250.62,266.07) -- (369.15,265.1) ;
%Shape: Ellipse [id:dp47439590836921897] 
\draw  [color={rgb, 255:red, 139; green, 6; blue, 24 }  ,draw opacity=1 ][fill={rgb, 255:red, 139; green, 6; blue, 24 }  ,fill opacity=1 ] (404.65,200.4) .. controls (404.65,199.44) and (405.28,198.67) .. (406.06,198.67) .. controls (406.83,198.67) and (407.46,199.44) .. (407.46,200.4) .. controls (407.46,201.36) and (406.83,202.13) .. (406.06,202.13) .. controls (405.28,202.13) and (404.65,201.36) .. (404.65,200.4) -- cycle ;
%Shape: Ellipse [id:dp3020150159153402] 
\draw  [color={rgb, 255:red, 139; green, 6; blue, 24 }  ,draw opacity=1 ][fill={rgb, 255:red, 139; green, 6; blue, 24 }  ,fill opacity=1 ] (286.51,202.33) .. controls (286.51,201.37) and (287.14,200.6) .. (287.92,200.6) .. controls (288.69,200.6) and (289.32,201.37) .. (289.32,202.33) .. controls (289.32,203.29) and (288.69,204.06) .. (287.92,204.06) .. controls (287.14,204.06) and (286.51,203.29) .. (286.51,202.33) -- cycle ;

% Text Node
\draw (256.52,154.01) node [anchor=north west][inner sep=0.75pt]  [font=\small]  {$\CB_L$};
% Text Node
\draw (375.17,150.65) node [anchor=north west][inner sep=0.75pt]  [font=\small]  {$\CB_{\CC}$};
% Text Node
\draw (342.5,183.55) node [anchor=north west][inner sep=0.75pt]    {$L$};
\draw (330.5,225.55) node [anchor=north west][inner sep=0.75pt]    {$\CZ(\CC)$};
% Text Node
\draw (398.06,167.07) node [anchor=north west][inner sep=0.75pt]  [font=\small]  {$F_{\CB_{\CC}}(L)$};
% Text Node
\draw (278.21,170) node [anchor=north west][inner sep=0.75pt] [font=\small]   {$F_{\CB_{L}}(L)$};
% Text Node
\draw (405.65,200.8) node [anchor=north west][inner sep=0.75pt]  [font=\small]  {$\psi $};
% Text Node
\draw (275,200.4) node [anchor=north west][inner sep=0.75pt]  [font=\small]  {$\chi $};

\end{tikzpicture}
    \caption{Fusion of line operators in $\CZ(\CC)$ on gapped boundaries.}
    \label{fig:F map}
\end{figure}

We will show that $F(L)$ is a sum of algebra objects in $\CC$ which correspond to physically equivalent gaugings. In other words, $F(L)$ is a sum of Morita-equivalent algebras. To see this recall that $\CT_{L}$ is the 1+1D TQFT obtained from the interval compactification of the SymTFT $\CZ(\CC)$ with gapped boundaries $\CB_{\CC}$ and $\CB_{L}$. Consider the $\CC$-module category $\CM_L$ describing the 1D gapped boundaries of $\CT_{L}$. Recall the definition of the algebra object $A_{\cm}$ in section \ref{sec:Csym TQFTs}. We will prove the following theorem. 
\begin{theorem}
\label{th: F(L)}
Let $L$ be a Lagrangian algebra in the SymTFT $\CZ(\CC)$. As an object in $\CC$
\be
\label{eq: non anomalous line from SymTFT}
F(L)=\bigoplus_{\cm\in \CM_L} A_{\cm}~.
\ee
\end{theorem}

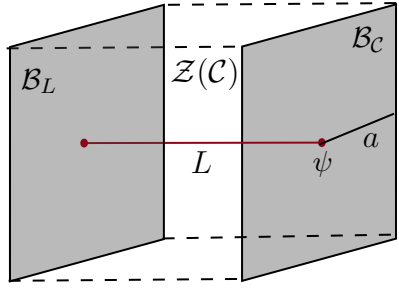
\begin{figure}[h!]
    \centering

\tikzset{every picture/.style={line width=0.75pt}} %set default line width to 0.75pt        

\begin{tikzpicture}[x=0.75pt,y=0.75pt,yscale=-1,xscale=1]
%uncomment if require: \path (0,330); %set diagram left start at 0, and has height of 330

%Shape: Parallelogram [id:dp84461910785525] 
\draw  [color={rgb, 255:red, 0; green, 0; blue, 0 }  ,draw opacity=1 ][fill={rgb, 255:red, 74; green, 74; blue, 74 }  ,fill opacity=0.38 ] (351.15,246.1) -- (351.17,128.24) -- (428,107) -- (427.98,224.86) -- cycle ;
%Straight Lines [id:da714765442747638] 
\draw [color={rgb, 255:red, 139; green, 6; blue, 24 }  ,draw opacity=1 ]   (273.72,176.78) -- (389.58,176.55) ;
%Shape: Ellipse [id:dp06189842406096169] 
\draw  [color={rgb, 255:red, 139; green, 6; blue, 24 }  ,draw opacity=1 ][fill={rgb, 255:red, 139; green, 6; blue, 24 }  ,fill opacity=1 ] (389.58,176.55) .. controls (389.58,175.59) and (390.21,174.82) .. (390.98,174.82) .. controls (391.76,174.82) and (392.39,175.59) .. (392.39,176.55) .. controls (392.39,177.51) and (391.76,178.28) .. (390.98,178.28) .. controls (390.21,178.28) and (389.58,177.51) .. (389.58,176.55) -- cycle ;
%Straight Lines [id:da9630485063154574] 
\draw  [dash pattern={on 4.5pt off 4.5pt}]  (233.64,129.14) -- (352.17,128.25) ;
%Straight Lines [id:da4046196775672194] 
\draw  [dash pattern={on 4.5pt off 4.5pt}]  (314.14,107.24) -- (432.66,106.33) ;
%Straight Lines [id:da09607122408834579] 
\draw  [dash pattern={on 4.5pt off 4.5pt}]  (311.45,224.76) -- (429.98,222.78) ;
%Straight Lines [id:da8486075282162301] 
\draw  [dash pattern={on 4.5pt off 4.5pt}]  (234.62,246.07) -- (353.15,245.1) ;
%Straight Lines [id:da590263320550159] 
\draw    (389.58,177.55) -- (427,162) ;
%Shape: Parallelogram [id:dp5459991906512744] 
\draw  [color={rgb, 255:red, 0; green, 0; blue, 0 }  ,draw opacity=1 ][fill={rgb, 255:red, 74; green, 74; blue, 74 }  ,fill opacity=0.38 ] (235.29,246.33) -- (235.31,128.47) -- (312.14,107.24) -- (312.12,225.09) -- cycle ;
%Shape: Ellipse [id:dp7297452295551219] 
\draw  [color={rgb, 255:red, 139; green, 6; blue, 24 }  ,draw opacity=1 ][fill={rgb, 255:red, 139; green, 6; blue, 24 }  ,fill opacity=1 ] (270.9,176.78) .. controls (270.9,175.83) and (271.53,175.05) .. (272.31,175.05) .. controls (273.09,175.05) and (273.72,175.83) .. (273.72,176.78) .. controls (273.72,177.74) and (273.09,178.52) .. (272.31,178.52) .. controls (271.53,178.52) and (270.9,177.74) .. (270.9,176.78) -- cycle ;

% Text Node
\draw (404.41,117.01) node [anchor=north west][inner sep=0.75pt]  [font=\small]  {$\mathcal{B}_{\CC}$};
% Text Node
\draw (314,132.4) node [anchor=north west][inner sep=0.75pt]    {$\CZ(\CC)$};
% Text Node
\draw (324,179.4) node [anchor=north west][inner sep=0.75pt]    {$L$};
% Text Node
\draw (385,179.4) node [anchor=north west][inner sep=0.75pt]  [font=\small]  {$\psi $};
% Text Node
\draw (410,170.4) node [anchor=north west][inner sep=0.75pt]    {$a$};
% Text Node
\draw (239.41,138.01) node [anchor=north west][inner sep=0.75pt]  [font=\small]  {$\mathcal{B}_{L}$};

\end{tikzpicture}
    \caption{Junctions between $L$ and $a$ on the gapped boundary $\CB_{\CC}$ correspond to twisted sector states in the Hilbert space $\CH_a$ of the TQFT $\CT_{L}$.}
    \label{fig:twisted sector}
\end{figure}

\vspace{0.2cm}
\noindent {\bf Proof:} Consider some simple line operator $a$ such that $N_{F(L)}^a\neq 0$. Then, the configuration in Fig. \ref{fig:twisted sector}, is non-trivial. 
Compactifying this diagram gives $N_{F(L)}^a$ operators at the end of the line operator $a$ in the TQFT $\CT_L$.\footnote{One could wonder whether the choice of the junction of $L$ with the gapped boundary $\CB_{L}$ affects this counting. Note that this is already taken into account in the argument since the number of possible point junctions between a line operator $x\in L$ and the gapped boundary $\CB_L$ is precisely $N_{L}^x$. The Lagrangian algebra object $L$ contains the term $N_{L}^x ~ x$, and on acting with $F$ we get $N_{L}^x F(x)$.} Now, using the state-operator correspondence we find that dim$(\CH_a)=N_{F(L)}^a$, where $\CH_a$ is the twisted Hilbert space for the line operator $a$.\footnote{We have assumed that interval compactification of the configuration in Fig. \ref{fig:twisted sector} for different $\psi$ produces all twisted sector local operators at the end of the line operator $a$ in $\CT_L$.} Consider the isomorphism \cite{Huang:2021zvu} 
\be
\label{eq: twisted Hilb a internal hom}
\CH_{a} \cong \bigoplus_{\cm \in \CM_L} \text{Hom}_{\CM_L}(\cm,a \times \cm)~.
\ee
Note that $\text{Hom}_{\CM_L}(\cm,a \times \cm)$ is non-empty if and only if $a \in A_{\cm}$. Indeed,
\be
|\text{Hom}_{\CM_L}(\cm,a \times \cm)|= N_{\cm}^{a}
\ee
where, $N_{\cm}^{a}$ is the number of copies of $a$ in $A_{\cm}$ (see definition \eqref{eq:Am definition}). Therefore,
\be
\sum_{\cm \in \CM_L}N_{\cm}^a = \sum_{\cm \in \CM_L}|\text{Hom}_{\CM_L}(\cm,a \times \cm)|= \text{dim}(\CH_a)= N_{F(L)}^a~.
\ee
Therefore, we find that there are $N_{F(L)}^a$ copies of the line operator $a$ in $\bigoplus_{\cm\in \CM_L} A_{\cm}$. $\hfill \square$
\vspace{0.2cm}

\begin{figure}[h!]
    \centering

\tikzset{every picture/.style={line width=0.75pt}} %set default line width to 0.75pt        

\begin{tikzpicture}[x=0.75pt,y=0.75pt,yscale=-1,xscale=1]

%Shape: Parallelogram [id:dp7009737618627823] 
\draw  [fill={rgb, 255:red, 74; green, 74; blue, 74 }  ,fill opacity=0.15 ] (227.4,72.83) -- (226.65,183.73) -- (142.81,135.32) -- (143.56,24.42) -- cycle ;
%Shape: Parallelogram [id:dp9238300653667753] 
\draw  [fill={rgb, 255:red, 74; green, 74; blue, 74 }  ,fill opacity=0.15 ] (226.84,74.01) -- (226.65,183.73) -- (140.38,264.53) -- (140.57,154.81) -- cycle ;
%Straight Lines [id:da18933110176244428] 
\draw [color={rgb, 255:red, 139; green, 6; blue, 24 }  ,draw opacity=1 ]   (188,92) -- (187.43,167.78) ;
%Straight Lines [id:da3619900216413029] 
\draw [color={rgb, 255:red, 74; green, 144; blue, 226 }  ,draw opacity=1 ][line width=1.5]    (227.3,184.22) -- (227.47,72.89) ;
%Straight Lines [id:da38555187791037104] 
\draw    (290,150) -- (370,150) ;
\draw [shift={(372,150)}, rotate = 180] [color={rgb, 255:red, 0; green, 0; blue, 0 }  ][line width=0.75]    (10.93,-3.29) .. controls (6.95,-1.4) and (3.31,-0.3) .. (0,0) .. controls (3.31,0.3) and (6.95,1.4) .. (10.93,3.29)   ;
%Straight Lines [id:da4739357041102966] 
\draw [color={rgb, 255:red, 74; green, 144; blue, 226 }  ,draw opacity=1 ][line width=1.5]    (498.98,201.86) -- (499,84) ;
%Shape: Ellipse [id:dp8436270035301053] 
\draw  [color={rgb, 255:red, 139; green, 6; blue, 24 }  ,draw opacity=1 ][fill={rgb, 255:red, 139; green, 6; blue, 24 }  ,fill opacity=1 ] (185.47,165.56) .. controls (185.47,164.32) and (186.35,163.33) .. (187.43,163.33) .. controls (188.51,163.33) and (189.39,164.32) .. (189.39,165.56) .. controls (189.39,166.79) and (188.51,167.78) .. (187.43,167.78) .. controls (186.35,167.78) and (185.47,166.79) .. (185.47,165.56) -- cycle ;
%Straight Lines [id:da7085338924099465] 
\draw    (189.39,164.56) -- (227.39,128.56) ;
%Straight Lines [id:da22880396435582095] 
\draw    (449,197) -- (498.99,142.93) ;
%Shape: Ellipse [id:dp46337964064590864] 
\draw  [color={rgb, 255:red, 139; green, 6; blue, 24 }  ,draw opacity=1 ][fill={rgb, 255:red, 139; green, 6; blue, 24 }  ,fill opacity=1 ] (186.04,94.23) .. controls (186.04,93) and (186.92,92) .. (188,92) .. controls (189.08,92) and (189.96,93) .. (189.96,94.23) .. controls (189.96,95.46) and (189.08,96.46) .. (188,96.46) .. controls (186.92,96.46) and (186.04,95.46) .. (186.04,94.23) -- cycle ;
%Shape: Ellipse [id:dp9146632749803072] 
\draw  [color={rgb, 255:red, 139; green, 6; blue, 24 }  ,draw opacity=1 ][fill={rgb, 255:red, 0; green, 0; blue, 0 }  ,fill opacity=1 ] (446.08,197) .. controls (446.08,195.77) and (446.96,194.77) .. (448.04,194.77) .. controls (449.12,194.77) and (450,195.77) .. (450,197) .. controls (450,198.23) and (449.12,199.23) .. (448.04,199.23) .. controls (446.96,199.23) and (446.08,198.23) .. (446.08,197) -- cycle ;
%Shape: Ellipse [id:dp4338505794070868] 
\draw  [color={rgb, 255:red, 74; green, 144; blue, 226 }  ,draw opacity=1 ][fill={rgb, 255:red, 74; green, 144; blue, 226 }  ,fill opacity=1 ] (225.43,129.78) .. controls (225.43,128.55) and (226.31,127.56) .. (227.39,127.56) .. controls (228.47,127.56) and (229.35,128.55) .. (229.35,129.78) .. controls (229.35,131.02) and (228.47,132.01) .. (227.39,132.01) .. controls (226.31,132.01) and (225.43,131.02) .. (225.43,129.78) -- cycle ;
%Shape: Ellipse [id:dp6683381197796407] 
\draw  [color={rgb, 255:red, 74; green, 144; blue, 226 }  ,draw opacity=1 ][fill={rgb, 255:red, 74; green, 144; blue, 226 }  ,fill opacity=1 ] (497.03,143.16) .. controls (497.03,141.93) and (497.91,140.93) .. (498.99,140.93) .. controls (500.07,140.93) and (500.95,141.93) .. (500.95,143.16) .. controls (500.95,144.39) and (500.07,145.39) .. (498.99,145.39) .. controls (497.91,145.39) and (497.03,144.39) .. (497.03,143.16) -- cycle ;

% Text Node
\draw (142,153.4) node [anchor=north west][inner sep=0.75pt]    {$\mathcal{B}_{\CC}$};
% Text Node
\draw (146,37.4) node [anchor=north west][inner sep=0.75pt]    {$\mathcal{B}_{L}$};
% Text Node
\draw (230,156.4) node [anchor=north west][inner sep=0.75pt]    {$r$};
% Text Node
\draw (438.41,87.01) node [anchor=north west][inner sep=0.75pt]  [font=\small]  {$T$};
% Text Node
\draw (510,173.4) node [anchor=north west][inner sep=0.75pt]    {$\cm_{r}$};
% Text Node
\draw (91,54.4) node [anchor=north west][inner sep=0.75pt]    {$\CZ( \CC)$};
% Text Node
\draw (173.39,128.96) node [anchor=north west][inner sep=0.75pt]    {$L$};
% Text Node
\draw (232,85.4) node [anchor=north west][inner sep=0.75pt]    {$r$};
% Text Node
\draw (509,89.4) node [anchor=north west][inner sep=0.75pt]    {$\cm_{r}$};
% Text Node
\draw (207.39,148.96) node [anchor=north west][inner sep=0.75pt]    {$a$};
% Text Node
\draw (471.39,173.96) node [anchor=north west][inner sep=0.75pt]    {$a$};
% Text Node
\draw (173,80.4) node [anchor=north west][inner sep=0.75pt]  [font=\small]  {$\psi $};
% Text Node
\draw (179,167.4) node [anchor=north west][inner sep=0.75pt]  [font=\small]  {$\chi $};
% Text Node
\draw (434,202.4) node [anchor=north west][inner sep=0.75pt]    {$\mathcal{O}_{( x,\psi ,\chi )}$};
% Text Node
\draw (231,119.4) node [anchor=north west][inner sep=0.75pt]    {$\alpha $};
% Text Node
\draw (503,133.4) node [anchor=north west][inner sep=0.75pt]    {$\alpha $};

\end{tikzpicture}
    \caption{$a$ is a line operator in $\CC$ which can form a non-trivial junction on the gapped interface $r$. On closing the wedge, we find that $a$ can form a junction on the corresponding gapped boundary $\cm_r\in \CM_L$.}
    \label{fig:lines fixing gapped interface}
\end{figure}
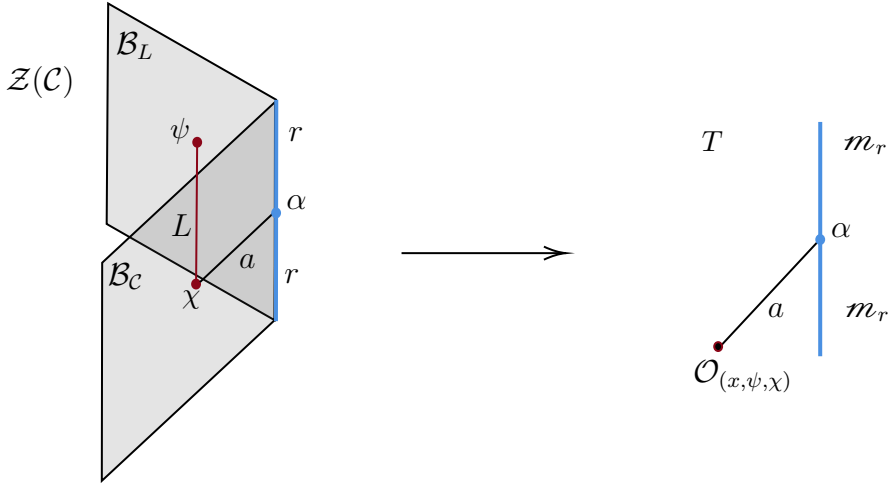

When $\CC$ is a modular tensor category, Theorem \ref{th: F(L)} follows from \cite{Schellekens:2001xv} (see also \cite[Proposition 4.3]{Kong:2007yv}). Recall from the discussion in section \ref{sec:Csym TQFTs} that each $A_{\cm}$ admits the structure of a haploid symmetric separable Frobenius algebra for each $\cm$. Therefore, their direct sums also admit the structure of a symmetric separable Frobenius algebra (for example, see \cite[Section 3.5]{Fuchs:2002cm}). Hence, $F(L)$ is a non-anomalous line operator in $\CC$. It is clear from \eqref{eq: non anomalous line from SymTFT} that, in general, the number of copies of the identity line $\trl$ in $F(L)$ is equal to the number of distinct boundary conditions of the TQFT $\CT_L$ since $A_{\cm}$ has exactly one identity operator for every $\cm \in \CM_{L}$. Therefore, in general, the algebra $F(L)$ is not haploid. In order to identify the haploid subalgebras in $F(L)$, we have to consider the action of the simple line operators in $F(L)$ on gapped interfaces between gapped boundaries $\CB_{\CC}$ and $\CB_{L}$ as in Fig. \ref{fig:action on gapped boundaries from SymTFT}. Choose a simple gapped interface $r$ between the gapped boundaries $\CB_{\CC}$ and $\CB_{L}$.\footnote{A gapped interface is simple if it does not host any non-trivial point operators on it.} Define $A_r$ to be the sub-object/line operator of $F(L)$ 
\be
A_{r}:=\sum_{a ~\in~ F(L)} N_r^{a} ~ a~,
\ee
where $N_r^a$ is the dimension of Hilbert space of operators at the junction of $a$ with the gapped interface $r$ (see Fig. \ref{fig:lines fixing gapped interface}). Compactifying this diagram, we clearly have that
\be
A_{r}= A_{\cm_r}~,
\ee
where $\cm_{r}\in \CM_L$ is the 1D gapped boundary of the 1+1D TQFT $\CT_{L}$ corresponding to the gapped interface $r$. 

Suppose $L_1,\dots,L_n$ is the full set of Lagrangian algebras in the SymTFT $\CZ(\CC)$. Then, $F(L_1),\dots,F(L_n)$ is the full set of equivalence classes of algebras in $\CC$. Therefore, Theorem \ref{th: F(L)} provides an alternative method to using NIM-reps to classify algebra objects in a fusion category \cite{grossman2012quantum,grossman2016brauer,Komargodski:2020mxz,Diatlyk:2023fwf}. In fact, as will explore in detail in section \ref{sec:transporting algebras}, if the SymTFT $\CZ(\CC)$ and its Lagrangian algebras are known, then Theorem \ref{th: F(L)} can be used to find algebra objects in all fusion categories which share the same SymTFT $\CZ(\CC)$.

\subsection{Generalized discrete torsion from SymTFT}

\label{sec:Generalized discrete torsion}

In the previous section, we studied the structure of $F_{\CB_{\CC}}(L)$ in terms of the algebras $A_{\cm}$, $\cm\in \CM_L$. This discussion only used the fact that $L$ is a line operator in the SymTFT $\CZ(\CC)$ which admits the structure of a Lagrangian algebra. In particular, so far, we did not use the explicit algebra structure on $L$. In this section, we will discuss how to determine the explicit algebra structure on $F_{\CB_{\CC}}(L)$ from that on $L$. 

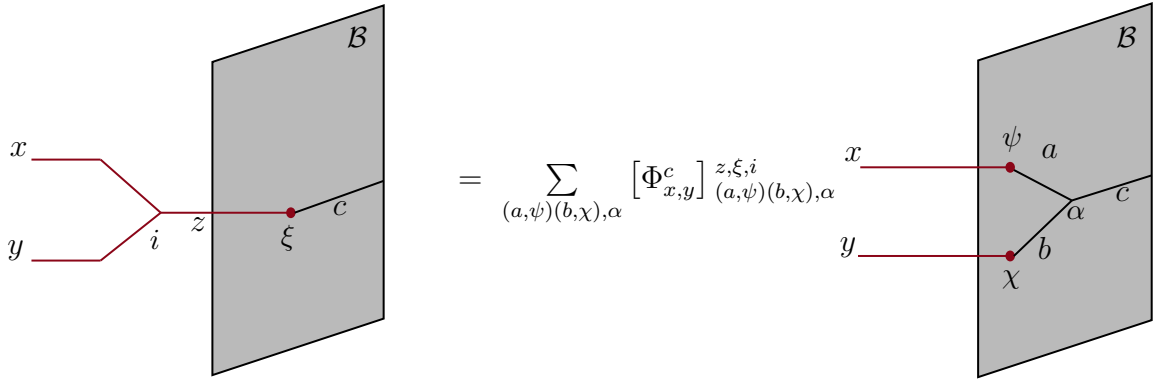
\begin{figure}[h!]
    \centering

\tikzset{every picture/.style={line width=0.75pt}} %set default line width to 0.75pt        

\begin{tikzpicture}[x=0.75pt,y=0.75pt,yscale=-1,xscale=0.95]

%Shape: Parallelogram [id:dp5071201530638992] 
\draw  [color={rgb, 255:red, 0; green, 0; blue, 0 }  ,draw opacity=1 ][fill={rgb, 255:red, 74; green, 74; blue, 74 }  ,fill opacity=0.38 ] (515.77,274) -- (515.8,114.69) -- (607,86.01) -- (606.97,245.32) -- cycle ;
%Straight Lines [id:da7370921953956616] 
\draw [color={rgb, 255:red, 139; green, 6; blue, 24 }  ,draw opacity=1 ]   (453.85,168.45) -- (534.58,168.45) ;
%Straight Lines [id:da17953444830556864] 
\draw [color={rgb, 255:red, 139; green, 6; blue, 24 }  ,draw opacity=1 ]   (452.66,213.05) -- (534.58,213.05) ;
%Shape: Ellipse [id:dp6460834498803231] 
\draw  [color={rgb, 255:red, 139; green, 6; blue, 24 }  ,draw opacity=1 ][fill={rgb, 255:red, 139; green, 6; blue, 24 }  ,fill opacity=1 ] (530.66,168.45) .. controls (530.66,167.22) and (531.54,166.22) .. (532.62,166.22) .. controls (533.7,166.22) and (534.58,167.22) .. (534.58,168.45) .. controls (534.58,169.68) and (533.7,170.68) .. (532.62,170.68) .. controls (531.54,170.68) and (530.66,169.68) .. (530.66,168.45) -- cycle ;
%Shape: Ellipse [id:dp1840279363543933] 
\draw  [color={rgb, 255:red, 139; green, 6; blue, 24 }  ,draw opacity=1 ][fill={rgb, 255:red, 139; green, 6; blue, 24 }  ,fill opacity=1 ] (530.66,213.05) .. controls (530.66,211.82) and (531.54,210.82) .. (532.62,210.82) .. controls (533.7,210.82) and (534.58,211.82) .. (534.58,213.05) .. controls (534.58,214.28) and (533.7,215.28) .. (532.62,215.28) .. controls (531.54,215.28) and (530.66,214.28) .. (530.66,213.05) -- cycle ;
%Straight Lines [id:da5988735794676354] 
\draw    (534.58,169.8) -- (565.03,185.14) ;
%Straight Lines [id:da788774950668632] 
\draw    (534.58,213.05) -- (565.03,185.14) ;
%Straight Lines [id:da03797084738159218] 
\draw    (565.03,185.14) -- (607,172.5) ;
%Shape: Parallelogram [id:dp5700377145786201] 
\draw  [color={rgb, 255:red, 0; green, 0; blue, 0 }  ,draw opacity=1 ][fill={rgb, 255:red, 74; green, 74; blue, 74 }  ,fill opacity=0.38 ] (113.43,272.98) -- (113.45,115.45) -- (203.43,87.03) -- (203.41,244.56) -- cycle ;
%Straight Lines [id:da3346362854106609] 
\draw [color={rgb, 255:red, 139; green, 6; blue, 24 }  ,draw opacity=1 ]   (18.36,164.57) -- (54.68,164.57) ;
%Straight Lines [id:da812283867141578] 
\draw [color={rgb, 255:red, 139; green, 6; blue, 24 }  ,draw opacity=1 ]   (18.36,215.36) -- (54.68,215.36) ;
%Shape: Ellipse [id:dp7404132838036065] 
\draw  [color={rgb, 255:red, 139; green, 6; blue, 24 }  ,draw opacity=1 ][fill={rgb, 255:red, 139; green, 6; blue, 24 }  ,fill opacity=1 ] (152.72,191.3) .. controls (152.72,190.08) and (153.59,189.1) .. (154.65,189.1) .. controls (155.72,189.1) and (156.58,190.08) .. (156.58,191.3) .. controls (156.58,192.52) and (155.72,193.51) .. (154.65,193.51) .. controls (153.59,193.51) and (152.72,192.52) .. (152.72,191.3) -- cycle ;
%Straight Lines [id:da7014536522942414] 
\draw [color={rgb, 255:red, 139; green, 6; blue, 24 }  ,draw opacity=1 ]   (54.68,164.57) -- (86.3,191.3) ;
%Straight Lines [id:da6957537530549055] 
\draw [color={rgb, 255:red, 139; green, 6; blue, 24 }  ,draw opacity=1 ]   (54.68,215.36) -- (86.3,191.3) ;
%Straight Lines [id:da06615819146937119] 
\draw    (156.58,191.3) -- (203.44,175.26) ;
%Straight Lines [id:da25677724132160895] 
\draw [color={rgb, 255:red, 139; green, 6; blue, 24 }  ,draw opacity=1 ]   (86.3,191.3) -- (156.58,191.3) ;

% Text Node
\draw (182.62,96.57) node [anchor=north west][inner sep=0.75pt]  [font=\small]  {$\CB$};
% Text Node
\draw (586.95,96.58) node [anchor=north west][inner sep=0.75pt]  [font=\small]  {$\CB$};
% Text Node
\draw (444.06,156.91) node [anchor=north west][inner sep=0.75pt]    {$x$};
% Text Node
\draw (5.34,154.29) node [anchor=north west][inner sep=0.75pt]    {$x$};
% Text Node
\draw (441.46,202.36) node [anchor=north west][inner sep=0.75pt]    {$y$};
% Text Node
\draw (4.16,203.75) node [anchor=north west][inner sep=0.75pt]    {$y$};
% Text Node
\draw (100.22,192.15) node [anchor=north west][inner sep=0.75pt]    {$z$};
% Text Node
\draw (547.58,155.44) node [anchor=north west][inner sep=0.75pt]    {$a$};
% Text Node
\draw (545.52,202.04) node [anchor=north west][inner sep=0.75pt]    {$b$};
% Text Node
\draw (586.34,177.07) node [anchor=north west][inner sep=0.75pt]    {$c$};
% Text Node
\draw (175.18,184.13) node [anchor=north west][inner sep=0.75pt]    {$c$};
% Text Node
\draw (526.75,145.95) node [anchor=north west][inner sep=0.75pt]  [font=\small]  {$\psi $};
% Text Node
\draw (526.85,218.86) node [anchor=north west][inner sep=0.75pt]  [font=\small]  {$\chi $};
% Text Node
\draw (147.8,196.13) node [anchor=north west][inner sep=0.75pt]  [font=\small]  {$\xi $};
% Text Node
\draw (560.58,186.84) node [anchor=north west][inner sep=0.75pt]  [font=\small]  {$\alpha$};
% Text Node
\draw (79,198.2) node [anchor=north west][inner sep=0.75pt]  [font=\small]  {$i$};
% Text Node
\draw (241,162.4) node [anchor=north west][inner sep=0.75pt]    {$=\ \sum\limits_{( a,\psi )( b,\chi ) ,\alpha}\left[ \Phi _{x,y}^{c}\right]{_{( a,\psi )( b,\chi ) ,\alpha}^{z,\xi,i}}$};

\end{tikzpicture}
    \caption{The complex numbers $\left[ \Phi _{x,y}^{c}\right]{_{( a,\psi )( b,\chi ) ,\alpha}^{z,\xi,i}}$ relate the fusion spaces of $\CZ(\CC)$ and $\CB$.}
    \label{fig:Phi definition}
\end{figure}

\begin{figure}
    \centering

\tikzset{every picture/.style={line width=0.75pt}} %set default line width to 0.75pt        

\begin{tikzpicture}[x=0.75pt,y=0.75pt,yscale=-0.9,xscale=0.9]
%uncomment if require: \path (0,881); %set diagram left start at 0, and has height of 881

%Shape: Parallelogram [id:dp5071201530638992] 
\draw  [color={rgb, 255:red, 0; green, 0; blue, 0 }  ,draw opacity=1 ][fill={rgb, 255:red, 74; green, 74; blue, 74 }  ,fill opacity=0.38 ] (153.88,362.21) -- (153.9,242.86) -- (233.6,221.36) -- (233.58,340.71) -- cycle ;
%Straight Lines [id:da7370921953956616] 
\draw [color={rgb, 255:red, 139; green, 6; blue, 24 }  ,draw opacity=1 ]   (96.29,261.45) -- (166.83,261.45) ;
%Straight Lines [id:da17953444830556864] 
\draw [color={rgb, 255:red, 139; green, 6; blue, 24 }  ,draw opacity=1 ]   (96.01,297.97) -- (167.59,297.97) ;
%Shape: Ellipse [id:dp6460834498803231] 
\draw  [color={rgb, 255:red, 139; green, 6; blue, 24 }  ,draw opacity=1 ][fill={rgb, 255:red, 139; green, 6; blue, 24 }  ,fill opacity=1 ] (163.41,261.45) .. controls (163.41,260.53) and (164.17,259.78) .. (165.12,259.78) .. controls (166.06,259.78) and (166.83,260.53) .. (166.83,261.45) .. controls (166.83,262.38) and (166.06,263.12) .. (165.12,263.12) .. controls (164.17,263.12) and (163.41,262.38) .. (163.41,261.45) -- cycle ;
%Shape: Ellipse [id:dp1840279363543933] 
\draw  [color={rgb, 255:red, 139; green, 6; blue, 24 }  ,draw opacity=1 ][fill={rgb, 255:red, 139; green, 6; blue, 24 }  ,fill opacity=1 ] (164.16,297.97) .. controls (164.16,297.05) and (164.93,296.3) .. (165.88,296.3) .. controls (166.82,296.3) and (167.59,297.05) .. (167.59,297.97) .. controls (167.59,298.89) and (166.82,299.64) .. (165.88,299.64) .. controls (164.93,299.64) and (164.16,298.89) .. (164.16,297.97) -- cycle ;
%Shape: Parallelogram [id:dp5700377145786201] 
\draw  [color={rgb, 255:red, 0; green, 0; blue, 0 }  ,draw opacity=1 ][fill={rgb, 255:red, 74; green, 74; blue, 74 }  ,fill opacity=0.38 ] (524.21,623.44) -- (524.23,505.42) -- (602.86,484.11) -- (602.84,602.13) -- cycle ;
%Straight Lines [id:da3346362854106609] 
\draw [color={rgb, 255:red, 139; green, 6; blue, 24 }  ,draw opacity=1 ]   (447.55,523.98) -- (502.1,561.9) ;
%Shape: Ellipse [id:dp7404132838036065] 
\draw  [color={rgb, 255:red, 139; green, 6; blue, 24 }  ,draw opacity=1 ][fill={rgb, 255:red, 139; green, 6; blue, 24 }  ,fill opacity=1 ] (558.55,562.24) .. controls (558.55,561.32) and (559.3,560.58) .. (560.23,560.58) .. controls (561.17,560.58) and (561.92,561.32) .. (561.92,562.24) .. controls (561.92,563.15) and (561.17,563.89) .. (560.23,563.89) .. controls (559.3,563.89) and (558.55,563.15) .. (558.55,562.24) -- cycle ;
%Straight Lines [id:da7014536522942414] 
\draw [color={rgb, 255:red, 139; green, 6; blue, 24 }  ,draw opacity=1 ]   (446.79,561.13) -- (474.75,580.81) ;
%Straight Lines [id:da6957537530549055] 
\draw [color={rgb, 255:red, 139; green, 6; blue, 24 }  ,draw opacity=1 ]   (447.87,598.84) -- (502.1,561.9) ;
%Straight Lines [id:da06615819146937119] 
\draw    (561.92,562.24) -- (602.86,550.22) ;
%Straight Lines [id:da25677724132160895] 
\draw [color={rgb, 255:red, 139; green, 6; blue, 24 }  ,draw opacity=1 ]   (500.51,562.24) -- (561.92,562.24) ;
%Shape: Parallelogram [id:dp428570746998566] 
\draw  [color={rgb, 255:red, 0; green, 0; blue, 0 }  ,draw opacity=1 ][fill={rgb, 255:red, 74; green, 74; blue, 74 }  ,fill opacity=0.38 ] (523.06,352.22) -- (523.08,234.2) -- (601.71,212.9) -- (601.69,330.92) -- cycle ;
%Straight Lines [id:da12204816031543986] 
\draw [color={rgb, 255:red, 139; green, 6; blue, 24 }  ,draw opacity=1 ]   (446.4,252.76) -- (500.94,290.69) ;
%Shape: Ellipse [id:dp6977808521638921] 
\draw  [color={rgb, 255:red, 139; green, 6; blue, 24 }  ,draw opacity=1 ][fill={rgb, 255:red, 139; green, 6; blue, 24 }  ,fill opacity=1 ] (557.39,291.02) .. controls (557.39,290.11) and (558.15,289.37) .. (559.08,289.37) .. controls (560.01,289.37) and (560.77,290.11) .. (560.77,291.02) .. controls (560.77,291.93) and (560.01,292.67) .. (559.08,292.67) .. controls (558.15,292.67) and (557.39,291.93) .. (557.39,291.02) -- cycle ;
%Straight Lines [id:da6534465064737073] 
\draw [color={rgb, 255:red, 139; green, 6; blue, 24 }  ,draw opacity=1 ]   (445.64,289.91) -- (473.67,271.72) ;
%Straight Lines [id:da9828010832203633] 
\draw [color={rgb, 255:red, 139; green, 6; blue, 24 }  ,draw opacity=1 ]   (446.72,327.62) -- (500.94,290.69) ;
%Straight Lines [id:da7416419109922913] 
\draw    (560.77,291.02) -- (601.71,279) ;
%Straight Lines [id:da7332694291069444] 
\draw [color={rgb, 255:red, 139; green, 6; blue, 24 }  ,draw opacity=1 ]   (499.36,291.02) -- (560.77,291.02) ;
%Straight Lines [id:da3039297058275816] 
\draw [color={rgb, 255:red, 139; green, 6; blue, 24 }  ,draw opacity=1 ]   (96.77,336.66) -- (168.34,336.66) ;
%Shape: Ellipse [id:dp2311680570659299] 
\draw  [color={rgb, 255:red, 139; green, 6; blue, 24 }  ,draw opacity=1 ][fill={rgb, 255:red, 139; green, 6; blue, 24 }  ,fill opacity=1 ] (164.92,336.66) .. controls (164.92,335.74) and (165.69,334.99) .. (166.63,334.99) .. controls (167.58,334.99) and (168.34,335.74) .. (168.34,336.66) .. controls (168.34,337.59) and (167.58,338.33) .. (166.63,338.33) .. controls (165.69,338.33) and (164.92,337.59) .. (164.92,336.66) -- cycle ;
%Straight Lines [id:da979908786493764] 
\draw [color={rgb, 255:red, 0; green, 0; blue, 0 }  ,draw opacity=1 ]   (166.47,261.99) -- (221.02,299.91) ;
%Straight Lines [id:da8661919757233254] 
\draw [color={rgb, 255:red, 0; green, 0; blue, 0 }  ,draw opacity=1 ]   (165.71,299.13) -- (193.74,280.95) ;
%Straight Lines [id:da19948357160674401] 
\draw [color={rgb, 255:red, 0; green, 0; blue, 0 }  ,draw opacity=1 ]   (166.79,336.85) -- (221.02,299.91) ;
%Straight Lines [id:da45636541872427405] 
\draw [color={rgb, 255:red, 0; green, 0; blue, 0 }  ,draw opacity=1 ]   (221.02,299.91) -- (233.9,294.49) ;
%Shape: Parallelogram [id:dp617606102167132] 
\draw  [color={rgb, 255:red, 0; green, 0; blue, 0 }  ,draw opacity=1 ][fill={rgb, 255:red, 74; green, 74; blue, 74 }  ,fill opacity=0.38 ] (357.14,144.4) -- (357.16,26.38) -- (435.79,5.08) -- (435.77,123.09) -- cycle ;
%Straight Lines [id:da05175208098332529] 
\draw [color={rgb, 255:red, 139; green, 6; blue, 24 }  ,draw opacity=1 ]   (281.23,51.13) -- (308.51,70.09) ;
%Shape: Ellipse [id:dp823673157417386] 
\draw  [color={rgb, 255:red, 139; green, 6; blue, 24 }  ,draw opacity=1 ][fill={rgb, 255:red, 139; green, 6; blue, 24 }  ,fill opacity=1 ] (369.34,70.42) .. controls (369.34,69.5) and (370.1,68.76) .. (371.03,68.76) .. controls (371.96,68.76) and (372.72,69.5) .. (372.72,70.42) .. controls (372.72,71.33) and (371.96,72.07) .. (371.03,72.07) .. controls (370.1,72.07) and (369.34,71.33) .. (369.34,70.42) -- cycle ;
%Straight Lines [id:da20478805845139558] 
\draw [color={rgb, 255:red, 139; green, 6; blue, 24 }  ,draw opacity=1 ]   (280.47,88.28) -- (308.51,70.09) ;
%Straight Lines [id:da3647389459359004] 
\draw [color={rgb, 255:red, 139; green, 6; blue, 24 }  ,draw opacity=1 ]   (283.83,118.25) -- (371.03,116.85) ;
%Straight Lines [id:da8080393100601955] 
\draw    (406.64,90.54) -- (435.79,82.02) ;
%Straight Lines [id:da4897458363863959] 
\draw [color={rgb, 255:red, 139; green, 6; blue, 24 }  ,draw opacity=1 ]   (308.51,70.09) -- (371.03,70.42) ;
%Shape: Ellipse [id:dp08141326035597074] 
\draw  [color={rgb, 255:red, 139; green, 6; blue, 24 }  ,draw opacity=1 ][fill={rgb, 255:red, 139; green, 6; blue, 24 }  ,fill opacity=1 ] (369.32,116.85) .. controls (369.32,115.93) and (370.08,115.18) .. (371.03,115.18) .. controls (371.97,115.18) and (372.74,115.93) .. (372.74,116.85) .. controls (372.74,117.77) and (371.97,118.52) .. (371.03,118.52) .. controls (370.08,118.52) and (369.32,117.77) .. (369.32,116.85) -- cycle ;
%Straight Lines [id:da7217877582067771] 
\draw    (372.72,70.42) -- (406.64,90.54) ;
%Straight Lines [id:da8270374444199559] 
\draw    (372.74,116.85) -- (406.64,90.54) ;
%Shape: Parallelogram [id:dp05922209576788795] 
\draw  [color={rgb, 255:red, 0; green, 0; blue, 0 }  ,draw opacity=1 ][fill={rgb, 255:red, 74; green, 74; blue, 74 }  ,fill opacity=0.38 ] (153.88,625.75) -- (153.9,506.4) -- (233.6,484.89) -- (233.58,604.25) -- cycle ;
%Straight Lines [id:da3257440657897954] 
\draw [color={rgb, 255:red, 139; green, 6; blue, 24 }  ,draw opacity=1 ]   (96.29,524.99) -- (166.83,524.99) ;
%Straight Lines [id:da20696627121890132] 
\draw [color={rgb, 255:red, 139; green, 6; blue, 24 }  ,draw opacity=1 ]   (96.01,561.51) -- (167.59,561.51) ;
%Shape: Ellipse [id:dp6076259855273993] 
\draw  [color={rgb, 255:red, 139; green, 6; blue, 24 }  ,draw opacity=1 ][fill={rgb, 255:red, 139; green, 6; blue, 24 }  ,fill opacity=1 ] (163.41,524.99) .. controls (163.41,524.07) and (164.17,523.32) .. (165.12,523.32) .. controls (166.06,523.32) and (166.83,524.07) .. (166.83,524.99) .. controls (166.83,525.92) and (166.06,526.66) .. (165.12,526.66) .. controls (164.17,526.66) and (163.41,525.92) .. (163.41,524.99) -- cycle ;
%Shape: Ellipse [id:dp6296927155587009] 
\draw  [color={rgb, 255:red, 139; green, 6; blue, 24 }  ,draw opacity=1 ][fill={rgb, 255:red, 139; green, 6; blue, 24 }  ,fill opacity=1 ] (164.16,561.51) .. controls (164.16,560.58) and (164.93,559.84) .. (165.88,559.84) .. controls (166.82,559.84) and (167.59,560.58) .. (167.59,561.51) .. controls (167.59,562.43) and (166.82,563.18) .. (165.88,563.18) .. controls (164.93,563.18) and (164.16,562.43) .. (164.16,561.51) -- cycle ;
%Straight Lines [id:da8381861706748448] 
\draw [color={rgb, 255:red, 139; green, 6; blue, 24 }  ,draw opacity=1 ]   (96.77,600.2) -- (168.34,600.2) ;
%Shape: Ellipse [id:dp30998956329557714] 
\draw  [color={rgb, 255:red, 139; green, 6; blue, 24 }  ,draw opacity=1 ][fill={rgb, 255:red, 139; green, 6; blue, 24 }  ,fill opacity=1 ] (164.92,600.2) .. controls (164.92,599.28) and (165.69,598.53) .. (166.63,598.53) .. controls (167.58,598.53) and (168.34,599.28) .. (168.34,600.2) .. controls (168.34,601.13) and (167.58,601.87) .. (166.63,601.87) .. controls (165.69,601.87) and (164.92,601.13) .. (164.92,600.2) -- cycle ;
%Straight Lines [id:da30117753510948686] 
\draw [color={rgb, 255:red, 0; green, 0; blue, 0 }  ,draw opacity=1 ]   (166.47,525.52) -- (221.02,563.45) ;
%Straight Lines [id:da03652127397841898] 
\draw [color={rgb, 255:red, 0; green, 0; blue, 0 }  ,draw opacity=1 ]   (165.71,562.67) -- (193.91,581.92) ;
%Straight Lines [id:da08823951632364502] 
\draw [color={rgb, 255:red, 0; green, 0; blue, 0 }  ,draw opacity=1 ]   (166.79,600.39) -- (221.02,563.45) ;
%Straight Lines [id:da01963711920737199] 
\draw [color={rgb, 255:red, 0; green, 0; blue, 0 }  ,draw opacity=1 ]   (221.02,563.45) -- (233.9,558.03) ;
%Shape: Parallelogram [id:dp8626402584132267] 
\draw  [color={rgb, 255:red, 0; green, 0; blue, 0 }  ,draw opacity=1 ][fill={rgb, 255:red, 74; green, 74; blue, 74 }  ,fill opacity=0.38 ] (339.71,844.5) -- (339.73,726.48) -- (418.36,705.18) -- (418.34,823.19) -- cycle ;
%Shape: Ellipse [id:dp4308412155897874] 
\draw  [color={rgb, 255:red, 139; green, 6; blue, 24 }  ,draw opacity=1 ][fill={rgb, 255:red, 139; green, 6; blue, 24 }  ,fill opacity=1 ] (351.91,749.62) .. controls (351.91,748.71) and (352.67,747.97) .. (353.6,747.97) .. controls (354.53,747.97) and (355.29,748.71) .. (355.29,749.62) .. controls (355.29,750.53) and (354.53,751.27) .. (353.6,751.27) .. controls (352.67,751.27) and (351.91,750.53) .. (351.91,749.62) -- cycle ;
%Straight Lines [id:da7719533053868681] 
\draw    (389.21,769.74) -- (418.36,761.22) ;
%Straight Lines [id:da9939304038218564] 
\draw [color={rgb, 255:red, 139; green, 6; blue, 24 }  ,draw opacity=1 ]   (271.78,749.62) -- (353.6,749.62) ;
%Shape: Ellipse [id:dp9429592349530695] 
\draw  [color={rgb, 255:red, 139; green, 6; blue, 24 }  ,draw opacity=1 ][fill={rgb, 255:red, 139; green, 6; blue, 24 }  ,fill opacity=1 ] (351.89,796.06) .. controls (351.89,795.13) and (352.66,794.39) .. (353.6,794.39) .. controls (354.55,794.39) and (355.31,795.13) .. (355.31,796.06) .. controls (355.31,796.98) and (354.55,797.73) .. (353.6,797.73) .. controls (352.66,797.73) and (351.89,796.98) .. (351.89,796.06) -- cycle ;
%Straight Lines [id:da05808388084280691] 
\draw    (355.29,749.62) -- (389.21,769.74) ;
%Straight Lines [id:da9735975965549029] 
\draw    (355.31,796.06) -- (389.21,769.74) ;
%Straight Lines [id:da27397451632113345] 
\draw [color={rgb, 255:red, 139; green, 6; blue, 24 }  ,draw opacity=1 ]   (267.99,776.71) -- (295.95,796.39) ;
%Straight Lines [id:da6092446924842493] 
\draw [color={rgb, 255:red, 139; green, 6; blue, 24 }  ,draw opacity=1 ]   (269.07,814.42) -- (295.95,796.39) ;
%Straight Lines [id:da2982624744381117] 
\draw [color={rgb, 255:red, 139; green, 6; blue, 24 }  ,draw opacity=1 ]   (296.02,796.06) -- (353.6,796.06) ;
%Straight Lines [id:da8573705714884304] 
\draw    (198,204.5) -- (258.55,146.88) ;
\draw [shift={(260,145.5)}, rotate = 136.42] [color={rgb, 255:red, 0; green, 0; blue, 0 }  ][line width=0.75]    (10.93,-3.29) .. controls (6.95,-1.4) and (3.31,-0.3) .. (0,0) .. controls (3.31,0.3) and (6.95,1.4) .. (10.93,3.29)   ;
%Straight Lines [id:da012328399177399918] 
\draw    (473,132.5) -- (523.79,198.91) ;
\draw [shift={(525,200.5)}, rotate = 232.59] [color={rgb, 255:red, 0; green, 0; blue, 0 }  ][line width=0.75]    (10.93,-3.29) .. controls (6.95,-1.4) and (3.31,-0.3) .. (0,0) .. controls (3.31,0.3) and (6.95,1.4) .. (10.93,3.29)   ;
%Straight Lines [id:da25626437141143144] 
\draw    (560,379.5) -- (559.02,473.5) ;
\draw [shift={(559,475.5)}, rotate = 270.6] [color={rgb, 255:red, 0; green, 0; blue, 0 }  ][line width=0.75]    (10.93,-3.29) .. controls (6.95,-1.4) and (3.31,-0.3) .. (0,0) .. controls (3.31,0.3) and (6.95,1.4) .. (10.93,3.29)   ;
%Straight Lines [id:da28423224002145864] 
\draw    (191,377.5) -- (190.02,471.5) ;
\draw [shift={(190,473.5)}, rotate = 270.6] [color={rgb, 255:red, 0; green, 0; blue, 0 }  ][line width=0.75]    (10.93,-3.29) .. controls (6.95,-1.4) and (3.31,-0.3) .. (0,0) .. controls (3.31,0.3) and (6.95,1.4) .. (10.93,3.29)   ;
%Straight Lines [id:da21400581810964592] 
\draw    (229,636.5) -- (279.79,702.91) ;
\draw [shift={(281,704.5)}, rotate = 232.59] [color={rgb, 255:red, 0; green, 0; blue, 0 }  ][line width=0.75]    (10.93,-3.29) .. controls (6.95,-1.4) and (3.31,-0.3) .. (0,0) .. controls (3.31,0.3) and (6.95,1.4) .. (10.93,3.29)   ;
%Straight Lines [id:da49910202781752244] 
\draw    (467,702.5) -- (527.55,644.88) ;
\draw [shift={(529,643.5)}, rotate = 136.42] [color={rgb, 255:red, 0; green, 0; blue, 0 }  ][line width=0.75]    (10.93,-3.29) .. controls (6.95,-1.4) and (3.31,-0.3) .. (0,0) .. controls (3.31,0.3) and (6.95,1.4) .. (10.93,3.29)   ;

% Text Node
\draw (583.41,489.48) node [anchor=north west][inner sep=0.75pt]  [font=\small]  {$\CB$};
% Text Node
\draw (212.11,230.18) node [anchor=north west][inner sep=0.75pt]  [font=\small]  {$\CB$};
% Text Node
\draw (86.98,250.9) node [anchor=north west][inner sep=0.75pt]    {$x$};
% Text Node
\draw (435.83,550.4) node [anchor=north west][inner sep=0.75pt]    {$y$};
% Text Node
\draw (85.46,288.05) node [anchor=north west][inner sep=0.75pt]    {$y$};
% Text Node
\draw (435.56,589.01) node [anchor=north west][inner sep=0.75pt]    {$z$};
% Text Node
\draw (480.61,572.7) node [anchor=north west][inner sep=0.75pt]  [font=\normalsize]  {$u$};
% Text Node
\draw (579.42,556.5) node [anchor=north west][inner sep=0.75pt]    {$d$};
% Text Node
\draw (158.23,263.61) node [anchor=north west][inner sep=0.75pt]  [font=\small]  {$\psi $};
% Text Node
\draw (159.07,299.44) node [anchor=north west][inner sep=0.75pt]  [font=\small]  {$\chi $};
% Text Node
\draw (554.49,563.65) node [anchor=north west][inner sep=0.75pt]  [font=\small]  {$\sigma $};
% Text Node
\draw (469.13,585) node [anchor=north west][inner sep=0.75pt]  [font=\small]  {$i$};
% Text Node
\draw (45.38,135.11) node [anchor=north west][inner sep=0.75pt]    {$\sum\limits_{( a,\psi )( b,\chi ) ,\alpha }\left[ \Phi _{x,y}^{e}\right]{_{( a,\psi )( b,\chi ) ,\alpha }^{v,\varphi ,k}}$};
% Text Node
\draw (437.34,514.03) node [anchor=north west][inner sep=0.75pt]    {$x$};
% Text Node
\draw (582.26,222.27) node [anchor=north west][inner sep=0.75pt]  [font=\small]  {$\CB$};
% Text Node
\draw (434.67,279.19) node [anchor=north west][inner sep=0.75pt]    {$y$};
% Text Node
\draw (434.41,317.79) node [anchor=north west][inner sep=0.75pt]    {$z$};
% Text Node
\draw (507,290.2) node [anchor=north west][inner sep=0.75pt]    {$w$};
% Text Node
\draw (576.26,285.29) node [anchor=north west][inner sep=0.75pt]    {$d$};
% Text Node
\draw (552.34,293.44) node [anchor=north west][inner sep=0.75pt]  [font=\small]  {$\sigma $};
% Text Node
\draw (436.19,242.81) node [anchor=north west][inner sep=0.75pt]    {$x$};
% Text Node
\draw (508.64,560.64) node [anchor=north west][inner sep=0.75pt]    {$w$};
% Text Node
\draw (482.64,261.92) node [anchor=north west][inner sep=0.75pt]    {$v$};
% Text Node
\draw (496.41,564.65) node [anchor=north west][inner sep=0.75pt]  [font=\small]  {$j$};
% Text Node
\draw (468.07,276.8) node [anchor=north west][inner sep=0.75pt]  [font=\small]  {$k$};
% Text Node
\draw (495.58,295.7) node [anchor=north west][inner sep=0.75pt]  [font=\small]  {$l$};
% Text Node
\draw (572.92,399.12) node [anchor=north west][inner sep=0.75pt]    {$\sum\limits_{k,v,l}\left[ F_{xyz}^{w}\right]_{k,v,l}^{i,u,j}$};
% Text Node
\draw (86.22,326.75) node [anchor=north west][inner sep=0.75pt]    {$z$};
% Text Node
\draw (159.31,341.13) node [anchor=north west][inner sep=0.75pt]  [font=\small]  {$\xi $};
% Text Node
\draw (175.44,293.73) node [anchor=north west][inner sep=0.75pt]    {$b$};
% Text Node
\draw (187.06,318.5) node [anchor=north west][inner sep=0.75pt]    {$c$};
% Text Node
\draw (175.96,252.36) node [anchor=north west][inner sep=0.75pt]    {$a$};
% Text Node
\draw (222.93,277.9) node [anchor=north west][inner sep=0.75pt]    {$d$};
% Text Node
\draw (207.02,273.26) node [anchor=north west][inner sep=0.75pt]    {$e$};
% Text Node
\draw (186.51,281.09) node [anchor=north west][inner sep=0.75pt]  [font=\small]  {$\alpha $};
% Text Node
\draw (213.78,303.76) node [anchor=north west][inner sep=0.75pt]  [font=\small]  {$\beta $};
% Text Node
\draw (416.36,12.9) node [anchor=north west][inner sep=0.75pt]  [font=\small]  {$\CB$};
% Text Node
\draw (269.51,77.56) node [anchor=north west][inner sep=0.75pt]    {$y$};
% Text Node
\draw (271.52,108.42) node [anchor=north west][inner sep=0.75pt]    {$z$};
% Text Node
\draw (383.07,102.23) node [anchor=north west][inner sep=0.75pt]    {$c$};
% Text Node
\draw (364.03,73.78) node [anchor=north west][inner sep=0.75pt]  [font=\small]  {$\varphi $};
% Text Node
\draw (271.02,41.18) node [anchor=north west][inner sep=0.75pt]    {$x$};
% Text Node
\draw (328.56,54) node [anchor=north west][inner sep=0.75pt]    {$v$};
% Text Node
\draw (302.14,75.17) node [anchor=north west][inner sep=0.75pt]  [font=\small]  {$k$};
% Text Node
\draw (364.63,122.38) node [anchor=north west][inner sep=0.75pt]  [font=\small]  {$\xi $};
% Text Node
\draw (417.64,67.43) node [anchor=north west][inner sep=0.75pt]    {$d$};
% Text Node
\draw (528.31,135.3) node [anchor=north west][inner sep=0.75pt]    {$\sum\limits_{( e,\varphi )( c,\xi ) ,i}\left[ \Phi _{v,z}^{d}\right]{_{( e,\varphi )( c,\xi ) ,\beta }^{w,\sigma ,l}}$};
% Text Node
\draw (388.09,64.34) node [anchor=north west][inner sep=0.75pt]    {$e$};
% Text Node
\draw (400.92,93.62) node [anchor=north west][inner sep=0.75pt]  [font=\small]  {$\beta $};
% Text Node
\draw (214.11,491.72) node [anchor=north west][inner sep=0.75pt]  [font=\small]  {$\CB$};
% Text Node
\draw (157.47,527.05) node [anchor=north west][inner sep=0.75pt]  [font=\small]  {$\psi $};
% Text Node
\draw (159.07,565.98) node [anchor=north west][inner sep=0.75pt]  [font=\small]  {$\chi $};
% Text Node
\draw (159.31,605.67) node [anchor=north west][inner sep=0.75pt]  [font=\small]  {$\xi $};
% Text Node
\draw (180.23,553.82) node [anchor=north west][inner sep=0.75pt]    {$b$};
% Text Node
\draw (191.87,525.73) node [anchor=north west][inner sep=0.75pt]    {$a$};
% Text Node
\draw (221.93,540.44) node [anchor=north west][inner sep=0.75pt]    {$d$};
% Text Node
\draw (187.26,581.55) node [anchor=north west][inner sep=0.75pt]  [font=\small]  {$\gamma $};
% Text Node
\draw (218.3,564.75) node [anchor=north west][inner sep=0.75pt]  [font=\small]  {$\delta $};
% Text Node
\draw (84.7,515.22) node [anchor=north west][inner sep=0.75pt]    {$x$};
% Text Node
\draw (83.19,552.37) node [anchor=north west][inner sep=0.75pt]    {$y$};
% Text Node
\draw (83.95,591.06) node [anchor=north west][inner sep=0.75pt]    {$z$};
% Text Node
\draw (53.62,397.82) node [anchor=north west][inner sep=0.75pt]    {$\sum\limits_{\alpha ,e,\beta }\left[ F_{abc}^{d}\right]_{\alpha ,e,\beta }^{\gamma ,f,\delta }$};
% Text Node
\draw (201.96,574.07) node [anchor=north west][inner sep=0.75pt]    {$f$};
% Text Node
\draw (53.77,721.35) node [anchor=north west][inner sep=0.75pt]    {$\sum\limits_{( b,\chi )( c,\xi ) ,\gamma }\left[ \Phi _{y,z}^{f}\right]{_{( b,\chi )( c,\xi ) ,\gamma }^{u,\rho ,i}}$};
% Text Node
\draw (365.64,784.53) node [anchor=north west][inner sep=0.75pt]    {$f$};
% Text Node
\draw (346.6,751.98) node [anchor=north west][inner sep=0.75pt]  [font=\small]  {$\psi $};
% Text Node
\draw (400.22,746.64) node [anchor=north west][inner sep=0.75pt]    {$d$};
% Text Node
\draw (489.46,709.8) node [anchor=north west][inner sep=0.75pt]    {$\sum\limits_{( a,\psi )( f,\rho ) ,\delta }\left[ \Phi _{x,u}^{d}\right]{_{( a,\psi )( f,\rho ) ,\delta }^{w,\sigma ,j}}$};
% Text Node
\draw (372.67,741.54) node [anchor=north west][inner sep=0.75pt]    {$a$};
% Text Node
\draw (383.49,772.82) node [anchor=north west][inner sep=0.75pt]  [font=\small]  {$\delta $};
% Text Node
\draw (398.42,712.79) node [anchor=north west][inner sep=0.75pt]  [font=\small]  {$\CB$};
% Text Node
\draw (260.47,739.31) node [anchor=north west][inner sep=0.75pt]    {$x$};
% Text Node
\draw (258.54,766.76) node [anchor=north west][inner sep=0.75pt]    {$y$};
% Text Node
\draw (258.27,803.82) node [anchor=north west][inner sep=0.75pt]    {$z$};
% Text Node
\draw (291.09,800.36) node [anchor=north west][inner sep=0.75pt]  [font=\small]  {$i$};
% Text Node
\draw (309.62,778.09) node [anchor=north west][inner sep=0.75pt]    {$u$};
% Text Node
\draw (347.6,798.81) node [anchor=north west][inner sep=0.75pt]  [font=\small]  {$\rho $};
% Text Node
\draw (176.45,589.69) node [anchor=north west][inner sep=0.75pt]    {$c$};

\end{tikzpicture}
    \caption{Associativity of fusion rules in $\CZ(\CC)$ and $\CB_L$ implies that the above diagram must commute.}
    \label{fig:Constraint on Phi}
\end{figure}
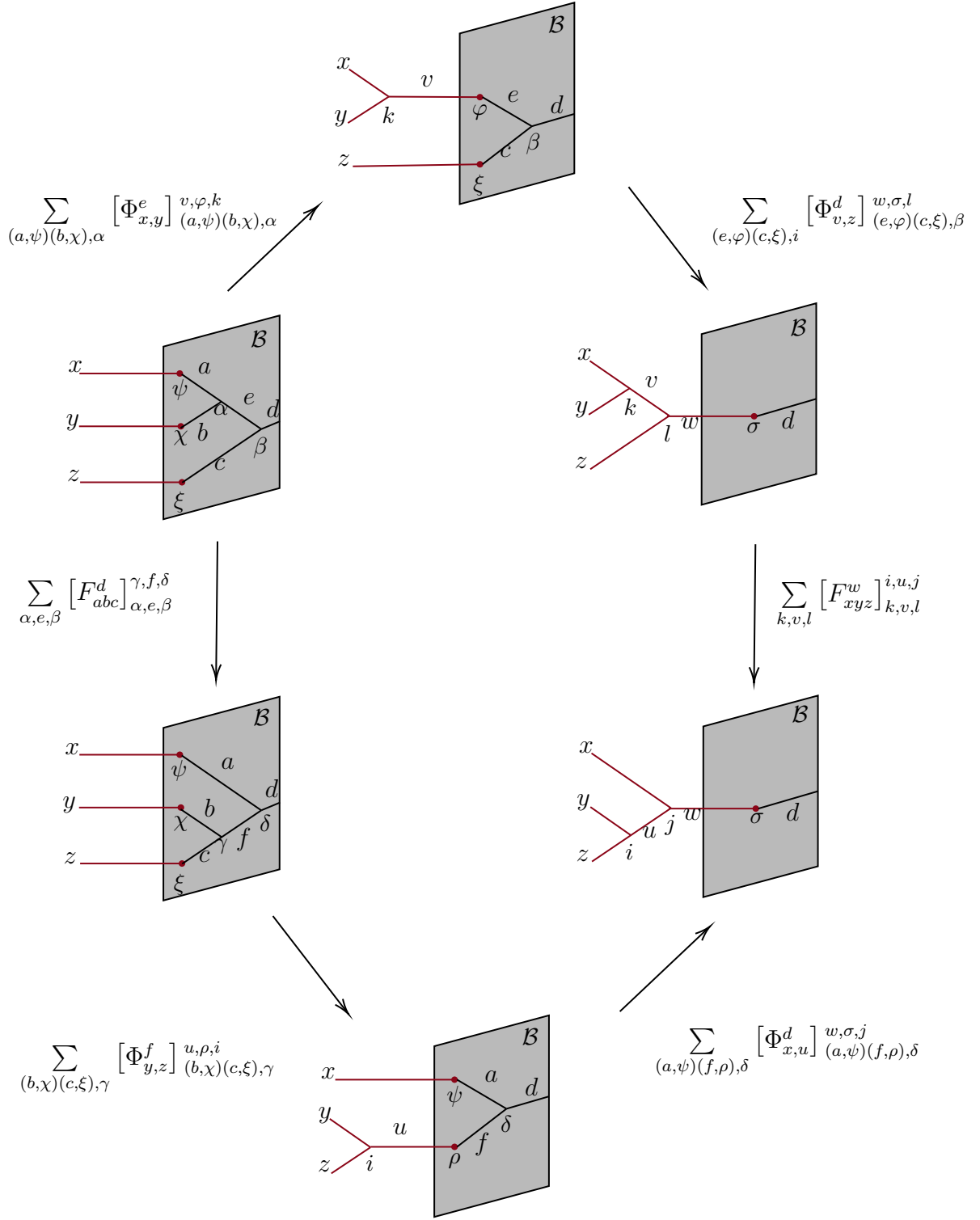

Recall the bulk-to-boundary map
\be
F_{\CB}: \mathcal{Z}(\CC) \to \CB~, 
\ee
introduced in \eqref{eq:bulk to boundary} where we will use $\CB$ to denote both a gapped boundary as well as the fusion category of line operators living on the gapped boundary. In order to describe how the full data of an algebra in $\mathcal{Z}(\CC)$ is mapped to $\CB$ under $F_{\CB}$, we have to determine how the fusion spaces in $\CZ(\CC)$ get mapped to fusion spaces in $\CB$. This is determined by an isomorphism 
\be
\label{eq: definition of Phi}
\Phi_{x,y}: F_{\CB}(x) \times F_{\CB}(y) \to F_{\CB}(x \times y)~.
\ee
By choosing a basis, we can write these isomorphisms explicitly in terms of some complex numbers as given in Fig. \ref{fig:Phi definition}.
The choice of isomorphisms $\Phi_{x,y}$ must be compatible with the associativity of fusion rules in both $\CZ(\CC)$ and $\CB$. This implies that $\Phi_{x,y}$ must satisfy the constraints in Fig. \ref{fig:Constraint on Phi}. The isomorphisms $\Phi$ make $F_{\CB}$ a strong monoidal functor (see, for example \cite{baez2004some,Kong:2009inh}). 

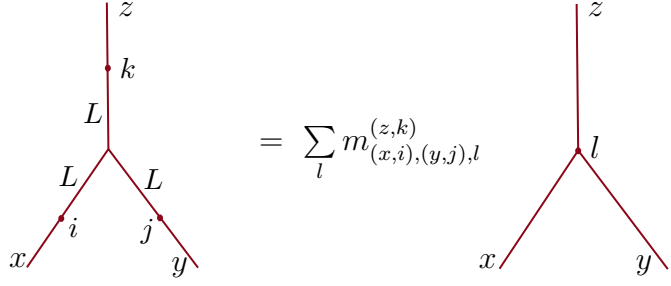
\begin{figure}[h!]
    \centering

\tikzset{every picture/.style={line width=0.75pt}} %set default line width to 0.75pt        

\begin{tikzpicture}[x=0.75pt,y=0.75pt,yscale=-0.7,xscale=0.7]
%uncomment if require: \path (0,496); %set diagram left start at 0, and has height of 496

%Straight Lines [id:da9415933665017819] 
\draw [color={rgb, 255:red, 139; green, 6; blue, 24 }  ,draw opacity=1 ][fill={rgb, 255:red, 139; green, 6; blue, 24 }  ,fill opacity=1 ]   (120.5,323) -- (178.6,238.01) ;
%Straight Lines [id:da21101691304436687] 
\draw [color={rgb, 255:red, 139; green, 6; blue, 24 }  ,draw opacity=1 ][fill={rgb, 255:red, 139; green, 6; blue, 24 }  ,fill opacity=1 ]   (242.5,323.25) -- (178.6,238.01) ;
%Straight Lines [id:da875525930995093] 
\draw [color={rgb, 255:red, 139; green, 6; blue, 24 }  ,draw opacity=1 ][fill={rgb, 255:red, 139; green, 6; blue, 24 }  ,fill opacity=1 ]   (178.6,238.01) -- (177.5,132.75) ;
%Shape: Ellipse [id:dp604318982156609] 
\draw  [color={rgb, 255:red, 139; green, 6; blue, 24 }  ,draw opacity=1 ][fill={rgb, 255:red, 139; green, 6; blue, 24 }  ,fill opacity=1 ] (213.99,287.47) .. controls (213.99,286.51) and (214.62,285.74) .. (215.4,285.74) .. controls (216.18,285.74) and (216.81,286.51) .. (216.81,287.47) .. controls (216.81,288.42) and (216.18,289.2) .. (215.4,289.2) .. controls (214.62,289.2) and (213.99,288.42) .. (213.99,287.47) -- cycle ;
%Shape: Ellipse [id:dp2008765308742959] 
\draw  [color={rgb, 255:red, 139; green, 6; blue, 24 }  ,draw opacity=1 ][fill={rgb, 255:red, 139; green, 6; blue, 24 }  ,fill opacity=1 ] (143.59,287.87) .. controls (143.59,286.91) and (144.22,286.14) .. (145,286.14) .. controls (145.78,286.14) and (146.41,286.91) .. (146.41,287.87) .. controls (146.41,288.82) and (145.78,289.6) .. (145,289.6) .. controls (144.22,289.6) and (143.59,288.82) .. (143.59,287.87) -- cycle ;
%Shape: Ellipse [id:dp014429472393691523] 
\draw  [color={rgb, 255:red, 139; green, 6; blue, 24 }  ,draw opacity=1 ][fill={rgb, 255:red, 139; green, 6; blue, 24 }  ,fill opacity=1 ] (176.79,179.87) .. controls (176.79,178.91) and (177.42,178.14) .. (178.2,178.14) .. controls (178.98,178.14) and (179.61,178.91) .. (179.61,179.87) .. controls (179.61,180.82) and (178.98,181.6) .. (178.2,181.6) .. controls (177.42,181.6) and (176.79,180.82) .. (176.79,179.87) -- cycle ;
%Straight Lines [id:da4393713821021201] 
\draw [color={rgb, 255:red, 139; green, 6; blue, 24 }  ,draw opacity=1 ][fill={rgb, 255:red, 139; green, 6; blue, 24 }  ,fill opacity=1 ]   (457.5,324) -- (513.6,239.01) ;
%Straight Lines [id:da22137399773841426] 
\draw [color={rgb, 255:red, 139; green, 6; blue, 24 }  ,draw opacity=1 ][fill={rgb, 255:red, 139; green, 6; blue, 24 }  ,fill opacity=1 ]   (577.5,324.25) -- (513.6,239.01) ;
%Straight Lines [id:da0693936045476865] 
\draw [color={rgb, 255:red, 139; green, 6; blue, 24 }  ,draw opacity=1 ][fill={rgb, 255:red, 139; green, 6; blue, 24 }  ,fill opacity=1 ]   (513.6,239.01) -- (512.5,133.75) ;
%Shape: Ellipse [id:dp6224123219203231] 
\draw  [color={rgb, 255:red, 139; green, 6; blue, 24 }  ,draw opacity=1 ][fill={rgb, 255:red, 139; green, 6; blue, 24 }  ,fill opacity=1 ] (512.19,239.28) .. controls (512.19,238.32) and (512.82,237.55) .. (513.6,237.55) .. controls (514.38,237.55) and (515.01,238.32) .. (515.01,239.28) .. controls (515.01,240.24) and (514.38,241.01) .. (513.6,241.01) .. controls (512.82,241.01) and (512.19,240.24) .. (512.19,239.28) -- cycle ;

% Text Node
\draw (105.73,310.72) node [anchor=north west][inner sep=0.75pt]    {$x$};
% Text Node
\draw (221.49,313.99) node [anchor=north west][inner sep=0.75pt]    {$y$};
% Text Node
\draw (183.31,130.18) node [anchor=north west][inner sep=0.75pt]    {$z$};
% Text Node
\draw (158.41,203.13) node [anchor=north west][inner sep=0.75pt]  [font=\small]  {$L$};
% Text Node
\draw (140.87,248.14) node [anchor=north west][inner sep=0.75pt]  [font=\small]  {$L$};
% Text Node
\draw (201.49,250.03) node [anchor=north west][inner sep=0.75pt]  [font=\small]  {$L$};
% Text Node
\draw (284,213.4) node [anchor=north west][inner sep=0.75pt]    {$=\ \sum\limits_{l} m_{( x,i) ,( y,j),l}^{( z,k)}$};
% Text Node
\draw (148,285.4) node [anchor=north west][inner sep=0.75pt]  [font=\small]  {$i$};
% Text Node
\draw (199.48,286.05) node [anchor=north west][inner sep=0.75pt]  [font=\small]  {$j$};
% Text Node
\draw (184.5,170.4) node [anchor=north west][inner sep=0.75pt]  [font=\small]  {$k$};
% Text Node
\draw (440.73,311.72) node [anchor=north west][inner sep=0.75pt]    {$x$};
% Text Node
\draw (552.49,311.99) node [anchor=north west][inner sep=0.75pt]    {$y$};
% Text Node
\draw (518.31,130.18) node [anchor=north west][inner sep=0.75pt]    {$z$};
% Text Node
\draw (520,226.4) node [anchor=north west][inner sep=0.75pt]    {$l$};

\end{tikzpicture}
    \caption{The multiplication on $L$ is determined by the complex numbers $m_{( x,i) ,( y,j) ,l}^{( z,k)}$.}
    \label{fig:multiplication m}
\end{figure}

\begin{figure}[h!]
    \centering

\tikzset{every picture/.style={line width=0.75pt}} %set default line width to 0.75pt        

\begin{tikzpicture}[x=0.75pt,y=0.75pt,yscale=-0.8,xscale=0.8]
%uncomment if require: \path (0,496); %set diagram left start at 0, and has height of 496

%Shape: Parallelogram [id:dp6610265419290424] 
\draw  [color={rgb, 255:red, 0; green, 0; blue, 0 }  ,draw opacity=1 ][fill={rgb, 255:red, 74; green, 74; blue, 74 }  ,fill opacity=0.38 ] (116.59,387.85) -- (116.63,155.8) -- (242.51,113.8) -- (242.46,345.85) -- cycle ;
%Straight Lines [id:da9415933665017819] 
\draw    (117,331.75) -- (178.6,241.01) ;
%Straight Lines [id:da21101691304436687] 
\draw    (242.5,326.25) -- (178.6,241.01) ;
%Straight Lines [id:da875525930995093] 
\draw    (178.6,241.01) -- (177.5,135.75) ;
%Shape: Ellipse [id:dp604318982156609] 
\draw  [color={rgb, 255:red, 0; green, 0; blue, 0 }  ,draw opacity=1 ][fill={rgb, 255:red, 0; green, 0; blue, 0 }  ,fill opacity=1 ] (213.99,290.47) .. controls (213.99,289.51) and (214.62,288.74) .. (215.4,288.74) .. controls (216.18,288.74) and (216.81,289.51) .. (216.81,290.47) .. controls (216.81,291.42) and (216.18,292.2) .. (215.4,292.2) .. controls (214.62,292.2) and (213.99,291.42) .. (213.99,290.47) -- cycle ;
%Shape: Ellipse [id:dp2008765308742959] 
\draw  [color={rgb, 255:red, 0; green, 0; blue, 0 }  ,draw opacity=1 ][fill={rgb, 255:red, 0; green, 0; blue, 0 }  ,fill opacity=1 ] (143.59,290.87) .. controls (143.59,289.91) and (144.22,289.14) .. (145,289.14) .. controls (145.78,289.14) and (146.41,289.91) .. (146.41,290.87) .. controls (146.41,291.82) and (145.78,292.6) .. (145,292.6) .. controls (144.22,292.6) and (143.59,291.82) .. (143.59,290.87) -- cycle ;
%Shape: Ellipse [id:dp014429472393691523] 
\draw  [color={rgb, 255:red, 0; green, 0; blue, 0 }  ,draw opacity=1 ][fill={rgb, 255:red, 0; green, 0; blue, 0 }  ,fill opacity=1 ] (176.79,182.87) .. controls (176.79,181.91) and (177.42,181.14) .. (178.2,181.14) .. controls (178.98,181.14) and (179.61,181.91) .. (179.61,182.87) .. controls (179.61,183.82) and (178.98,184.6) .. (178.2,184.6) .. controls (177.42,184.6) and (176.79,183.82) .. (176.79,182.87) -- cycle ;
%Shape: Parallelogram [id:dp29844963841780925] 
\draw  [color={rgb, 255:red, 0; green, 0; blue, 0 }  ,draw opacity=1 ][fill={rgb, 255:red, 74; green, 74; blue, 74 }  ,fill opacity=0.38 ] (451.59,385.85) -- (451.63,153.8) -- (577.51,111.8) -- (577.46,343.85) -- cycle ;
%Straight Lines [id:da4393713821021201] 
\draw    (452,329.75) -- (513.6,239.01) ;
%Straight Lines [id:da22137399773841426] 
\draw    (577.5,324.25) -- (513.6,239.01) ;
%Straight Lines [id:da0693936045476865] 
\draw    (513.6,239.01) -- (512.5,133.75) ;
%Shape: Ellipse [id:dp6224123219203231] 
\draw  [color={rgb, 255:red, 0; green, 0; blue, 0 }  ,draw opacity=1 ][fill={rgb, 255:red, 0; green, 0; blue, 0 }  ,fill opacity=1 ] (512.19,239.28) .. controls (512.19,238.32) and (512.82,237.55) .. (513.6,237.55) .. controls (514.38,237.55) and (515.01,238.32) .. (515.01,239.28) .. controls (515.01,240.24) and (514.38,241.01) .. (513.6,241.01) .. controls (512.82,241.01) and (512.19,240.24) .. (512.19,239.28) -- cycle ;

% Text Node
\draw (223.7,125.1) node [anchor=north west][inner sep=0.75pt]  [font=\normalsize]  {$\CB$};
% Text Node
\draw (117.73,295.72) node [anchor=north west][inner sep=0.75pt]    {$a$};
% Text Node
\draw (221.49,310.99) node [anchor=north west][inner sep=0.75pt]    {$b$};
% Text Node
\draw (180.31,149.18) node [anchor=north west][inner sep=0.75pt]    {$c$};
% Text Node
\draw (130.41,208.13) node [anchor=north west][inner sep=0.75pt]  [font=\small]  {$F( L)$};
% Text Node
\draw (115.87,250.14) node [anchor=north west][inner sep=0.75pt]  [font=\small]  {$F( L)$};
% Text Node
\draw (199.49,253.03) node [anchor=north west][inner sep=0.75pt]  [font=\small]  {$F( L)$};
% Text Node
\draw (281,218.4) node [anchor=north west][inner sep=0.75pt]    {$=\ \sum\limits_{\delta } \mu _{( a,\alpha ) ,( b,\beta ),\delta}^{( c,\gamma )  }$};
% Text Node
\draw (148,288.4) node [anchor=north west][inner sep=0.75pt]  [font=\small]  {$\alpha $};
% Text Node
\draw (199.48,289.05) node [anchor=north west][inner sep=0.75pt]  [font=\small]  {$\beta $};
% Text Node
\draw (191.5,172.4) node [anchor=north west][inner sep=0.75pt]  [font=\small]  {$\gamma $};
% Text Node
\draw (558.7,123.1) node [anchor=north west][inner sep=0.75pt]  [font=\normalsize]  {$\CB$};
% Text Node
\draw (464.73,278.72) node [anchor=north west][inner sep=0.75pt]    {$a$};
% Text Node
\draw (539.49,282.99) node [anchor=north west][inner sep=0.75pt]    {$b$};
% Text Node
\draw (518.31,179.18) node [anchor=north west][inner sep=0.75pt]    {$c$};
% Text Node
\draw (520,226.4) node [anchor=north west][inner sep=0.75pt]    {$\delta $};

\end{tikzpicture}
    \caption{A multiplication on $F(L)$ is determined by the complex numbers $\mu _{( a,\alpha ) ,( b,\beta ),\delta}^{( c,\gamma )  }$.}
    \label{fig:multiplication mu}
\end{figure}
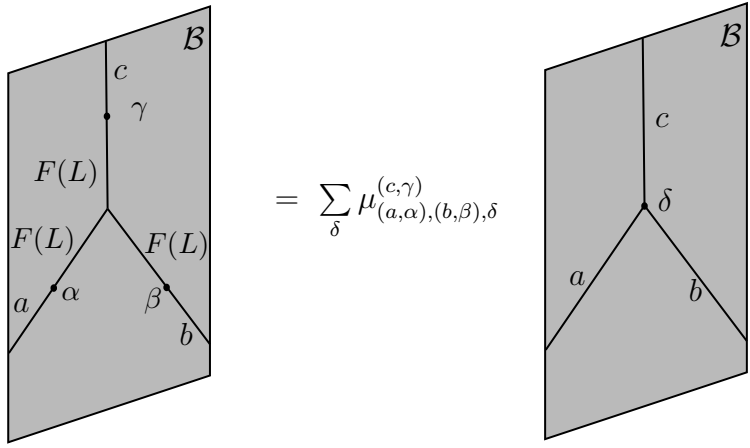

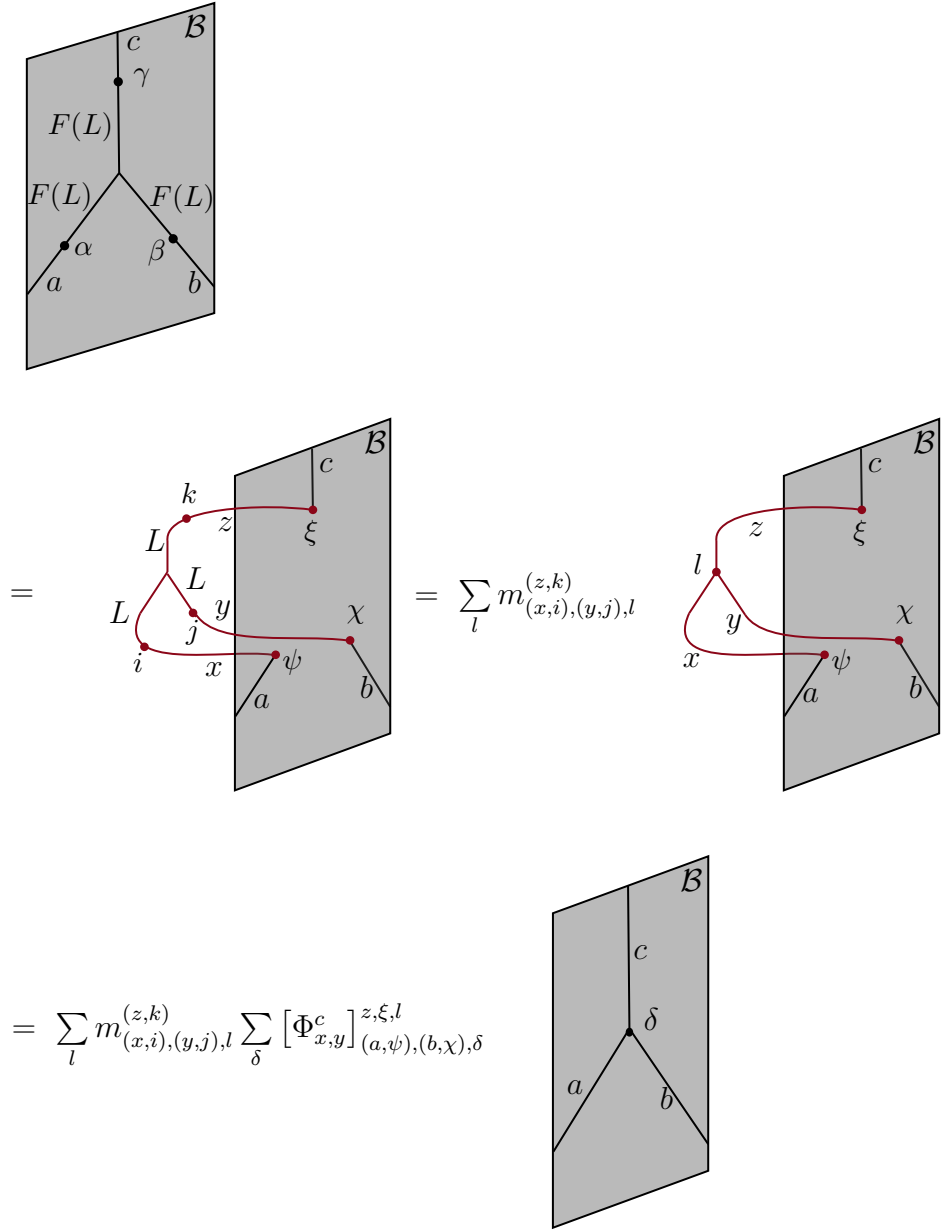
\begin{figure}[h!]
    \centering

\tikzset{every picture/.style={line width=0.75pt}} %set default line width to 0.75pt        

\begin{tikzpicture}[x=0.75pt,y=0.75pt,yscale=-1,xscale=1]
%uncomment if require: \path (0,678); %set diagram left start at 0, and has height of 678

%Straight Lines [id:da2087902824644159] 
\draw    (459.8,292.4) -- (459.38,259.97) ;
%Curve Lines [id:da2926194867988392] 
\draw [color={rgb, 255:red, 139; green, 6; blue, 24 }  ,draw opacity=1 ]   (373.6,342.8) .. controls (360,373.2) and (415.8,360.45) .. (439.55,363.65) ;
%Straight Lines [id:da17428000090694695] 
\draw    (498.42,389.85) -- (478,356.4) ;
%Shape: Parallelogram [id:dp12176200743007692] 
\draw  [color={rgb, 255:red, 0; green, 0; blue, 0 }  ,draw opacity=1 ][fill={rgb, 255:red, 74; green, 74; blue, 74 }  ,fill opacity=0.38 ] (420.86,431.85) -- (420.89,273.65) -- (498.42,245) -- (498.4,403.21) -- cycle ;
%Straight Lines [id:da6599718082487387] 
\draw    (421.12,394.6) -- (442.2,362) ;
%Straight Lines [id:da694684917771885] 
\draw [color={rgb, 255:red, 139; green, 6; blue, 24 }  ,draw opacity=1 ]   (387.2,306.4) -- (387.2,322) ;
%Shape: Circle [id:dp8940451595392945] 
\draw  [color={rgb, 255:red, 139; green, 6; blue, 24 }  ,draw opacity=1 ][fill={rgb, 255:red, 139; green, 6; blue, 24 }  ,fill opacity=1 ] (458.15,290.75) .. controls (458.15,289.84) and (458.89,289.1) .. (459.8,289.1) .. controls (460.71,289.1) and (461.45,289.84) .. (461.45,290.75) .. controls (461.45,291.66) and (460.71,292.4) .. (459.8,292.4) .. controls (458.89,292.4) and (458.15,291.66) .. (458.15,290.75) -- cycle ;
%Shape: Circle [id:dp2358935728081022] 
\draw  [color={rgb, 255:red, 139; green, 6; blue, 24 }  ,draw opacity=1 ][fill={rgb, 255:red, 139; green, 6; blue, 24 }  ,fill opacity=1 ] (439.55,363.65) .. controls (439.55,362.74) and (440.29,362) .. (441.2,362) .. controls (442.11,362) and (442.85,362.74) .. (442.85,363.65) .. controls (442.85,364.56) and (442.11,365.3) .. (441.2,365.3) .. controls (440.29,365.3) and (439.55,364.56) .. (439.55,363.65) -- cycle ;
%Shape: Circle [id:dp19992108586102064] 
\draw  [color={rgb, 255:red, 139; green, 6; blue, 24 }  ,draw opacity=1 ][fill={rgb, 255:red, 139; green, 6; blue, 24 }  ,fill opacity=1 ] (476.7,356.4) .. controls (476.7,355.49) and (477.44,354.75) .. (478.35,354.75) .. controls (479.26,354.75) and (480,355.49) .. (480,356.4) .. controls (480,357.31) and (479.26,358.05) .. (478.35,358.05) .. controls (477.44,358.05) and (476.7,357.31) .. (476.7,356.4) -- cycle ;
%Straight Lines [id:da6489760000491996] 
\draw [color={rgb, 255:red, 139; green, 6; blue, 24 }  ,draw opacity=1 ]   (373.6,342.8) -- (387.2,322) ;
%Shape: Parallelogram [id:dp348174606211172] 
\draw  [color={rgb, 255:red, 0; green, 0; blue, 0 }  ,draw opacity=1 ][fill={rgb, 255:red, 74; green, 74; blue, 74 }  ,fill opacity=0.38 ] (305.59,651.85) -- (305.61,493.65) -- (383.15,465) -- (383.12,623.21) -- cycle ;
%Straight Lines [id:da5485333537311946] 
\draw    (305.84,613.6) -- (343.79,551.74) ;
%Straight Lines [id:da49535280205518584] 
\draw    (383.15,609.85) -- (343.79,551.74) ;
%Straight Lines [id:da11156514473725188] 
\draw    (343.79,551.74) -- (343.11,479.97) ;
%Shape: Ellipse [id:dp7767519310418336] 
\draw  [color={rgb, 255:red, 0; green, 0; blue, 0 }  ,draw opacity=1 ][fill={rgb, 255:red, 0; green, 0; blue, 0 }  ,fill opacity=1 ] (342.44,553.47) .. controls (342.44,552.51) and (343.04,551.74) .. (343.79,551.74) .. controls (344.53,551.74) and (345.14,552.51) .. (345.14,553.47) .. controls (345.14,554.42) and (344.53,555.2) .. (343.79,555.2) .. controls (343.04,555.2) and (342.44,554.42) .. (342.44,553.47) -- cycle ;
%Straight Lines [id:da5673313288875034] 
\draw [color={rgb, 255:red, 139; green, 6; blue, 24 }  ,draw opacity=1 ]   (401.6,343.6) -- (387.2,322) ;
%Shape: Circle [id:dp33074553289842656] 
\draw  [color={rgb, 255:red, 139; green, 6; blue, 24 }  ,draw opacity=1 ][fill={rgb, 255:red, 139; green, 6; blue, 24 }  ,fill opacity=1 ] (385.55,322) .. controls (385.55,321.09) and (386.29,320.35) .. (387.2,320.35) .. controls (388.11,320.35) and (388.85,321.09) .. (388.85,322) .. controls (388.85,322.91) and (388.11,323.65) .. (387.2,323.65) .. controls (386.29,323.65) and (385.55,322.91) .. (385.55,322) -- cycle ;
%Curve Lines [id:da42573848779532275] 
\draw [color={rgb, 255:red, 139; green, 6; blue, 24 }  ,draw opacity=1 ]   (401.6,343.6) .. controls (412.6,360) and (452.95,353.2) .. (476.7,356.4) ;
%Curve Lines [id:da15851202206172854] 
\draw [color={rgb, 255:red, 139; green, 6; blue, 24 }  ,draw opacity=1 ]   (387.2,306.4) .. controls (385.4,289.2) and (436.05,287.55) .. (459.8,290.75) ;
%Straight Lines [id:da8881498331448328] 
\draw    (185.8,292.4) -- (185.38,259.97) ;
%Curve Lines [id:da28931581131380224] 
\draw [color={rgb, 255:red, 139; green, 6; blue, 24 }  ,draw opacity=1 ]   (99.6,342.8) .. controls (86,373.2) and (141.8,360.45) .. (165.55,363.65) ;
%Straight Lines [id:da3675610678888179] 
\draw    (224.42,389.85) -- (204,356.4) ;
%Shape: Parallelogram [id:dp5026623404087279] 
\draw  [color={rgb, 255:red, 0; green, 0; blue, 0 }  ,draw opacity=1 ][fill={rgb, 255:red, 74; green, 74; blue, 74 }  ,fill opacity=0.38 ] (146.86,431.85) -- (146.89,273.65) -- (224.42,245) -- (224.4,403.21) -- cycle ;
%Straight Lines [id:da8477182797285383] 
\draw    (147.12,394.6) -- (168.2,362) ;
%Straight Lines [id:da3371198720042269] 
\draw [color={rgb, 255:red, 139; green, 6; blue, 24 }  ,draw opacity=1 ]   (113.2,306.4) -- (113.2,322) ;
%Shape: Circle [id:dp5162462527290326] 
\draw  [color={rgb, 255:red, 139; green, 6; blue, 24 }  ,draw opacity=1 ][fill={rgb, 255:red, 139; green, 6; blue, 24 }  ,fill opacity=1 ] (184.15,290.75) .. controls (184.15,289.84) and (184.89,289.1) .. (185.8,289.1) .. controls (186.71,289.1) and (187.45,289.84) .. (187.45,290.75) .. controls (187.45,291.66) and (186.71,292.4) .. (185.8,292.4) .. controls (184.89,292.4) and (184.15,291.66) .. (184.15,290.75) -- cycle ;
%Shape: Circle [id:dp6400098514358489] 
\draw  [color={rgb, 255:red, 139; green, 6; blue, 24 }  ,draw opacity=1 ][fill={rgb, 255:red, 139; green, 6; blue, 24 }  ,fill opacity=1 ] (165.55,363.65) .. controls (165.55,362.74) and (166.29,362) .. (167.2,362) .. controls (168.11,362) and (168.85,362.74) .. (168.85,363.65) .. controls (168.85,364.56) and (168.11,365.3) .. (167.2,365.3) .. controls (166.29,365.3) and (165.55,364.56) .. (165.55,363.65) -- cycle ;
%Shape: Circle [id:dp6047320443622873] 
\draw  [color={rgb, 255:red, 139; green, 6; blue, 24 }  ,draw opacity=1 ][fill={rgb, 255:red, 139; green, 6; blue, 24 }  ,fill opacity=1 ] (202.7,356.4) .. controls (202.7,355.49) and (203.44,354.75) .. (204.35,354.75) .. controls (205.26,354.75) and (206,355.49) .. (206,356.4) .. controls (206,357.31) and (205.26,358.05) .. (204.35,358.05) .. controls (203.44,358.05) and (202.7,357.31) .. (202.7,356.4) -- cycle ;
%Straight Lines [id:da1985089447574644] 
\draw [color={rgb, 255:red, 139; green, 6; blue, 24 }  ,draw opacity=1 ]   (99.6,342.8) -- (113.2,322) ;
%Straight Lines [id:da21661963755851343] 
\draw [color={rgb, 255:red, 139; green, 6; blue, 24 }  ,draw opacity=1 ]   (127.6,344.6) -- (113.2,323) ;
%Curve Lines [id:da17175947985681728] 
\draw [color={rgb, 255:red, 139; green, 6; blue, 24 }  ,draw opacity=1 ]   (127.6,343.6) .. controls (138.6,360) and (178.95,353.2) .. (202.7,356.4) ;
%Curve Lines [id:da11535497350052604] 
\draw [color={rgb, 255:red, 139; green, 6; blue, 24 }  ,draw opacity=1 ]   (113.2,306.4) .. controls (111.4,289.2) and (162.05,287.55) .. (185.8,290.75) ;
%Shape: Circle [id:dp9079857541625648] 
\draw  [color={rgb, 255:red, 139; green, 6; blue, 24 }  ,draw opacity=1 ][fill={rgb, 255:red, 139; green, 6; blue, 24 }  ,fill opacity=1 ] (100.05,359.65) .. controls (100.05,358.74) and (100.79,358) .. (101.7,358) .. controls (102.61,358) and (103.35,358.74) .. (103.35,359.65) .. controls (103.35,360.56) and (102.61,361.3) .. (101.7,361.3) .. controls (100.79,361.3) and (100.05,360.56) .. (100.05,359.65) -- cycle ;
%Shape: Circle [id:dp27785219334822586] 
\draw  [color={rgb, 255:red, 139; green, 6; blue, 24 }  ,draw opacity=1 ][fill={rgb, 255:red, 139; green, 6; blue, 24 }  ,fill opacity=1 ] (124.3,342.6) .. controls (124.3,341.69) and (125.04,340.95) .. (125.95,340.95) .. controls (126.86,340.95) and (127.6,341.69) .. (127.6,342.6) .. controls (127.6,343.51) and (126.86,344.25) .. (125.95,344.25) .. controls (125.04,344.25) and (124.3,343.51) .. (124.3,342.6) -- cycle ;
%Shape: Circle [id:dp5622717742710011] 
\draw  [color={rgb, 255:red, 139; green, 6; blue, 24 }  ,draw opacity=1 ][fill={rgb, 255:red, 139; green, 6; blue, 24 }  ,fill opacity=1 ] (121.05,295.15) .. controls (121.05,294.24) and (121.79,293.5) .. (122.7,293.5) .. controls (123.61,293.5) and (124.35,294.24) .. (124.35,295.15) .. controls (124.35,296.06) and (123.61,296.8) .. (122.7,296.8) .. controls (121.79,296.8) and (121.05,296.06) .. (121.05,295.15) -- cycle ;
%Shape: Parallelogram [id:dp8929331055896793] 
\draw  [color={rgb, 255:red, 0; green, 0; blue, 0 }  ,draw opacity=1 ][fill={rgb, 255:red, 74; green, 74; blue, 74 }  ,fill opacity=0.38 ] (42.97,220) -- (43,64) -- (136.59,35.81) -- (136.56,191.8) -- cycle ;
%Straight Lines [id:da9139310763654576] 
\draw    (43.28,182.3) -- (89.08,121.31) ;
%Straight Lines [id:da9850088509560109] 
\draw    (136.58,178.6) -- (89.08,121.31) ;
%Straight Lines [id:da6064986764568228] 
\draw    (89.08,121.31) -- (88.26,50.55) ;
%Shape: Ellipse [id:dp4453723425990759] 
\draw  [color={rgb, 255:red, 0; green, 0; blue, 0 }  ,draw opacity=1 ][fill={rgb, 255:red, 0; green, 0; blue, 0 }  ,fill opacity=1 ] (86.82,75.41) .. controls (86.82,74.39) and (87.63,73.56) .. (88.63,73.56) .. controls (89.63,73.56) and (90.43,74.39) .. (90.43,75.41) .. controls (90.43,76.42) and (89.63,77.25) .. (88.63,77.25) .. controls (87.63,77.25) and (86.82,76.42) .. (86.82,75.41) -- cycle ;
%Shape: Ellipse [id:dp5907819738980172] 
\draw  [color={rgb, 255:red, 0; green, 0; blue, 0 }  ,draw opacity=1 ][fill={rgb, 255:red, 0; green, 0; blue, 0 }  ,fill opacity=1 ] (60.02,157.91) .. controls (60.02,156.89) and (60.83,156.06) .. (61.83,156.06) .. controls (62.83,156.06) and (63.64,156.89) .. (63.64,157.91) .. controls (63.64,158.92) and (62.83,159.75) .. (61.83,159.75) .. controls (60.83,159.75) and (60.02,158.92) .. (60.02,157.91) -- cycle ;
%Shape: Ellipse [id:dp29425188769017707] 
\draw  [color={rgb, 255:red, 0; green, 0; blue, 0 }  ,draw opacity=1 ][fill={rgb, 255:red, 0; green, 0; blue, 0 }  ,fill opacity=1 ] (114.23,154.41) .. controls (114.23,153.39) and (115.03,152.56) .. (116.03,152.56) .. controls (117.03,152.56) and (117.84,153.39) .. (117.84,154.41) .. controls (117.84,155.42) and (117.03,156.25) .. (116.03,156.25) .. controls (115.03,156.25) and (114.23,155.42) .. (114.23,154.41) -- cycle ;

% Text Node
\draw (235,320.4) node [anchor=north west][inner sep=0.75pt]    {$=\ \sum\limits_{l} m_{( x,i) ,( y,j) ,l}^{( z,k)}$};
% Text Node
\draw (367.72,469.97) node [anchor=north west][inner sep=0.75pt]  [font=\normalsize]  {$\CB$};
% Text Node
\draw (310.98,575.99) node [anchor=north west][inner sep=0.75pt]    {$a$};
% Text Node
\draw (357.43,579.3) node [anchor=north west][inner sep=0.75pt]    {$b$};
% Text Node
\draw (344.38,508.53) node [anchor=north west][inner sep=0.75pt]    {$c$};
% Text Node
\draw (349.42,539.72) node [anchor=north west][inner sep=0.75pt]    {$\delta $};
% Text Node
\draw (484,249.97) node [anchor=north west][inner sep=0.75pt]  [font=\normalsize]  {$\CB$};
% Text Node
\draw (428.65,378.79) node [anchor=north west][inner sep=0.75pt]    {$a$};
% Text Node
\draw (481.7,372.9) node [anchor=north west][inner sep=0.75pt]    {$b$};
% Text Node
\draw (461.38,263.37) node [anchor=north west][inner sep=0.75pt]    {$c$};
% Text Node
\draw (369,361.2) node [anchor=north west][inner sep=0.75pt]    {$x$};
% Text Node
\draw (390.4,342) node [anchor=north west][inner sep=0.75pt]    {$y$};
% Text Node
\draw (401.6,295.4) node [anchor=north west][inner sep=0.75pt]    {$z$};
% Text Node
\draw (443.2,359.4) node [anchor=north west][inner sep=0.75pt]  [font=\small]  {$\psi $};
% Text Node
\draw (475.2,337.4) node [anchor=north west][inner sep=0.75pt]  [font=\small]  {$\chi $};
% Text Node
\draw (374.2,311.6) node [anchor=north west][inner sep=0.75pt]  [font=\small]  {$l$};
% Text Node
\draw (454,294.2) node [anchor=north west][inner sep=0.75pt]  [font=\small]  {$\xi $};
% Text Node
\draw (34,537.4) node [anchor=north west][inner sep=0.75pt]    {$=\ \sum\limits_{l} m_{( x,i) ,( y,j) ,l}^{( z,k)}\sum\limits_{\delta }\left[ \Phi _{x,y}^{c}\right]_{( a,\psi ) ,( b,\chi ) ,\delta }^{ z,\xi  ,l}$};
% Text Node
\draw (210,249.97) node [anchor=north west][inner sep=0.75pt]  [font=\normalsize]  {$\CB$};
% Text Node
\draw (154.65,381.79) node [anchor=north west][inner sep=0.75pt]    {$a$};
% Text Node
\draw (207.7,372.9) node [anchor=north west][inner sep=0.75pt]    {$b$};
% Text Node
\draw (187.38,263.37) node [anchor=north west][inner sep=0.75pt]    {$c$};
% Text Node
\draw (130.5,365.7) node [anchor=north west][inner sep=0.75pt]    {$x$};
% Text Node
\draw (135.3,335) node [anchor=north west][inner sep=0.75pt]    {$y$};
% Text Node
\draw (136.6,292.9) node [anchor=north west][inner sep=0.75pt]    {$z$};
% Text Node
\draw (169.2,359.4) node [anchor=north west][inner sep=0.75pt]  [font=\small]  {$\psi $};
% Text Node
\draw (201.2,337.4) node [anchor=north west][inner sep=0.75pt]  [font=\small]  {$\chi $};
% Text Node
\draw (180,294.2) node [anchor=north west][inner sep=0.75pt]  [font=\small]  {$\xi $};
% Text Node
\draw (94.5,361.4) node [anchor=north west][inner sep=0.75pt]  [font=\small]  {$i$};
% Text Node
\draw (121,344.4) node [anchor=north west][inner sep=0.75pt]  [font=\small]  {$j$};
% Text Node
\draw (118.5,275.9) node [anchor=north west][inner sep=0.75pt]  [font=\small]  {$k$};
% Text Node
\draw (100,298.9) node [anchor=north west][inner sep=0.75pt]    {$L$};
% Text Node
\draw (82.5,335.4) node [anchor=north west][inner sep=0.75pt]    {$L$};
% Text Node
\draw (120.5,317.9) node [anchor=north west][inner sep=0.75pt]    {$L$};
% Text Node
\draw (33,329.4) node [anchor=north west][inner sep=0.75pt]    {$=$};
% Text Node
\draw (120.04,40.57) node [anchor=north west][inner sep=0.75pt]  [font=\normalsize]  {$\CB$};
% Text Node
\draw (51.14,172.04) node [anchor=north west][inner sep=0.75pt]    {$a$};
% Text Node
\draw (121.86,168.85) node [anchor=north west][inner sep=0.75pt]    {$b$};
% Text Node
\draw (91.25,51.09) node [anchor=north west][inner sep=0.75pt]    {$c$};
% Text Node
\draw (51.95,89.04) node [anchor=north west][inner sep=0.75pt]  [font=\small]  {$F( L)$};
% Text Node
\draw (41.63,124.28) node [anchor=north west][inner sep=0.75pt]  [font=\small]  {$F( L)$};
% Text Node
\draw (102.59,125.22) node [anchor=north west][inner sep=0.75pt]  [font=\small]  {$F( L)$};
% Text Node
\draw (65.05,154.7) node [anchor=north west][inner sep=0.75pt]  [font=\small]  {$\alpha $};
% Text Node
\draw (101.85,153.43) node [anchor=north west][inner sep=0.75pt]  [font=\small]  {$\beta $};
% Text Node
\draw (94.69,67.02) node [anchor=north west][inner sep=0.75pt]  [font=\small]  {$\gamma $};

\end{tikzpicture}
    \caption{The multiplication on $F(L)$ in terms of $m$ and $\Phi$. In the first equality we used Hom$(F(L),a)=\bigoplus_{x\in \CZ(\CC)} \text{Hom}(L,x) \times \text{Hom}(x,a)$, so that $\alpha=(\psi,x,i),\;\beta=(x,y,j),\;\gamma=(\xi,z,k)$. In the second equality we used Fig. \ref{fig:multiplication m} and in the third equality we used Fig. \ref{fig:Phi definition}}.
    \label{fig: map between multiplications}
\end{figure}

Consider a Lagrangian algebra $L$ in $\CZ(\CC)$. Let $m: L \times L \to L$ be the multiplication on $L$. By choosing a basis, we can write the multiplication $m$ explicitly as in Fig. \ref{fig:multiplication m}.
A multiplication on $F(L)$ is determined by the complex numbers as defined in Fig. \ref{fig:multiplication mu}.
In fact, $F(L)$ is an associative algebra in $\CB$ given by the multiplication
\be
\mu: m \circ \Phi~.
\ee
By choosing a basis, this can be explicitly depicted as in the Fig. \ref{fig: map between multiplications}. 
Comparing Fig. \ref{fig:multiplication mu} and Fig. \ref{fig: map between multiplications}, we get the equation
\be
\label{eq:relating multiplications}
\mu _{( a,(\psi ,x,i)) ,( b,(\chi ,y,j)) ,\delta }^{( c,(\xi ,z,k))}=\sum _{l} m_{( x,i) ,( y,j) ,l}^{(z,k)}  \left[ \Phi _{x,y}^{c}\right]{_{( a,\psi )( b,\chi ) ,\delta }^{z,\xi ,l}}~.
\ee

We are in particular interested in the perpendicular fusion of line operators on the canonical gapped boundary $\CB_{\CC}$. In this case, we have the forgetful functor $F\equiv F_{\CB_{\CC}}$
\be
F((a,e_a))=a.
\ee
Therefore, we have
\bea
&& F((a,e_a)) \times F((b,e_b))= a \times b~,\\
&& F((a,e_a) \times (b,e_b))=F((a\times b,e_{a\times b}))= a\times b~.
\eea
%\bea
%&& F_{\CB_{\CC}}((a,e_a)) \times F_{\CB_{\CC}}((b,e_b))= a \times b~,\\
%&& F_{\CB_{\CC}}((a,e_a) \times (b,e_b))=F_{\CB_{\CC}}((a\times b,e_{a\times b}))= a\times b~.
%\eea
and the isomorphisms $\Phi_{x,y}$ thus can be chosen to be trivial. In other words, there exists a basis in which the complex numbers $\left[ \Phi _{x,y}^{c}\right]{_{( a,\psi )( b,\chi ) ,\alpha }^{z,\xi ,k}}$ are all equal to $1$. This significantly simplifies the expression \eqref{eq:relating multiplications}.
\vspace{0.2cm}

\noindent \textbf{Remark:} The discussion above implies that the bulk-to-boundary map $F$ can also be used to transport non-Lagrangian algebras in $\CZ(\CC)$ to $\CC$. If $E$ is a general algebra in $\CZ(\CC)$, then \eqref{eq:relating multiplications} specifies a multiplication on $F(E)$. In general, $F(E)$ is not haploid. 

\subsection{Boundary-to-bulk map: From $A\in \CC$ to $L \in \CZ(\CC)$}

\label{sec:A to L}

Given an algebra $A$ with multiplication $\mu$ in $\CC$ the corresponding Lagrangian algebra with object $L$ and multiplication $m$ in $\CZ(\CC)$ can be found through a two-step process. Let $K: \CC \to \CZ(\CC)$ be the boundary-to-bulk map whose action on the line operators is given by
\be
K(a):=\bigoplus_{(x,e_x)\in \CZ(\CC)} N_{a}^{x} (x,e_x)
\ee
where $N_{a}^{x}$ counts the number of point operators at the junction of $a$ and $x$ in $\CC$ \cite[Section 8.1]{muger2003subfactors}.\footnote{$K$ is sometimes called the induction functor \cite[Section 9.2]{etingof2016tensor}.} $K(a)$ has the interpretation as the sum of line operators of the SymTFT $\CZ(\CC)$ which can form a junction with $a\in \CC$ on the gapped boundary $\CB_{\CC}$.  Given an algebra $A$ in $\CC$, consider the line operator $K(A)$ in $\CC$. In general, $K(A)$ does not admit the structure of a commutative algebra. This is in particular because $K(A)$ may contain non-bosonic simple line operators in $\CZ(\CC)$. However, since $A$ is an algebra in $\CC$, $K(A)$ does admit the structure of an algebra in $\CZ(\CC)$. Indeed, the map $K$ gives an algebra structure on $K(A)$. Similar to the case of $F$, to determine how $K$ transports algebras from $\CC$ to $\CZ(\CC)$, we have to determine how $K$ relates the fusion spaces of $\CC$ and $\CZ(\CC)$. We need to specify maps
\be
\Xi_{a,b}: K(a) \times K(b) \to K(a\times b)~.
\ee
$\Xi_{a,b}$ can be determined as follows. The bulk-to-boundary map $F$ and boundary-to-bulk map $K$ together satisfies \cite[Section 8.1]{muger2003subfactors}
\be
\text{Hom}_{\CC}(F((a,e_a)),b) \simeq \text{Hom}_{\CZ(\CC)}((a,e_a),K(b))~.
\ee
Let $\Theta_{(a,e_a),b}$ be this isomorphism. Let $\text{id}_{\CZ(\CC)}$ be the identity map on $\CZ(\CC)$. Consider the natural transformation
\be
\delta: \text{id}_{\CZ(\CC)} \to K \circ F~,
\ee
specified by
\be
\delta_{(a,e_a)}:= \Theta_{(a,e_a),F((a,e_a))}(\text{id}_{F(a,e_a)}) :  (a,e_a) \to K \circ F ((a,e_a))~.
\ee
$\delta_{(a,e_a)}$ is non-trivial because the action of the composition $K \circ F$ on $\CZ(\CC)$ is not the same as the identity map id$_{\CZ(\CC)}$. Similarly, we have 
\be
\rho: F \circ K \to \text{id}_{\CC}~,
\ee
specified by
\be
\rho_a:= \Theta_{K(a),a}^{-1}(\text{id}_{K(a)}):  F\circ K(a) \to a~.
\ee
$\rho_{(a,e_a)}$ is non-trivial because the action of the composition $F \circ K$ on $\CC$ is not the same as the identity map id$_{\CC}$. Using the data introduced above, we can write
\be
\Xi_{a,b}: K(a) \times K(b) \xrightarrow{\delta} K\circ F (K(a) \times K(b)) \xrightarrow{\Phi^{-1}} K\circ (F\circ K(a) \times F\circ K(a)) \xrightarrow{K(\rho)} K(a\times b)~. 
\ee
where $\Phi$ is the map introduced in \eqref{eq: definition of Phi} which relates the fusion spaces of $\CZ(\CC)$ and $\CC$ under $F$.\footnote{In other words, $K$ is the adjoint of the forgetful functor $F$, and therefore $K$ itself is a lax monoidal functor (see, for example, \cite[Lemma 2.7]{Kong:2009inh}, \cite{nlab:monoidal_adjunction}).}\footnote{Since $K$ is a two-sided adjoint of $F_{\CB_{\CC}}$, it is in fact both a lax and colax monoidal functor (see, for example, \cite{Kong:2009inh}). $K$ is generally not strong monoidal (\cite{445787} (see also \cite{Kong:2009inh})).  However, $K$ still maps Frobenius algebras to Frobenius algebras. This latter statement can be shown using \cite[Theorem 3.3]{yadav2024frobenius}.}
Therefore, $\Xi$ can be used to determine the multiplication on $K(A)$ given the multiplication on $A$ precisely as in \eqref{eq:relating multiplications}. However, note that unlike $\Phi$, $\Xi$ is not trivial in general. Therefore, finding the multiplication on $K(A)$ requires finding $\Xi$ explicitly.

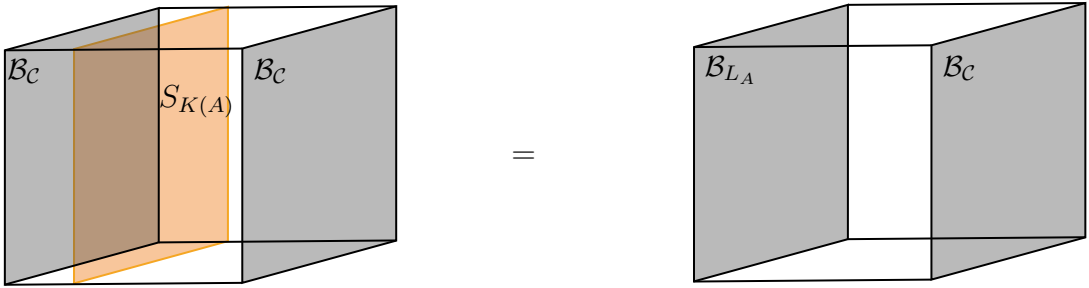
\begin{figure}[h!]
    \centering

\tikzset{every picture/.style={line width=0.75pt}} %set default line width to 0.75pt        

\begin{tikzpicture}[x=0.75pt,y=0.75pt,yscale=-1,xscale=1]
%uncomment if require: \path (0,191); %set diagram left start at 0, and has height of 191

%Shape: Parallelogram [id:dp03485931016275601] 
\draw  [color={rgb, 255:red, 245; green, 166; blue, 35 }  ,draw opacity=1 ][fill={rgb, 255:red, 248; green, 198; blue, 155 }  ,fill opacity=1 ] (103.15,160.37) -- (103.17,42.51) -- (180,21.27) -- (179.98,139.13) -- cycle ;
%Shape: Parallelogram [id:dp84461910785525] 
\draw  [color={rgb, 255:red, 0; green, 0; blue, 0 }  ,draw opacity=1 ][fill={rgb, 255:red, 74; green, 74; blue, 74 }  ,fill opacity=0.38 ] (187.15,160.1) -- (187.17,42.24) -- (264,21) -- (263.98,138.86) -- cycle ;
%Shape: Parallelogram [id:dp9593515224484056] 
\draw  [color={rgb, 255:red, 0; green, 0; blue, 0 }  ,draw opacity=1 ][fill={rgb, 255:red, 74; green, 74; blue, 74 }  ,fill opacity=0.38 ] (68.63,161) -- (68.65,43.14) -- (145.47,21.9) -- (145.45,139.76) -- cycle ;
%Straight Lines [id:da2261827344809514] 
\draw    (68.64,43.14) -- (187.17,42.25) ;
%Straight Lines [id:da1624712471301223] 
\draw    (145.47,21.9) -- (264,21) ;
%Straight Lines [id:da3858563893008591] 
\draw    (145.45,139.76) -- (263.98,138.78) ;
%Straight Lines [id:da030473465674629385] 
\draw    (68.62,161.07) -- (187.15,160.1) ;
%Shape: Parallelogram [id:dp439184162646864] 
\draw  [color={rgb, 255:red, 0; green, 0; blue, 0 }  ,draw opacity=1 ][fill={rgb, 255:red, 74; green, 74; blue, 74 }  ,fill opacity=0.38 ] (531.15,158.47) -- (531.17,40.61) -- (608,19.37) -- (607.98,137.23) -- cycle ;
%Shape: Parallelogram [id:dp770687065115593] 
\draw  [color={rgb, 255:red, 0; green, 0; blue, 0 }  ,draw opacity=1 ][fill={rgb, 255:red, 74; green, 74; blue, 74 }  ,fill opacity=0.38 ] (412.63,159.37) -- (412.65,41.51) -- (489.47,20.27) -- (489.45,138.13) -- cycle ;
%Straight Lines [id:da8907416660616856] 
\draw    (412.64,41.51) -- (531.17,40.62) ;
%Straight Lines [id:da6055128234107137] 
\draw    (489.47,20.27) -- (608,19.37) ;
%Straight Lines [id:da7223452270934689] 
\draw    (489.45,138.13) -- (607.98,137.15) ;
%Straight Lines [id:da6739046987853499] 
\draw    (412.62,159.44) -- (531.15,158.47) ;

% Text Node
\draw (68.52,46.01) node [anchor=north west][inner sep=0.75pt]  [font=\small]  {$\mathcal{B}_{\CC}$};
% Text Node
\draw (191.41,46.01) node [anchor=north west][inner sep=0.75pt]  [font=\small]  {$\mathcal{B}_{\CC}$};
% Text Node
\draw (144,59.4) node [anchor=north west][inner sep=0.75pt]    {$S_{K( A)}$};
% Text Node
\draw (416.52,44.38) node [anchor=north west][inner sep=0.75pt]  [font=\small]  {$\mathcal{B}_{L_A}$};
% Text Node
\draw (535.41,44.38) node [anchor=north west][inner sep=0.75pt]  [font=\small]  {$\mathcal{B}_{\CC}$};
% Text Node
\draw (320,92.4) node [anchor=north west][inner sep=0.75pt]    {$=$};

\end{tikzpicture}
    \caption{Parallel fusion of the surface operator $S_{K(A)}$ on the canonical gapped boundary $\CB_{\CC}$ produces a new gapped boundary $\CB_{L_A}$.}
    \label{fig: surface fusion on gapped boundary}
\end{figure}

As we discussed above, though the map $K$ specifies a multiplication on $K(A)$, it is not the Lagrangian algebra that we are looking for since $K(A)$ is not even commutative in general. However, $K(A)$ can be higher-gauged to obtain a surface operator $S_{K(A)}$ of the SymTFT \cite{Roumpedakis:2022aik}. Consider the gapped boundary $\CB_{L_A}$ of the SymTFT obtained from fusing the surface $S_{K(A)}$ on the canonical gapped boundary $\CB_{\CC}$ as in Fig. \ref{fig: surface fusion on gapped boundary}. 

\begin{figure}
    \centering

\tikzset{every picture/.style={line width=0.75pt}} %set default line width to 0.75pt        

\begin{tikzpicture}[x=0.75pt,y=0.75pt,yscale=-1,xscale=1]
%uncomment if require: \path (0,410); %set diagram left start at 0, and has height of 410

%Shape: Parallelogram [id:dp03485931016275601] 
\draw  [color={rgb, 255:red, 245; green, 166; blue, 35 }  ,draw opacity=1 ][fill={rgb, 255:red, 248; green, 198; blue, 155 }  ,fill opacity=1 ] (440.15,171.37) -- (440.17,53.51) -- (517,32.27) -- (516.98,150.13) -- cycle ;
%Shape: Parallelogram [id:dp84461910785525] 
\draw  [color={rgb, 255:red, 0; green, 0; blue, 0 }  ,draw opacity=1 ][fill={rgb, 255:red, 74; green, 74; blue, 74 }  ,fill opacity=0.38 ] (503.15,171.1) -- (503.17,53.24) -- (580,32) -- (579.98,149.86) -- cycle ;
%Shape: Parallelogram [id:dp9593515224484056] 
\draw  [color={rgb, 255:red, 0; green, 0; blue, 0 }  ,draw opacity=1 ][fill={rgb, 255:red, 74; green, 74; blue, 74 }  ,fill opacity=0.38 ] (384.63,172) -- (384.65,54.14) -- (461.47,32.9) -- (461.45,150.76) -- cycle ;
%Straight Lines [id:da2261827344809514] 
\draw    (384.64,54.14) -- (503.17,53.25) ;
%Straight Lines [id:da1624712471301223] 
\draw    (461.47,32.9) -- (580,32) ;
%Straight Lines [id:da3858563893008591] 
\draw    (461.45,150.76) -- (579.98,149.78) ;
%Straight Lines [id:da030473465674629385] 
\draw    (384.62,172.07) -- (503.15,171.1) ;
%Shape: Parallelogram [id:dp439184162646864] 
\draw  [color={rgb, 255:red, 0; green, 0; blue, 0 }  ,draw opacity=1 ][fill={rgb, 255:red, 74; green, 74; blue, 74 }  ,fill opacity=0.38 ] (187.15,172.47) -- (187.17,54.61) -- (264,33.37) -- (263.98,151.23) -- cycle ;
%Shape: Parallelogram [id:dp770687065115593] 
\draw  [color={rgb, 255:red, 0; green, 0; blue, 0 }  ,draw opacity=1 ][fill={rgb, 255:red, 74; green, 74; blue, 74 }  ,fill opacity=0.38 ] (68.63,173.37) -- (68.65,55.51) -- (145.47,34.27) -- (145.45,152.13) -- cycle ;
%Straight Lines [id:da8907416660616856] 
\draw    (68.64,55.51) -- (187.17,54.62) ;
%Straight Lines [id:da6055128234107137] 
\draw    (145.47,34.27) -- (264,33.37) ;
%Straight Lines [id:da7223452270934689] 
\draw    (145.45,152.13) -- (263.98,151.15) ;
%Straight Lines [id:da6739046987853499] 
\draw    (68.62,173.44) -- (187.15,172.47) ;
%Straight Lines [id:da3478049213877189] 
\draw    (263.5,93) -- (225.58,103.92) ;
%Straight Lines [id:da43036693864139497] 
\draw [color={rgb, 255:red, 139; green, 6; blue, 24 }  ,draw opacity=1 ]   (107.05,103.82) -- (225.58,102.92) ;
%Shape: Ellipse [id:dp6619704136893642] 
\draw  [color={rgb, 255:red, 139; green, 6; blue, 24 }  ,draw opacity=1 ][fill={rgb, 255:red, 139; green, 6; blue, 24 }  ,fill opacity=1 ] (225.58,102.92) .. controls (225.58,101.97) and (226.21,101.19) .. (226.98,101.19) .. controls (227.76,101.19) and (228.39,101.97) .. (228.39,102.92) .. controls (228.39,103.88) and (227.76,104.65) .. (226.98,104.65) .. controls (226.21,104.65) and (225.58,103.88) .. (225.58,102.92) -- cycle ;
%Shape: Ellipse [id:dp5947925357782091] 
\draw  [color={rgb, 255:red, 139; green, 6; blue, 24 }  ,draw opacity=1 ][fill={rgb, 255:red, 139; green, 6; blue, 24 }  ,fill opacity=1 ] (105.64,103.82) .. controls (105.64,102.87) and (106.27,102.09) .. (107.05,102.09) .. controls (107.83,102.09) and (108.45,102.87) .. (108.45,103.82) .. controls (108.45,104.78) and (107.83,105.55) .. (107.05,105.55) .. controls (106.27,105.55) and (105.64,104.78) .. (105.64,103.82) -- cycle ;
%Straight Lines [id:da2706492044290113] 
\draw    (580.5,93) -- (545.39,102.92) ;
%Straight Lines [id:da5137120378860905] 
\draw [color={rgb, 255:red, 139; green, 6; blue, 24 }  ,draw opacity=1 ]   (474.58,103.82) -- (544.58,102.92) ;
%Shape: Ellipse [id:dp024369925704809448] 
\draw  [color={rgb, 255:red, 139; green, 6; blue, 24 }  ,draw opacity=1 ][fill={rgb, 255:red, 139; green, 6; blue, 24 }  ,fill opacity=1 ] (542.58,102.92) .. controls (542.58,101.97) and (543.21,101.19) .. (543.98,101.19) .. controls (544.76,101.19) and (545.39,101.97) .. (545.39,102.92) .. controls (545.39,103.88) and (544.76,104.65) .. (543.98,104.65) .. controls (543.21,104.65) and (542.58,103.88) .. (542.58,102.92) -- cycle ;
%Shape: Ellipse [id:dp729239098527448] 
\draw  [color={rgb, 255:red, 139; green, 6; blue, 24 }  ,draw opacity=1 ][fill={rgb, 255:red, 139; green, 6; blue, 24 }  ,fill opacity=1 ] (473.76,103.82) .. controls (473.76,102.87) and (474.39,102.09) .. (475.17,102.09) .. controls (475.95,102.09) and (476.58,102.87) .. (476.58,103.82) .. controls (476.58,104.78) and (475.95,105.55) .. (475.17,105.55) .. controls (474.39,105.55) and (473.76,104.78) .. (473.76,103.82) -- cycle ;
%Straight Lines [id:da7231500822276927] 
\draw [color={rgb, 255:red, 139; green, 6; blue, 24 }  ,draw opacity=1 ] [dash pattern={on 4.5pt off 4.5pt}]  (425.83,103.72) -- (473.76,103.82) ;
%Shape: Ellipse [id:dp20610543474349885] 
\draw  [color={rgb, 255:red, 139; green, 6; blue, 24 }  ,draw opacity=1 ][fill={rgb, 255:red, 139; green, 6; blue, 24 }  ,fill opacity=1 ] (424.42,103.45) .. controls (424.42,102.49) and (425.05,101.72) .. (425.83,101.72) .. controls (426.61,101.72) and (427.24,102.49) .. (427.24,103.45) .. controls (427.24,104.41) and (426.61,105.18) .. (425.83,105.18) .. controls (425.05,105.18) and (424.42,104.41) .. (424.42,103.45) -- cycle ;

% Text Node
\draw (386.64,57.54) node [anchor=north west][inner sep=0.75pt]  [font=\small]  {$\mathcal{B}_{\CC}$};
% Text Node
\draw (507.41,57.01) node [anchor=north west][inner sep=0.75pt]  [font=\small]  {$\mathcal{B}_{\CC}$};
% Text Node
\draw (72.52,58.38) node [anchor=north west][inner sep=0.75pt]  [font=\small]  {$\mathcal{B}_{L_A}$};
% Text Node
\draw (191.41,58.38) node [anchor=north west][inner sep=0.75pt]  [font=\small]  {$\mathcal{B}_{\CC}$};
% Text Node
\draw (321,103.4) node [anchor=north west][inner sep=0.75pt]    {$=$};
% Text Node
\draw (463,62.4) node [anchor=north west][inner sep=0.75pt]    {$S_{K( A)}$};
% Text Node
\draw (246.54,101.36) node [anchor=north west][inner sep=0.75pt]    {$a\in A$};
% Text Node
\draw (118,106.4) node [anchor=north west][inner sep=0.75pt]    {$x\in K( A)$};
% Text Node
\draw (560.54,96.86) node [anchor=north west][inner sep=0.75pt]    {$a$};
% Text Node
\draw (487.58,103.22) node [anchor=north west][inner sep=0.75pt]    {$x$};

\end{tikzpicture}
    \caption{The line operator $x\in K(A)$ can end on the gapped boundary $\CB_{L_A}$ if the action of $S_{K(A)}$ on $x$ produces the trivial line operator.}
    \label{fig: Lagrangian algebra from left center}
\end{figure}
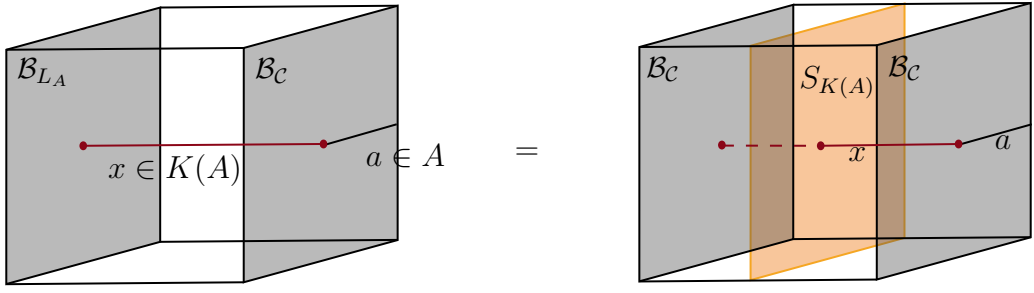

\begin{figure}[h!]
    \centering

\tikzset{every picture/.style={line width=0.75pt}} %set default line width to 0.75pt        

\begin{tikzpicture}[x=0.75pt,y=0.75pt,yscale=-1,xscale=1]

%Curve Lines [id:da032744426288151884] 
\draw [color={rgb, 255:red, 139; green, 6; blue, 24 }  ,draw opacity=1 ]   (434.76,92.3) .. controls (442.52,95.75) and (458.04,101.67) .. (457.17,110.54) ;
%Curve Lines [id:da05472227269127916] 
\draw [color={rgb, 255:red, 139; green, 6; blue, 24 }  ,draw opacity=1 ]   (413.21,111.04) .. controls (414.08,101.67) and (421.83,98.21) .. (434.76,92.3) ;
%Straight Lines [id:da17351413254591552] 
\draw [color={rgb, 255:red, 139; green, 6; blue, 24 }  ,draw opacity=1 ]   (434.76,81.04) -- (434.76,92.3) ;
%Straight Lines [id:da4257074023439641] 
\draw [color={rgb, 255:red, 139; green, 6; blue, 24 }  ,draw opacity=1 ]   (457.17,110.54) -- (457.17,187.96) ;
%Straight Lines [id:da9973120059685249] 
\draw [color={rgb, 255:red, 139; green, 6; blue, 24 }  ,draw opacity=1 ][fill={rgb, 255:red, 245; green, 166; blue, 35 }  ,fill opacity=1 ]   (413.21,111.04) -- (413.21,187.96) ;
%Curve Lines [id:da2538700245753329] 
\draw [color={rgb, 255:red, 139; green, 6; blue, 24 }  ,draw opacity=1 ]   (252.89,91.53) .. controls (260.65,94.98) and (272.71,100.54) .. (275.3,109.77) ;
%Curve Lines [id:da9393892688570242] 
\draw [color={rgb, 255:red, 139; green, 6; blue, 24 }  ,draw opacity=1 ]   (231.34,110.27) .. controls (232.2,100.9) and (239.96,97.45) .. (252.89,91.53) ;
%Straight Lines [id:da5810838496548381] 
\draw [color={rgb, 255:red, 139; green, 6; blue, 24 }  ,draw opacity=1 ]   (253,80) -- (252.89,91.53) ;
%Curve Lines [id:da009248254396189814] 
\draw [color={rgb, 255:red, 139; green, 6; blue, 24 }  ,draw opacity=1 ]   (230.48,187.33) .. controls (229,135) and (279,138) .. (275.3,109.77) ;
%Curve Lines [id:da012457160800589229] 
\draw [color={rgb, 255:red, 139; green, 6; blue, 24 }  ,draw opacity=1 ]   (231.34,110.27) .. controls (228.75,128.19) and (241.68,135.1) .. (247.72,139.71) ;
%Curve Lines [id:da03972340604752145] 
\draw [color={rgb, 255:red, 139; green, 6; blue, 24 }  ,draw opacity=1 ]   (254.61,143.55) .. controls (264.96,151.23) and (272.71,163.52) .. (272.71,186.57) ;

% Text Node
\draw (448.69,190.38) node [anchor=north west][inner sep=0.75pt]  [font=\small]  {$K( A)$};
% Text Node
\draw (370.42,189.41) node [anchor=north west][inner sep=0.75pt]  [font=\small]  {$C_{L}( K( A))$};
% Text Node
\draw (416.69,58.34) node [anchor=north west][inner sep=0.75pt]  [font=\small]  {$K( A)$};
% Text Node
\draw (186.1,189.95) node [anchor=north west][inner sep=0.75pt]  [font=\small]  {$C_{L}( K( A))$};
% Text Node
\draw (264.68,191.15) node [anchor=north west][inner sep=0.75pt]  [font=\small]  {$K( A)$};
% Text Node
\draw (236.54,58.03) node [anchor=north west][inner sep=0.75pt]  [font=\small]  {$K( A)$};
% Text Node
\draw (322.67,129.04) node [anchor=north west][inner sep=0.75pt]    {$=$};

\end{tikzpicture}
    \caption{Diagrammatic definition of left center of $K(A)$.}
    \label{fig:leftcenter}
\end{figure}
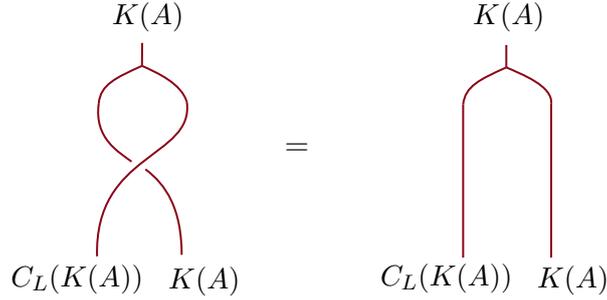

The Lagrangian algebra object $L_A$ corresponding to $A$ is given by the set of line operators in $K(A)$ that can end on $\CB_{L_A}$. From Fig. \ref{fig: Lagrangian algebra from left center} it is clear the a line operator $x\in K(A)$ can end on $\CB_{L_A}$ if 
\be
\label{eq:S acts to give trivial}
S_{K(A)}\cdot x = 1 + \dots
\ee
In other words, the action of the surface operator $S_{K(A)}$ on $x$ produces the trivial line. As discussed in \cite[Section 5.2.2]{Buican:2023bzl}, the set of line operators which satisfy \eqref{eq:S acts to give trivial} is given by the left center $C_L(K(A))$ of $K(A)$. The left center, $C_L(A)$ of $A$, is the maximal sub-object of $A$ such that 
\be
\label{eq:left centre}
m_{x,i}^{y} ~ R_{i,x}^{y}= m_{i,x}^y~,
\ee
where $m$ is the multiplication in $A$, and $R_{i,x}^{y}$ is the braiding of the lines $x \in A$ and $i\in C_{L}(A)$ in $\CZ(\CC)$. Diagrammatically, this is Fig. \ref{fig:leftcenter}.
Therefore, the Lagrangian algebra $L_A$ corresponding to the algebra $A$ is given by the left center of $K(A)$\footnote{This agrees with the abstract description of full center of an algebra given in \cite{Frohlich:2003hm,Kong:2007yv,davydov2010centre,Davydov:2013lma}.}. 

To make the discussion above more transparent, let us consider the example $\CC=\text{Vec}_{\DZ_2}$. Suppose the two line operators in $\CC$ are labeled by $e,g$, where $g$ is order two. We have two algebra objects in this case
\be
A_1=e~,~A_2=e+g~.
\ee
The SymTFT $\CZ(\CC)$ is the $\DZ_2$ Dijkgraaf-Witten theory with simple line operators
\be
([e],\mathds{1})~,~ ([e],\pi)~,~([g],\mathds{1})~,~([g],\pi)~.  
\ee
where $\mathds{1},\pi$ are the two irreducible representations of $\DZ_2$. $([e],\mathds{1}), ([e],\pi)$ and $([g],\mathds{1})$ are bosons while $([g],\pi)$ is a fermion. The boundary-to-bulk map $K$ is given by
\be
K(e)=([e],\mathds{1})+ ([e],\pi)~,~ K(g)=([g],\mathds{1})+ ([g],\pi)~.
\ee
Therefore, we get
\be
K(A_1)= ([e],\mathds{1})+ ([e],\pi)~.
\ee
This is already a commutative algebra because both $([e],\mathds{1})$ and $([e],\pi)$ are bosons, and they are closed under fusion. Indeed, $K(A_1)$ is precisely the Lagrangian algebra corresponding to $A_1$.

Now, consider 
\be
K(A_2)=([e],\mathds{1})+ ([e],\pi)+([g],\mathds{1})+ ([g],\pi)~.
\ee
Clearly, $K(A_2)$ is not a commutative algebra because of the fermion $([g],\pi)$. However, it admits the structure of an algebra and we can define a surface operator $S_{K(A_2)}$ by higher-gauging $A_2$ on a 2-manifold. The action of $S_{K(A_2)}$ on the line operators is given by \cite[Section 6.3.1]{Roumpedakis:2022aik}
\bea
&&S_{K(A_2)} \cdot ([e],\mathds{1})= S_{K(A_2)} \cdot ([g],\mathds{1}) = ([e],\mathds{1}) + ([e], \pi)~,\\
&& S_{K(A_2)} \cdot ([e],\pi)= S_{K(A_2)} \cdot ([g],\pi) = 0~.
\eea
The set of lines on which the surface operator acts to produce the trivial line is 
\be
([e],\mathds{1}) \text{ and } ([g],\mathds{1})~.
\ee
Therefore, we get a commutative algebra $A=([e],\mathds{1}) + ([g],\mathds{1})$ which is precisely the Lagrangian algebra object corresponding to $A_2$.

\subsection{Invertible symmetry}

\label{sec:invertible symmetry}

When the symmetry is invertible, the category $\CC$ is equivalent to Vec$_{G}^{\omega}$ for some group $G$ and anomaly given by a 3-cocycle $\omega \in H^3(G,U(1))$. When $\omega$ is non-trivial, the symmetry $G$ cannot be gauged. Given a subgroup $H \subseteq G$, we can ask whether $H$ is non-anomalous. This is true if and only if 
\be
\omega|_{H} = 1 \in H^3(H,U(1))~.
\ee
Given a non-anomalous group $H$, we can gauge it with a choice of discrete torsion $\sigma \in H^2(H,U(1))$. The algebra objects in this case will be denoted as $A(H,\sigma)$. Clearly, if $H$ is non-anomalous, so is the subgroup $gHg^{-1}$ for any $g \in G$. In fact, the algebras $A(H,\sigma)$ and $A(gHg^{-1},g\sigma g^{-1})$ are Morita equivalent for any $g \in G$ where $g\sigma g^{-1}(h,k) := \sigma(ghg^{-1},gkg^{-1})$.  Therefore, the equivalence classes $[A(H,\sigma)]$ are given by conjugacy classes of subgroups of $G$ on which $\omega$ trivializes and corresponding 2-cocycles.

So far we discussed how to determine whether $H \subset G$ is anomalous purely from a 1+1D perspective. Given $H \subseteq G$, how do we determine whether $H$ is anomalous from the SymTFT $\CZ(\text{Vec}_{G}^{\omega})\equiv \text{DW}(G,\omega)$? We learned that for $H \subseteq G$ to be non-anomalous there must be a Lagrangian algebra $L_{A(H,\sigma)}$ in $\CZ(\text{Vec}_{G}^{\omega})$ corresponding to it. To determine  this Lagrangian algebra object in terms of the simple objects of $\CZ(\text{Vec}_{G}^{\omega})$, we can use the theory of characters for objects in $\CZ(\text{Vec}_{G}^{\omega})$ \cite{davydov2010modular,davydov2017lagrangian}. Let us define 
\be
\omega(g,h|k):= \omega(g,h,k)^{-1}\omega(g,hgh^{-1},h)\omega(gh k (gh)^{-1},g,h)^{-1}~.
\ee
Let $([g],\pi_{g})$ be a simple object in $\CZ(\text{Vec}_{G}^{\omega})$, where $[g]$ is a conjugacy class in $G$ and $\pi_g$ is a projective representation of the centralizer $C_g$ of $g$ satisfying 
\be
\pi_g(h) \pi_g(k)= \omega(h,k|g) \pi_g(hk)~ ~ \forall h,k \in C_g~.
\ee
Note that for $h,k\in C_g$, the phases $\omega(h,k|f)$ is a 2-cocycle in $H^2(C_g,U(1))$.
The character of $([g],\pi_g)$ is defined as
\be
\label{eq:character of simple object}
\chi_{([g],\pi_g)}(h,k):= \begin{cases} \chi_{\pi_h}(k) & \mbox{if } h \in [g]~,\\ 0 & \mbox{otherwise}~, \end{cases} 
\ee
where $h,k$ is a pair of commuting elements in $G$ and $\chi_{\pi_g}(k)$ is the character of the representation $\pi_h$ of the centralizer $C_h$ of $h$ evaluated on $k \in C_h$. See Appendix \ref{ap:proof of theorem} for the definition of the character of a general line operator in the SymTFT and its properties. 

The Lagrangian algebra object $L_{A(H,\sigma)}$ can be determined by first computing the character 
\be
\label{eq:character of L}
\chi_{\CZ(A(H,\sigma))}(k,l):= \sum_{y \in Y} \frac{\omega(y^{-1}ly,y^{-1}|k)\sigma(y^{-1}ky,y^{-1}ly)}{\omega(y^{-1},l|k)\sigma(y^{-1}ly,y^{-1}ky)}~,
\ee
where 
\bea
Y&:=& \{y\in G, y^{-1}ky\in H, y^{-1}ly\in H\}/H \subset G/H~,
\eea
and then decomposing it in terms of the characters $\chi_{([g],\pi_g)}$. 

So far, we discussed how to get the Lagrangian algebra object $L_{A(H,\sigma)}$ corresponding to an algebra $A(H,\sigma)$ in Vec$_{G}^{\omega}$. Given a subgroup $H\subseteq G$, we would like to determine if $H$ is non-anomalous using the SymTFT $\CZ(\text{Vec}_{G}^{\omega})$. To this end, let us consider the following set 
\be
C(H):=\{[h], h \in H\}~.
\ee
where $[h]$ is the conjugacy class in $G$ with representative $h$. As a set, $C(H)$ contains all elements in the subgroup $H$ as well as elements of any subgroup conjugate to $H$. 

\begin{theorem}
\label{th: non-anomalous}
$H\subseteq G$ is non-anomalous if and only if there exists a Lagrangian algebra $L$ in $\CZ(\text{Vec}_{G}^{\omega})$ such that
\be
([h],\pi_h)\in L ~~~ \forall ~ [h] \in C(H)~,
\ee
for some representation $\pi_h$ of the centralizer of $h$. 
\end{theorem}
\vspace{0.2cm}
\noindent {\bf Proof:} See Appendix \ref{ap:proof of theorem}.

\vspace{0.2cm}

The above theorem shows that from a given object $L$  of $\CZ(\text{Vec}_{G}^{\omega})$ that admits the structure of a Lagrangian algebra we can identify a conjugacy class of subgroups $[H]$ such that any element of $[H]$ is a non-anomalous subgroup of $G$. 

\subsubsection{Lagrangian subcategory}

In the previous section, we proved a theorem which can be used to determine whether a subgroup $H\subseteq G$ is non-anomalous. To completely specify the equivalence class of algebras $[A(H,\sigma)]$, we also need to find the 2-cocycle $\sigma$. In general, this requires not just the decomposition of $L$ into simple objects, but also the multiplication of the algebra $L$. Indeed, the same object $L$ can admit distinct Lagrangian algebra structures \cite{davydov2014bogomolov} which then correspond to two non-equivalent algebra structures on some subgroup $H \subseteq G$. In general, the 2-cocycle $\sigma$ can be determined using \eqref{eq:relating multiplications}. However, when the simple line operators which form $L$ are closed under fusion, the 2-cocycle $\sigma$ can be easily obtained. 
In this case, these simple line operators form a Lagrangian subcategory of $\CZ(\text{Vec}_G^{\omega})$. Suppose we have
\be
\label{eq:Lagrangian subcategory}
L= \sum_{([g],\pi_g) \in \CZ(\text{Vec}_{G}^{\omega})} N_L^{([g],\pi_g)} ~ ([g],\pi_g)~.
\ee
Since the non-trivial simple line operators in $L$ are closed under fusion, the conjugacy classes $[g]$ such that $N_L^{([g],\pi_g)}\neq 0$ must all be closed under fusion. The Lagrangian algebra $L$ corresponds to an algebra object
\be
A= \sum_{([g]\pi_g) \in \CZ(\text{Vec}_{G}^{\omega})} N_L^{([g],\pi_g)} ~ A_g~,
\ee
where $A_g= \sum_{h\in[g]} h$. Since the non-trivial conjugacy classes of simple objects in $L$ are closed under fusion, we find that $A$ is in fact a sum over elements of a normal subgroup of $G$ given by 
\be
\label{eq:Normal subgroup from Lagrangian}
N= \bigcup_{N^{([g],\pi_g)}_L\neq 0}[g] ~,
\ee
Moreover, since the fusion of line operators in $\CZ(\text{Vec}_{G}^{\omega})$ is commutative, $N$ is an abelian normal subgroup. Therefore, the subset of gapped boundaries of the SymTFT $\CZ(\text{Vec}_{G}^{\omega})$ determined by Lagrangian subcategories correspond to non-anomalous abelian normal subgroups of $G$. In fact, there is a one-to-one correspondence between Lagrangian subcategories of $\CZ(\text{Vec}_{G}^{\omega})$ and algebra objects $A(N,\sigma)$ in Vec$_{G}^{\omega}$ where $N \subseteq G$ is an abelian normal subgroup and $\sigma \in H^2(N,U(1))$ \cite{naidu2008lagrangian}. 

The 2-cocycle $\sigma\in H^2(N,U(1))$ corresponding to the Lagrangian subcategory \eqref{eq:Lagrangian subcategory} can be determined using the relation
\be
\frac{\sigma(h_1,h_2)}{\sigma(h_2,h_1)}=\frac{\beta_g(l,h_2)\beta_g(lh_2,l^{-1})}{\beta_g(l,l^{-1})} ~ \frac{\chi_{\pi_g}(l^{-1}h_2l)}{\text{dim}(\pi_g)}
\ee
where 
\be
\beta_g(h,k):=\omega(g,h,k)\omega(h,h^{-1}gh,k)^{-1}\omega(h,k,(hk)^{-1}ghk)~,
\ee
$h_1,l$ and $g$ satisfies $h_1=lgl^{-1}$ for some $l\in G$ and $g\in N$. (See \cite[Theorem 4.12]{naidu2008lagrangian} for more details)

\section{Examples}

\label{sec:Examples}

In this section, we will go through various explicit examples to understand the general discussion in the previous sections. We will start with the familiar case of invertible symmetries. This is a good starting point to understand Theorems \ref{th: F(L)}
and \ref{th: non-anomalous}. Then we will consider the case of a modular tensor category $\CC$, before moving on to the case of fusion categories. 

\subsection{Invertible symmetries}

\subsubsection{$G=\DZ_4$}

When $G=\DZ_4$, there is no anomaly since $H^2(\DZ_4,U(1))\cong \DZ_1$. Therefore, we have the category Vec$_{\DZ_4}$ whose SymTFT is the $\DZ_4$ Dijkgraaf-Witten theory. It has three Lagrangian algebras given by 
\bea
&& L_1= \{(e,\mathds{1}), (e,\pi_1), (e,\pi_2), (e,\pi_3)\}~,\\
&& L_2 = \{(e,\mathds{1}),(g,\mathds{1}),(g^2,\mathds{1}),(g^3,\mathds{1})\}~, \\
&& L_3 = \{(e,\mathds{1}),(g^2,\mathds{1}),(e,\pi_2),(g^2,\pi_2)\}~.
\eea
where $g$ is the generator of $\DZ_4$ and $g^4=e$, and $\pi_j$ are the representations of $\DZ_4$ given by $\pi_j(g)=e^{\frac{2 \pi i j}{4}}$. Using Theorem \ref{th: non-anomalous}, we can read off the non-anomalous subgroups of $\DZ_4$ from the simple objects in the Lagrangian algebras above. We get
\be
L_1 \to \DZ_1, ~ L_2 \to \DZ_4, ~ L_3 \to \DZ_2~.
\ee
Alternatively, using Theorem \ref{th: F(L)}, the algebra objects which can be gauged in the $\DZ_4$ group can be obtained by fusing the Lagrangian algebra objects $L_i$ in the bulk TQFT with the gapped boundary $\CB_{L_1}$ corresponding to $L_1$. Fusing $L_1$ with the gapped boundary $\CB_{L_1}$, we get
\be
e + e + e + e~.
\ee
We see that the resulting algebra is not haploid because of multiple copies of the identity line in it.
Fusing $L_2$ with the gapped boundary $\CB_{L_1}$, we get
\be
e + g + g^2 + g^3~.
\ee
In this case, we get a haploid algebra.
Fusing $L_3$ with the gapped boundary $\CB_{L_1}$, we get
\be
e + g^2 + e + g^2~.
\ee
In this case we get a non-haploid algebra. A haploid subalgebra is $e + g^2$. 

\subsubsection{$G=\DZ_2 \times \DZ_2$}

Consider the $\DZ_2 \times \DZ_2$ symmetry without anomaly. The SymTFT is given by the $\DZ_2 \times \DZ_2=\langle g_1,g_2 \rangle$ Dijkgraaf-Witten theory. The line operators in this SymTFT can be labeled as
\bea
&&([e],\mathds{1}), ~ ([e],\omega_1), ~ ([e],\omega_2), ~ ([e],\omega_1\omega_2)~,\cr
&&([g_1],\mathds{1}), ~ ([g_1],\omega_1), ~ ([g_1],\omega_2), ~ ([g_1],\omega_1\omega_2)~,\cr
&&([g_2],\mathds{1}), ~ ([g_2],\omega_1), ~ ([g_2],\omega_2), ~ ([g_2],\omega_1\omega_2)~,\cr
&&([g_1g_2],\mathds{1}), ~ ([g_1g_2],\omega_1), ~ ([g_1g_2],\omega_2), ~ ([g_1g_2],\omega_1\omega_2)~,
\eea
where $\mathds{1},\omega_1,\omega_2$ and $\omega_1\omega_2$ label the four irreducible representations of $\DZ_2 \times \DZ_2$. This SymTFT has six gapped boundaries determined by the Lagrangian algebra objects
\bea
&&L_1= ([e],\mathds{1}), ~ ([e],\omega_1), ~ ([e],\omega_2), ~ ([e],\omega_1\omega_2)~,\cr
&&L_2= ([e],\mathds{1})+([g_1],\mathds{1})+([g_2],\mathds{1})+([g_1g_2],\mathds{1})~,\cr
&&L_3=([e],\mathds{1})+([g_1],\omega_2)+([g_2],\omega_1)+([g_1g_2],\omega_1\omega_2)~,\cr
&&L_4=([e],\mathds{1})+ ([e],\omega_2) + ([g_1],\mathds{1})+([g_1],\omega_2)~,\cr
&&L_5=([e],\mathds{1})+ ([e],\omega_1) ([g_2],\mathds{1})+([g_2],\omega_1)~,\cr
&&L_6=([e],\mathds{1})+ ([e],\omega_1\omega_2) ([g_1g_2],\mathds{1})+([g_1g_2],\omega_1\omega_2)~.
\eea
The canonical gapped boundary corresponding to the symmetry $\DZ_2 \times \DZ_2$ is $\CB_{L_1}$. Fusing the eight Lagrangian algebra objects on the gapped boundary $\CB_{L_{1}}$, we get 
\bea
&& F_{\CB_{L_1}}(L_1)= e + e+ e+e~,\\
&& F_{\CB_{L_1}}(L_2)= e + g_1+ g_2+g_1g_2~,\\
&& F_{\CB_{L_1}}(L_3)= e + g_1+ g_2+g_1g_2~,\\
&& F_{\CB_{L_1}}(L_4)= e + e+ g_1+g_1~,\\
&& F_{\CB_{L_1}}(L_5)= e + e+ g_2+g_2~,\\
&& F_{\CB_{L_1}}(L_6)= e + e+ g_1g_2+g_1g_2~.
\eea
Clearly, both Lagrangian algebras $L_2$ and $L_3$ correspond to the same algebra object
\be
A= e + g_1 + g_2 + g_1g_2~.
\ee
Since distinct Lagrangian algebras must correspond to non-equivalent gaugings in 1+1D, the algebra object $A$ has two distinct multiplications on it.  These multiplications can be determined using the multiplications of the Lagrangian algebras $L_2$ and $L_3$. 

Recall that in the multiplicity-free case, the multiplication in a Lagrangian algebra $L$ is determined by the complex numbers
\be
m_{xy}^{z} ~,
\ee
where $x,y,z$ are three simple line operators in $L$. These have to satisfy the constraints \eqref{eq: m associativity} and \eqref{eq:m commutativity}. For the $\DZ_2\times \DZ_2$ discrete gauge theory, there is a gauge in which the $F$ symbols are all trivial. The $R$ matrices can be written as 
\be
R_{([g],\pi_g),([h],\pi_h)}^{([gh],\pi_g\times \pi_h)}= \chi_{\pi_h}(g)~.
\ee
where $\chi_{\pi_g}$ is the character of the irreducible representation $\pi_g$. Using this expression for the $R$ matrix and constraint \eqref{eq:m commutativity}, we find that for the Lagrangian algebra $L_2$, we get
\be
m_{xy}^{z}=1 ~\forall ~x,y,z \in L_2~.
\ee
Therefore, using \eqref{eq:relating multiplications}, we find that the multiplication on the algebra $A$ corresponding to $L_2$ is trivial. 

Using the constraint \eqref{eq:m commutativity}, we find that for the Lagrangian algebra $L_3$, we get
\be
m_{([g_1],\omega_2),([g_2],\omega_1)}^{([g_1g_2],\omega_1\omega_2)}=-1
\ee
and all other $m_{xy}^{z}$ are trivial. Therefore, using \eqref{eq:relating multiplications}, we find that the multiplication on the algebra $A$ corresponding to $L_3$ is given by
\be
m_{g_1,g_2}^{g_1g_2}=-1~,
\ee
and all the other $m_{xy}^z$ are trivial. Therefore, the non-trivial multiplication on the algebra object $A$ corresponds to the Lagrangian algebra $L_3$. 

\subsubsection{$G=S_3$}

Consider the group 
\be
S_3 = \{r,s | r^3 =s^2=e; srs=r^{-1}\}~.
\ee
The conjugacy classes in this group are
\be
[e], ~~ [r]=\{r,r^2\}, ~~ [s]=\{s,rs,r^2s\}~.
\ee
$[e]$ has centralizer isomorphic to $S_3$ with the trivial representation $\mathds{1}_e$, non-trivial 1-dimensional representation $\pi_1$ and non-trivial two dimensional irreducible representation $\pi_2$. $[r]$ has a centralizer isomorphic to $\DZ_3$ with irreducible representations $\mathds{1}_r, \omega, \omega^2$. Finally, $[s]$ has a centralizer isomorphic to $\DZ_2$ which has the trivial representation $\mathds{1}_{s}$ and non-trivial representation $\gamma$. The full set of line operators in the $S_3$ DW theory are given by 
\be
([e],\mathds{1}_e), ([e],\pi_1), ([e],\pi_2), ([r], \mathds{1}_r), ([r],\omega), ([r], \omega^2), ([s],\mathds{1}_s), ([s],\gamma)~.
\ee
The 1+1D gapped and gapless phases in this case and their relation to SymTFTs were studied in detail in \cite{Bhardwaj:2023idu,Bhardwaj:2023fca,Bhardwaj:2023bbf,Bhardwaj:2024qrf}.

There are four Lagrangian algebras given by \cite{davydov2017lagrangian,Cong:2017ffh}
\bea
&& L_1:= ([e],\mathds{1}_e) + ([e],\pi_1) +  2 ([e],\pi_2)~,\\
&& L_2:= ([e],\mathds{1}_e) + ([e],\pi_2) +([s],\mathds{1}_s)~,\\
&& L_3:= ([e],\mathds{1}_e) + ([e],\pi_1) + 2([r],\mathds{1}_r)~,\\
&& L_4:= ([e],\mathds{1}_e) + ([s],\mathds{1}_s) +([r],\mathds{1}_r)~.
\eea
Let us consider the fusion of all line operators on the $\CB_{L_1}$ gapped boundary given by the map
\be
F_{\CB_{L_1}}: \CZ(\text{Vec}_{S_3}) \to \text{Vec}_{S_3}~.
\ee
In the following, we will denote $F_{\CB_{L_1}}$ simply as $F$. We get (see, for example, \cite[Section 2.6]{Cong:2017ffh})
\bea
&&F(([e],\mathds{1}_e))= F(([e],\pi_1))= e~,~ F(([e],\pi_2))= 2e~, \\
&& F(([r],\mathds{1}_r))= F(([r],\omega))= e~,~ F(([r],\omega^2))= r + r^2~, \\
&& F(([s],\mathds{1}_s))= F(([s],\gamma))= s  + sr + sr^2~.
\eea
Note that $F$ preserves quantum dimensions. Using this, we can find the fusion of all the Lagrangian algebras on $\CB_{L_1}$. Fusing $L_1$ on the gapped boundary $\CB_{L_1}$  results in the line operator
\be
e + e + 4e~.
\ee
Fusing $L_2$ on the gapped boundary $\CB_{L_1}$  results in the non-anomalous line operator
\be
e + 2 e + s + sr + sr^2~.
\ee
This is a sum of haploid subalgebras 
\be
A_1= e+s~,~ A_2=e+sr~,~A_3=e+sr^2~,
\ee
which are all Morita equivalent. Fusing $L_3$ on the gapped boundary $\CB_{L_1}$  results in the non-anomalous line operator
\be
e + e + 2 r + 2 r^2~.
\ee
This is two copies of the haploid algebra 
\be
e+r+r^2~.
\ee
Fusing $L_4$ on the gapped boundary $\CB_{L_1}$  results in the non-anomalous line operator
\be
e + s + sr + sr^2 + r + r^2~.
\ee

\subsection{When $\CC$ is modular}

When $\CC$ is a modular tensor category the SymTFT has a particularly simple form \cite{muger2003subfactors}.
\be
\CZ(\CC)\cong \CC \times \bar \CC~,
\ee
where $\bar \CC$ is the category with the opposite braiding. Therefore, the simple line operators in $\CZ(\CC)$ can be labeled as
\be
(a,b), ~ a,b\in \CC~.
\ee
However, for using the bulk-to-boundary map $F:\CZ(\CC)\to \CC$, it is convenient to relabel the simple line operators in the form $(a,e_a)$ for some object $a\in \CC$ and half-braiding $e_a$. This can be done using the explicit isomorphism between $\CZ(\CC)$ and $\CC \times \bar \CC$. We have \cite[Theorem 7.10]{muger2003subfactors}
\be
\label{eq:iso from Muger}
(a,b) \mapsto (a\times b, e_{a \times b})~,
\ee
where 
\be
e_{a\times b}(c):=(R_{a,c}^{-1} \times \text{id}_{b}) \circ (\text{id}_a \times R_{c,b}^{-1})~, 
\ee
with $R_{a,b}: a\times b\xrightarrow{\sim} b\times a$ being the braiding on $\CC$. 

\subsubsection{Fibonacci Category}

Let $\CC$ be the Fibonacci MTC with the non-trivial simple line operator $\tau$. The SymTFT in this case is 
\be
\CZ(\CC)= \CC \times \bar \CC~.
\ee
Therefore, the simple line operators in the SymTFT can be labeled as
\be
(\trl,\trl), (\trl,\tau), (\tau,\trl), (\tau, \tau)~.
\ee
The line operators $(\trl,\trl)$ and $(\tau,\tau)$ are bosonic. For identifying gaugeable algebras in $\CC$, we have to describe the simple objects in the notation $(a,e_a)$ where $a\in \CC$ and $e_a$ is the half-braiding. Using \eqref{eq:iso from Muger} we get\footnote{The line operators in $\CZ(\CC)$ can also be directly obtained in the form $(a,e_a)$ using the string-net construction \cite[Section VI. B]{Levin:2004mi}.}
\be
(\trl,e_{\trl}), (\tau, e_{\tau}), (\tau, e'_{\tau}), (\trl+\tau,e_{\trl+\tau})~.
\ee
We have  a unique Lagrangian  algebra 
\be
L= (\trl,e_{\trl}) + (\trl+\tau,e_{\trl+\tau})~.
\ee
Fusing $L$ with the gapped boundary $\CB_{L}$, we find the algebra object 
\be
\trl + \trl + \tau~,
\ee
which is non-haploid. We can decompose this algebra into haploid subalgebras 
\be
A_1=\trl, A_2= \trl + \tau~.
\ee
The fact that both of these algebras correspond to the same gapped boundary implies that they are Morita equivalent. Indeed, gauging $\trl+ \tau$ is a self-duality of a 1+1D QFT with $\CC$ symmetry \cite{Choi:2023xjw,Diatlyk:2023fwf}. 

\subsubsection{Ising Category}

Let $\CC$ be the Ising MTC with non-trivial line operators $\psi,\sigma$. The SymTFT in this case is 
\be
\CZ(\CC)= \CC \times \bar \CC~.
\ee
Therefore, the simple line operators in the SymTFT can be labeled as
\bea
&& (\trl,\trl), (\trl,\psi), (\trl,\sigma)~,\\ && (\psi,\trl), (\psi,\psi), (\psi,\sigma)~,\\ 
&& (\sigma,\trl), (\sigma,\psi), (\sigma,\sigma)~.
\eea
The line operators $(\trl,\trl),(\psi,\psi)$ and $(\sigma,\sigma)$ are bosonic. Using \eqref{eq:iso from Muger} we can relabel the simple line operators as
\bea
&& (\trl,e^{(1)}_{\trl}), (\psi,e^{(1)}_{\psi}), (\sigma,e^{(1)}_{\sigma})\\ && (\psi,e^{(2)}_{\psi}), (\trl,e^{(2)}_{\trl}), (\sigma,e^{(2)}_{\sigma})\\ 
&& (\sigma,e^{(2)}_{\sigma}), (\sigma,e^{(3)}_{\sigma}), (\trl+ \psi,e_{\trl+\psi})~.
\eea
The only Lagrangian algebra is 
\be
L=(\trl,e^{(1)}_{\trl})+(\trl,e^{(2)}_{\trl})+(\trl+\psi,e_{\trl+\psi})~.
\ee
Fusing $L$ with the gapped boundary $\CB_{L}$ we get the algebra object
\be
\trl+ \trl+ \trl+ \psi~.
\ee
This algebra is a sum of three haploid subalgebras
\be
A_1=A_2=\trl, ~ A_3= \trl+\psi~.
\ee
This is consistent with the fact that $A_3$ is a Morita trivial algebra \cite{Fuchs:2004dz,Choi:2023xjw,Diatlyk:2023fwf}.

\subsection{Non-invertible symmetries}

\subsubsection{Rep$(S_3)$}

For Rep$(S_3)$ the SymTFT is dual to $\CZ(\text{Vec}_{S_3})$ for which we have already considered all the Lagrangian algebras. Recall the simple line operators 
\be
([e],\mathds{1}_e), ([e],\pi_1), ([e],\pi_2), ([r], \mathds{1}_r), ([r],\omega), ([r], \omega^2), ([s],\mathds{1}_s), ([s],\gamma)~.
\ee
and gapped boundaries
\bea
&& L_1:= ([e],\mathds{1}_e) + ([e],\pi_1) +  2 ([e],\pi_2)~,\\
&& L_2:= ([e],\mathds{1}_e) + ([e],\pi_2) +([s],\mathds{1}_s)~,\\
&& L_3:= ([e],\mathds{1}_e) + ([e],\pi_1) + 2([r],\mathds{1}_r)~,\\
&& L_4:= ([e],\mathds{1}_e) + ([s],\mathds{1}_s) +([r],\mathds{1}_r)~.
\eea
For identifying the algebra objects in Rep$(S_3)$, we have to choose a gapped boundary of $\CZ(\text{Vec}_{S_3})$ such that the boundary line operators form Rep$(S_3)$. For example, we can choose the gapped boundary given by 
\be
L_2= ([e],\mathds{1}_e) + ([e],\pi_2) +([s],\mathds{1}_s)~.
\ee
Let $\mathds{1}, \pi_1,\pi_2$ be the three irreducible representations of $S_3$. Fusing the simple line operators in $\CZ(\text{Vec}_{S_3})$ on $\CB_{L_2}$, we get (see, for example, \cite[Section 2.6]{Cong:2017ffh})
\bea
&&F_{\CB_{L_2}}(([e],\mathds{1}_e))=\mathds{1}~,~  F_{\CB_{L_2}}(([e],\pi_1))= \pi_1~,~ F_{\CB_{L_2}}(([e],\pi_2))= \mathds{1} + \pi_1~, \\
&& F_{\CB_{L_2}}(([r],\mathds{1}_r))=F_{\CB_{L_2}}(([r],\omega))=F_{\CB_{L_2}}(([r],\omega^2))= \pi_2 ~,\\
&& F_{\CB_{L_2}}(([s],\mathds{1}_s))= \mathds{1} + \pi_2~,~ F_{\CB_{L_2}}(([s],\gamma))= \pi_1 + \pi_2~.
\eea
Fusing $L_1$ on the gapped boundary $\CB_{L_2}$  results in the non-anomalous line operator
\be
3 \mathds{1} + 3 \pi_1 ~.
\ee
It decomposes into three copies of the haploid algebra $\mathds{1}+\pi_1$.
Fusing $L_2$ on the gapped boundary $\CB_{L_2}$  results in the non-anomalous line operator
\be
2  \mathds{1} + \mathds{1} + \pi_1 + \pi_2~.
\ee
It decomposes into the haploid algebras
\be
A_1=\mathds{1}~,~A_2=\mathds{1}+\pi_1+\pi_2~.
\ee
Note that this implies that gauging $A_2$ is a self-duality as $A_2$ is a Morita trivial algebra. Fusing $L_3$ on the gapped boundary $\CB_{L_2}$  results in the non-anomalous line operator
\be
\mathds{1} + \pi_1 + 2 \pi_2~.
\ee
This is a haploid algebra corresponding to the unique fibre functor admitted by Rep$(S_3)$. Fusing $L_4$ on the gapped boundary $\CB_{L_2}$  results in the non-anomalous line operator
\be
2 \mathds{1} + 2 \pi_2~.
\ee
It decomposes into two copies of the haploid algebra object $\mathds{1}+\pi_2$.
The algebra objects obtained above agree with the classification of algebras in Rep$(S_3)$ given in \cite{Perez-Lona:2023djo}.

\subsection{Lagrangian algebras in $\CZ(\text{Vec}_{D_8})$ from algebras in Vec$_{D_8}$} 

\label{sec:D8 example}

Given the non-anomalous line operators in a fusion category $\CC$, we can use Theorem \ref{th: F(L)} to put strong constraints on consistent Lagrangian algebra objects in its SymTFT. In this section, we will show that all Lagrangian algebra objects in the SymTFT $\CZ(\text{Vec}_{D_8})$, where $D_8$ is the dihedral group of 8 elements, can be completely fixed using Theorem \ref{th: F(L)}. 

Consider the invertible symmetry Vec$_{D_{8}}$ where
\be
D_8 :=\langle r,s ~ | ~ r^4=s^2=e, srs=r^3 \rangle~.
\ee
Since this $D_8$ is chosen to be non-anomalous, any subgroup of $D_8$ can be gauged. The physically inequivalent gaugings correspond to conjugacy classes of subgroups of $D_8$ and a choice of discrete torsion. The conjugacy classes of subgroups are as follows. The normal subgroups are 
\be
\{e\}~,~ \{e,r^2\}~,~ \{e,r,r^2,r^3\}~,~ \{e,r^2,s,r^2s\}~,~ \{e,r^2,rs,r^3s\}~,~ D_8~. 
\ee
We will denote these groups as $\DZ_1,\DZ_2^{(r^2)},\DZ_4^{(r)},\DZ_2^{(r^2)}\times \DZ_2^{(s)}, \DZ_2^{(r^2)} \times \DZ_2^{(rs)},D_8$, respectively. There are four non-normal subgroups which form the following two conjugacy classes of subgroups
\be
\{\{e,s\},\{e,r^2s\}\}~,~ \{\{e,rs\},\{e,r^3s\}\}~.
\ee
All groups in these two conjugacy classes are isomorphic to $\DZ_2$, and we will denote them by the representatives $\DZ_{2}^{(s)}$ and $\DZ_2^{(rs)}$, respectively. A physically inequivalent gaugings in Vec$_{D_8}$ can be labeled as $(G,\sigma)$ where $G$ is a representative of a conjugacy class of subgroups in $D_8$ and $\sigma\in H^2(G,U(1))$ is the discrete torsion. Using the explicit data above, we get the following 11 physically inequivalent gaugings in Vec$_{D_8}$
\be
\label{eq: D8 algebras}
(\DZ_1,1)~,~(\DZ_2^{(r^2)},1)~,~(\DZ_4^{(r)},1)~,~(\DZ_2^{(r^2)}\times\DZ_2^{(s)},1)~,~(\DZ_2^{(r^2)}\times\DZ_2^{(s)},\alpha)~,~
\ee
\be
(\DZ_2^{(r^2)}\times\DZ_2^{(rs)},1)~,~(\DZ_2^{(r^2)}\times\DZ_2^{(rs)},\beta)~,~ (D_8,1)~,~(D_8,\gamma)~,~(\DZ_2^{(s)},1)~,~(\DZ_2^{(rs)},1)~. 
\ee
where $\alpha,\beta,\gamma$ denote the choice of non-trivial discrete torsion in $H^2(\DZ_2^{(r^2)}\times\DZ_2^{(s)},U(1))\cong\DZ_2,$ 
$H^2(\DZ_2^{(r^2)}\times\DZ_2^{(rs)},U(1))\cong\DZ_2$ and $H^2(D_8,U(1))\cong\DZ_2$, respectively. We will relate these $11$ algebra objects to gapped boundaries of the SymTFT of Vec$_{D_8}$.

The SymTFT $\CZ(\text{Vec}_{D_{8}})$ is the Dijkgraaf-Witten theory with gauge group $D_8$. The conjugacy classes of $D_8$ are
\be
[e]=\{e\}~,~ [r^2]=\{r^2\}~,~ [s]=\{s,r^2s\}~,~ [rs]=\{rs,r^3s\}~,~ [r]=\{a^2,a^3\}~.
\ee
The centralizers of the representative of these conjugacy classes are
\be
D_8~,~ D_8~,~ \{e,r^2,s,r^2s\}\cong \DZ_2\times \DZ_2~,~ \{e,r^2,rs,r^3s\}\cong \DZ_2 \times \DZ_2~,~ \{e,r,r^2,r^3\}\cong \DZ_4~,
\ee
respectively. The simple line operators in $\CZ(\text{Vec}_{D_{8}})$, their quantum dimensions and spins are given by 
\begin{center}
\begin{tabular}{ |c|c|c|c|c|c|c|} 
\hline
Line operator & $([e],\mathds{1})$ &$([e],\pi_1)$& $([e],\pi_2)$, & $([e],\pi_3)$& $([e],\pi_4)$ \\ \hline
$d_x$ & 1 & 1 & 1 & 1 & 2 \\ \hline
$\theta_{x}$ & 1 & 1 & 1 & 1 & 1 \\ \hline
\end{tabular}
\end{center}
\begin{center}
\begin{tabular}{ |c|c|c|c|c|c|c|} 
\hline
Line operator & $([r^2],\mathds{1})$ &$([r^2],\pi_1)$& $([r^2],\pi_2)$, & $([r^2],\pi_3)$& $([r^2],\pi_4)$ \\ \hline
$d_x$ & 1 & 1 & 1 & 1 & 2 \\ \hline
$\theta_{x}$ & 1 & 1 & 1 & 1 & -1 \\ \hline
\end{tabular}
\end{center}
\begin{center}
\begin{tabular}{|c|c|c|c|c|c|} 
\hline
Line operator & $([s],\mathds{1}_s)$ &$([s],\omega_1)$& $([s],\omega_2)$, & $([s],\omega_1\omega_2)$\\ \hline
$d_x$ & 2 & 2 & 2 & 2\\ \hline
$\theta_{x}$ & 1 & -1 & 1 & -1 \\ \hline
\end{tabular}
\end{center}
\begin{center}
\begin{tabular}{|c|c|c|c|c|c|} 
\hline
Line operator & $([rs],\mathds{1}_{rs})$ &$([rs],\tilde \omega_1)$& $([rs],\tilde \omega_2)$, & $([rs],\tilde \omega_1\tilde \omega_2)$\\ \hline
$d_x$ & 2 & 2 & 2 & 2\\ \hline
$\theta_{x}$ & 1 & -1 & 1 & -1 \\ \hline
\end{tabular}
\end{center}
\begin{center}
\begin{tabular}{|c|c|c|c|c|c|} 
\hline
Line operator & $([r],\mathds{1}_{r})$ &$([r],\tilde \omega)$& $([r],\tilde \omega^2)$, & $([r],\tilde \omega^3)$\\ \hline
$d_x$ & 2 & 2 & 2 & 2\\ \hline
$\theta_{x}$ & 1 & i & -1 & i \\ \hline
\end{tabular}
\end{center}
where $\mathds{1}$, $\pi_1$, $\pi_2$, $\pi_3$ are the 1-dimensional irreducible representations of $D_8$ and $\pi_4$ is the 2-dimensional one; $\pi_1$, $\pi_2$ and $\pi_3$ are the irreducible representations with kernels $\DZ_4^{(r)},\DZ_2^{(r^2)}\times \DZ_2^{(s)}$ and $\DZ_2^{(r^2)}\times \DZ_2^{(rs)}$, respectively; $\mathds{1}_{s}$, $\omega_1$, $\omega_2$ and $\omega_1\omega_2$ are the four irreducible representations of the $\DZ_2 \times \DZ_2$ centralizer of $s$; $\mathds{1}_{rs}$, $\tilde \omega_1$, $\tilde \omega_2$ and $\tilde \omega_1\tilde \omega_2$ are the four irreducible representations of the $\DZ_2 \times \DZ_2$ centralizer of $rs$; $\mathds{1}_{r}$, $\omega$, $\omega^2$ and $\omega^3$ are the four irreducible representations of the $\DZ_4$ centralizer of $r$. The fusion rules of these line operators and their modular $S$ matrix are given in \cite[Appendices A,B]{Iqbal:2023wvm}. 

A line operator in $\CZ(\text{Vec}_{D_8})$ which admits the structure of a Lagrangian algebra must be a direct sum of the bosonic line operators in the list above. In fact, using Theorem \ref{th: non-anomalous}, we can identify all Lagrangian algebra objects in the SymTFT.
For example, consider the algebra $(\DZ_2^{(rs)},1)$. We have two Morita equivalent classes of algebras
\be
A_1:= e+ rs~,~ A_2= e+ r^3s. 
\ee
We need to choose a Lagrangian algebra object $L$ in the SymTFT such that $F(L)$ produces the above two algebras. Note that the Lagrangian algebra $L$ has quantum dimension $8$. Since $F:\CZ(\text{Vec}_{D_{8}})\to \text{Vec}_{D_8}$ preserves quantum dimensions, we must have
\be
F(L)= 2 A_1 + 2 A_2~, ~ \text{or } F(L)= A_1 + 3 A_2~,~ \text{or } F(L)= 3A_1 + A_2~.
\ee
Therefore, $L$ must be of the form
\bea
L=&&([e],\mathds{1}) + N_L^{([e],\pi_1)}([e],\pi_1) + N_L^{([e],\pi_2)}([e],\pi_2)  + N_L^{([e],\pi_3)}([e],\pi_3) + N_L^{([e],\pi_4)}([e],\pi_4)+  \cr
&&N_L^{([rs],\mathds{1}_{rs})} ([rs],\mathds{1}_{rs}) + N_L^{([rs],\tilde \omega_2)} ([rs],\tilde \omega_2)~.
\eea
where $N_{L}^{*}$ are non-negative integers which need to be determined. Since the conjugacy class $[rs]=\{rs,r^3s\}$, $F(L)$ contains equal number of $rs$ and $r^3s$ terms. This implies that $F(L)=A_1+3A_2$ and $F(L)=3A_1+A_2$  are inconsistent. Therefore, we must have $F(L)=2A_1+2A_2$.

Suppose $N_L^{([rs],\mathds{1}_{rs})}\neq 0$. We will now show that $N_L^{([rs],\mathds{1}_{rs})} = 1$. To see this, suppose $N_L^{([rs],\mathds{1}_{rs})} >1$. We have the fusion rule \cite[Appendix B]{Iqbal:2023wvm}
\be
([rs],\mathds{1}_{rs}) \times ([rs],\mathds{1}_{rs})= ([e],\pi_4) + ([r^2],\pi_4)~.
\ee
Using the constraint \eqref{eq:Lagrangian algebra constraints} we find that
\be
N_L^{([e],\pi_4)}\geq 4~.
\ee
However, this contradicts the requirement that $F(L)=2A_1+2A_2$ contains exactly $4$ trivial line operators. Therefore, we must have $N_L^{([rs],\mathds{1}_{rs})} = 1$. Since $F(L)$ must contain two copies of $rs$, we also find that $N_L^{([rs],\tilde \omega_2)} = 1$. Now, consider the fusion rule
\be
([rs],\mathds{1}_{rs}) \times ([rs],\tilde \omega_2)=([e],\pi_4) + ([r^2],\pi_4)~.
\ee
Since $([r^2],\pi_4)$ is a fermion, using \eqref{eq:Lagrangian algebra constraints} we find that $N_L^{([e],\pi_4)}=1$. Combining the results above, we have determined that the Lagrangian algebra must be of the form
\bea
L=&&([e],\mathds{1}) + N_L^{([e],\pi_1)}([e],\pi_1) + N_L^{([e],\pi_2)}([e],\pi_2)  + N_L^{([e],\pi_3)}([e],\pi_3) + ([e],\pi_4)+\cr 
&&([rs],\mathds{1}_{rs}) +  ([rs],\tilde \omega_2)~.
\eea
$L$ must have quantum dimension $8$. Therefore, only one among $N_L^{([e],\pi_1)}$, $N_L^{([e],\pi_2)}$, $N_L^{([e],\pi_3)}$ can be non-zero. Using the fusion rules, we know that $([e],\pi_3)$ is the only line operator such that 
\be
([rs],\tilde \omega_2) \times ([e],\pi_2) \text{ contains } ([rs],\mathds{1}_{rs}) ~,
\ee
and 
\be
([rs],\mathds{1}_{rs}) \times ([e],\pi_2) \text{ contains } ([rs],\tilde \omega_2) ~,
\ee
which implies that $N_L^{([e],\pi_3)}=1$. Therefore, the Lagrangian algebra object corresponding to the algebra $(\DZ_{2}^{(rs)},1)$ is
\bea
L=([e],\mathds{1}) + ([e],\pi_3) + ([e],\pi_4)+([rs],\mathds{1}_{rs}) +  ([rs],\tilde \omega_2)~.
\eea
Similar analysis can be used to determine all Lagrangian algebra objects in the SymTFT $\CZ(\text{Vec}_{D_{8}})$. They are summarized in the table below. 

\bea
\arraycolsep=1.4pt\def\arraystretch{1.5}
\begin{array}{|c|c|}
\hline
\text{Algebra objects in Vec}_{D_8} & \text{Lagrangian algebra objects in } \CZ(\text{Vec}_{D_8}) \\ \hline
(\DZ_1,1) & ([e],\mathds{1})+([e],\pi_1) + ([e],\pi_2) + ([e],\pi_3) + 2 ([e],\pi_4) \\
\hline
(\DZ_2^{(r^2)},1) & 
\begin{tabular}{@{}c@{}}
$([e],\mathds{1})+([e],\mathds{\pi_1}) + ([e],\pi_2) + ([e],\pi_3)$ \\ + $([r^2],\mathds{1})+ ([r^2],\pi_1) + ([r^2],\pi_2) + ([r^2],\pi_3)$
\end{tabular}
\\
\hline
(\DZ_4^{(r)},1) & ([e],\mathds{1})+ ([e],\pi_1)+ 2 ([r],\mathds{1}_r) + ([r^2],\mathds{1}) + ([r^2],\pi_1) \\
\hline
(\DZ_2^{(r^2)} \times \DZ_2^{(s)},1) & ([e],\mathds{1})+([e],\pi_2) + 2([s],\mathds{1}) + ([r^2],\mathds{1}) + ([r^2],\pi_2) \\
\hline
(\DZ_2^{(r^2)} \times \DZ_2^{(s)},\alpha) & ([e],\mathds{1})+([e],\pi_2) + 2([s],\omega_2) + ([r^2],\pi_1) + ([r^2],\pi_3) \\
\hline
(\DZ_2^{(r^2)} \times \DZ_2^{(rs)},1) & ([e],\mathds{1})+([e],\pi_3) + 2([rs],\mathds{1}_{rs}) + ([r^2],\mathds{1}) + ([r^2],\pi_3) \\
\hline
(\DZ_2^{(r^2)} \times \DZ_2^{(rs)},\beta) & ([e],\mathds{1})+([e],\pi_3) + 2([rs],\tilde \omega_2) + ([r^2],\pi_1) + ([r^2],\pi_2) \\
\hline
(D_8,1) & ([e],\mathds{1})+([r^2],\mathds{1}) + ([s],\mathds{1}_s) + ([r],\mathds{1}_r) + ([rs],\mathds{1}_{rs})\\
\hline
(D_8,\gamma) & ([e],\mathds{1})+([r^2],\pi_1) + ([s],\omega_2) + ([r],\mathds{1}_r) + ([rs],\tilde \omega_2)\\ 
\hline
(\DZ_2^{(s)},1)  & ([e],\mathds{1}) + ([e],\pi_2) + ([e],\pi_4) + ([s],\mathds{1}_s) + ([s],\omega_2) \\
\hline
(\DZ_2^{(rs)},1) & ([e],\mathds{1})+ ([e],\pi_3) + ([e],\pi_4) +([rs],\mathds{1}_{rs}) + ([rs],\tilde \omega_2) \\
\hline
\end{array}
\eea

\section{Applications}

\label{sec:Applications}

\subsection{Higher-gauging data from the action of surface operators}

Consider a 2+1D TQFT whose line operators are described by a modular tensor category $\CC$. Let $S$ be a topological surface operator whose action on the line operators is given by 
\be
\label{eq:surface operator action}
S\cdot a= \sum_{b\in \CC} n_{S\cdot a}^b ~b~.
\ee
$S$ implements a non-invertible symmetry if the sum above is non-trivial. All surface operators in a 2+1D TQFT can be constructed through higher-gauging \cite{Kapustin:2010if,Fuchs:2012dt,Roumpedakis:2022aik}. Given \eqref{eq:surface operator action}, we can ask: Which algebra $A \in \CC$ should be higher-gauged to construct the surface operator $S$? When the higher-gauging involves only invertible line operators, this was answered in \cite{Buican:2023bzl} by providing an explicit map between the action of the surface operator and the abelian group of lines $+$ discrete torsion data for higher-gauging. In this section, we give a general answer to this question for arbitrary higher-gaugings. 

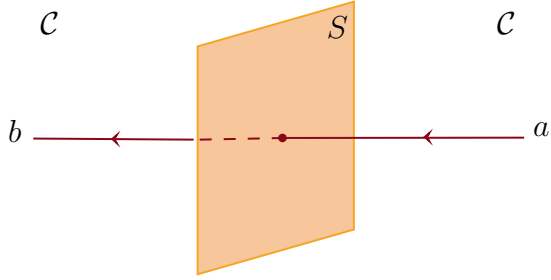
\begin{figure}[h!]
    \centering

\tikzset{every picture/.style={line width=0.75pt}} %set default line width to 0.75pt        

\begin{tikzpicture}[x=0.75pt,y=0.75pt,yscale=-1,xscale=1]
%uncomment if require: \path (0,202); %set diagram left start at 0, and has height of 202

%Shape: Parallelogram [id:dp07678548712448985] 
\draw  [color={rgb, 255:red, 245; green, 166; blue, 35 }  ,draw opacity=1 ][fill={rgb, 255:red, 248; green, 198; blue, 155 }  ,fill opacity=1 ] (291.99,166.47) -- (292.01,51.85) -- (369.99,29.35) -- (369.97,143.97) -- cycle ;
%Straight Lines [id:da8618720090293183] 
\draw [color={rgb, 255:red, 139; green, 6; blue, 24 }  ,draw opacity=1 ]   (335.99,97.91) -- (455,97.76) ;
%Straight Lines [id:da36219470975522705] 
\draw [color={rgb, 255:red, 139; green, 6; blue, 24 }  ,draw opacity=1 ] [dash pattern={on 4.5pt off 4.5pt}]  (293,98.76) -- (335.99,97.91) ;
%Straight Lines [id:da18382329342522152] 
\draw [color={rgb, 255:red, 139; green, 6; blue, 24 }  ,draw opacity=1 ]   (209.97,98.21) -- (290,98.76) ;
%Shape: Circle [id:dp7171868847369509] 
\draw  [color={rgb, 255:red, 139; green, 6; blue, 24 }  ,draw opacity=1 ][fill={rgb, 255:red, 139; green, 6; blue, 24 }  ,fill opacity=1 ] (332.69,97.91) .. controls (332.69,97) and (333.43,96.26) .. (334.34,96.26) .. controls (335.25,96.26) and (335.99,97) .. (335.99,97.91) .. controls (335.99,98.82) and (335.25,99.56) .. (334.34,99.56) .. controls (333.43,99.56) and (332.69,98.82) .. (332.69,97.91) -- cycle ;
\draw  [color={rgb, 255:red, 139; green, 6; blue, 24 }  ,draw opacity=1 ][fill={rgb, 255:red, 139; green, 6; blue, 24 }  ,fill opacity=1 ] (407.57,99.44) -- (405.36,97.41) -- (407.61,95.48) -- (406.47,97.44) -- cycle ;
\draw  [color={rgb, 255:red, 139; green, 6; blue, 24 }  ,draw opacity=1 ][fill={rgb, 255:red, 139; green, 6; blue, 24 }  ,fill opacity=1 ] (251.57,100.44) -- (249.36,98.41) -- (251.61,96.48) -- (250.47,98.44) -- cycle ;

% Text Node
\draw (355,35.4) node [anchor=north west][inner sep=0.75pt]    {$S$};
% Text Node
\draw (458,88.4) node [anchor=north west][inner sep=0.75pt]    {$a$};
% Text Node
\draw (196,87.4) node [anchor=north west][inner sep=0.75pt]    {$b$};
% Text Node
\draw (212,33.4) node [anchor=north west][inner sep=0.75pt]    {$\CC$};
% Text Node
\draw (440,33.4) node [anchor=north west][inner sep=0.75pt]    {$\CC$};

\end{tikzpicture}
    \caption{Action of a topological surface operator $S$ on line operators.}
    \label{fig: action of surface operator}
\end{figure}

\begin{figure}[h!]
    \centering

\tikzset{every picture/.style={line width=0.75pt}} %set default line width to 0.75pt        

\begin{tikzpicture}[x=0.75pt,y=0.75pt,yscale=-1,xscale=1]
%uncomment if require: \path (0,208); %set diagram left start at 0, and has height of 208

%Shape: Parallelogram [id:dp84461910785525] 
\draw  [color={rgb, 255:red, 0; green, 0; blue, 0 }  ,draw opacity=1 ][fill={rgb, 255:red, 74; green, 74; blue, 74 }  ,fill opacity=0.38 ] (520.15,172.1) -- (520.17,54.24) -- (597,33) -- (596.98,150.86) -- cycle ;
%Straight Lines [id:da714765442747638] 
\draw [color={rgb, 255:red, 139; green, 6; blue, 24 }  ,draw opacity=1 ]   (424,104) -- (558.58,102.55) ;
%Shape: Ellipse [id:dp06189842406096169] 
\draw  [color={rgb, 255:red, 139; green, 6; blue, 24 }  ,draw opacity=1 ][fill={rgb, 255:red, 139; green, 6; blue, 24 }  ,fill opacity=1 ] (558.58,102.55) .. controls (558.58,101.59) and (559.21,100.82) .. (559.98,100.82) .. controls (560.76,100.82) and (561.39,101.59) .. (561.39,102.55) .. controls (561.39,103.51) and (560.76,104.28) .. (559.98,104.28) .. controls (559.21,104.28) and (558.58,103.51) .. (558.58,102.55) -- cycle ;
%Shape: Parallelogram [id:dp9333827300182604] 
\draw  [color={rgb, 255:red, 245; green, 166; blue, 35 }  ,draw opacity=1 ][fill={rgb, 255:red, 248; green, 198; blue, 155 }  ,fill opacity=1 ] (142.99,171.69) -- (143.01,57.07) -- (220.99,34.57) -- (220.97,149.2) -- cycle ;
%Straight Lines [id:da9729776620886782] 
\draw [color={rgb, 255:red, 139; green, 6; blue, 24 }  ,draw opacity=1 ]   (186.99,103.13) -- (306,102.98) ;
%Straight Lines [id:da9892479272335981] 
\draw [color={rgb, 255:red, 139; green, 6; blue, 24 }  ,draw opacity=1 ] [dash pattern={on 4.5pt off 4.5pt}]  (144,103.98) -- (186.99,103.13) ;
%Straight Lines [id:da06925945610814954] 
\draw [color={rgb, 255:red, 139; green, 6; blue, 24 }  ,draw opacity=1 ]   (60.97,103.44) -- (141,103.98) ;
%Shape: Circle [id:dp23117152477136882] 
\draw  [color={rgb, 255:red, 139; green, 6; blue, 24 }  ,draw opacity=1 ][fill={rgb, 255:red, 139; green, 6; blue, 24 }  ,fill opacity=1 ] (183.69,103.13) .. controls (183.69,102.22) and (184.43,101.48) .. (185.34,101.48) .. controls (186.25,101.48) and (186.99,102.22) .. (186.99,103.13) .. controls (186.99,104.04) and (186.25,104.78) .. (185.34,104.78) .. controls (184.43,104.78) and (183.69,104.04) .. (183.69,103.13) -- cycle ;
\draw  [color={rgb, 255:red, 139; green, 6; blue, 24 }  ,draw opacity=1 ][fill={rgb, 255:red, 139; green, 6; blue, 24 }  ,fill opacity=1 ] (258.57,104.67) -- (256.36,102.64) -- (258.61,100.71) -- (257.47,102.66) -- cycle ;
\draw  [color={rgb, 255:red, 139; green, 6; blue, 24 }  ,draw opacity=1 ][fill={rgb, 255:red, 139; green, 6; blue, 24 }  ,fill opacity=1 ] (102.57,105.67) -- (100.36,103.64) -- (102.61,101.71) -- (101.47,103.66) -- cycle ;
\draw  [color={rgb, 255:red, 139; green, 6; blue, 24 }  ,draw opacity=1 ][fill={rgb, 255:red, 139; green, 6; blue, 24 }  ,fill opacity=1 ] (480.36,101.66) -- (482.59,103.68) -- (480.36,105.63) -- (481.47,103.66) -- cycle ;

% Text Node
\draw (579.41,40.01) node [anchor=north west][inner sep=0.75pt]  [font=\small]  {$\mathcal{B}$};
% Text Node
\draw (425,36.4) node [anchor=north west][inner sep=0.75pt]    {$\CC\times \overline{\CC}$};
% Text Node
\draw (421,108.4) node [anchor=north west][inner sep=0.75pt]    {$( a,\overline{b})$};
% Text Node
\draw (340,96.4) node [anchor=north west][inner sep=0.75pt]    {$\Longrightarrow $};
% Text Node
\draw (206,40.62) node [anchor=north west][inner sep=0.75pt]    {$S$};
% Text Node
\draw (309,93.62) node [anchor=north west][inner sep=0.75pt]    {$a$};
% Text Node
\draw (47,92.62) node [anchor=north west][inner sep=0.75pt]    {$b$};
% Text Node
\draw (63,38.62) node [anchor=north west][inner sep=0.75pt]    {$\CC$};
% Text Node
\draw (291,38.62) node [anchor=north west][inner sep=0.75pt]    {$\CC$};

\end{tikzpicture}
    \caption{Folding trick to produce a gapped boundary from a surface operator.}
    \label{fig:folding surface action}
\end{figure}
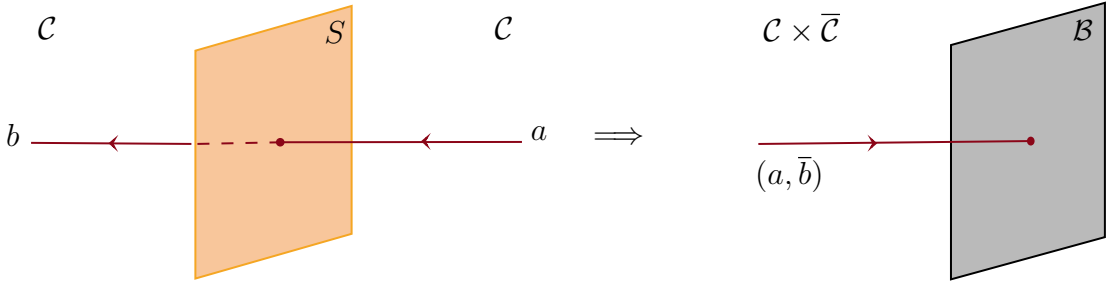

If $n_{S\cdot a}^b\neq 0$, then $a$ and $b$ can form a non-trivial junction on the surface operator (see Fig. \ref{fig: action of surface operator}).
Upon folding Fig. \ref{fig: action of surface operator} the surface operator becomes a gapped boundary of the TQFT $\CC \times \bar \CC$ on which the line operator $(a,\bar b)$ can end (see Fig. \ref{fig:folding surface action}).
Therefore, the action of the surface operator \eqref{eq:surface operator action} specifies a Lagrangian algebra object
\be
L_S=\sum_{a,b\in \CC } n_{S\dot a}^b ~ (a,\bar b)~.
\ee
In fact, the action of $S$ on the fusion spaces of the line operators also determines the Lagrangian algebra structure on $L_S$ (see \cite{Buican:2023bzl} for more details). Consider the fusion of $L_S$ on the canonical gapped boundary $\CB_{\CC}$ of $\CC \times \bar \CC$. From Theorem \ref{th: F(L)}, we know that $F(L_S)$ is a sum of Morita equivalent algebras in $\CC$. This is precisely the algebra which can be higher-gauged to construct the surface $S$. 
\vspace{0.2cm}

\noindent \textbf{Remark:} For a modular tensor category $\CC$, one can define the functors $T: \CC \times \bar \CC \to \CC$ and $\CC \to \CC \times \bar \CC$ as follows \cite{Kong:2007yv,Kong:2009inh}
\bea
&& T: (a,b) \to a\times b~, \\
&& R: a \to \sum_{c \in \CC} ~ (a \times \bar c, c)~.
\eea
These functors can also be used to transport algebras between $\CC$ and $\CC \times \bar\CC$. In fact, the functor $T$ is the same as the forgetful functor $F: \CZ(\CC) \to \CC$ under the equivalence $\CZ(\CC) \cong \CC \boxtimes \bar \CC$. Nevertheless, it maybe useful to use $F$ instead of $T$ since the isomorphisms $\Phi_{x,y}$ studied in section \ref{sec:Generalized discrete torsion} are trivial for $F$ while it is non-trivial for $T$.  

\subsection{A necessary condition for trivial anomaly}

Let $A$ be a non-simple line operator in $\CC$. Suppose $A$ is non-anomalous. Then we know that there must be a Lagrangian algebra $L$ corresponding to $A$. Using Theorem \ref{th: F(L)}, we must have 
\be
A \subseteq F(L) ~.
\ee
More explicitly, if
\be
L=\sum_{(a,e_a)\in \CZ(\CC)} N_L^{(a,e_a)} ~ (a,e_a)~,
\ee
then, we must have 
\be
\label{eq:necessary condition}
\sum_{(a,e_a)\in \CZ(\CC)}N_L^{(a,e_a)} N_{a}^b \neq 0 ~ \forall ~ b \in A~.
\ee
where $N_a^b$ is non-zero if the line operators $a$ and $b$ can form a topological junction. For an arbitrarily chosen $A$, we may not have bosonic line operators $(a,e_a) \in \CZ(\CC)$ satisfying \eqref{eq:necessary condition} implying that $A$ must be anomalous. In the next section, we will use this necessary condition to prove a statement about $\CC$-symmetric trivially gapped phases.

\subsubsection{Trivially gapped phase $\iff$ magnetic Lagrangian algebra}

In \cite{Zhang:2023wlu} the authors argue that a fusion category $\CC$ admits a trivially gapped phase only if the SymTFT $\CZ(\CC)$ admits a magnetic Lagrangian algebra $L_m$ defined by the property that the only simple line operator $L_m$ which can also end on the canonical gapped boundary $\CB_{\CC}$ is the trivial line operator. This is clear from the sandwich picture which shows that the local operators of the 1+1D TQFT $\CT_L$ correspond to the line operators which can end both on $\CB_{L_m}$ and $\CB_{\CC}$ (see \eqref{eq:number of local operators}). In this section, we will use Theorem \ref{th: F(L)} to show that this is a necessary and sufficient condition for the existence of a $\CC$-symmetric trivially gapped phase.

Recall the largest algebra object that a fusion category $\CC$ can admit is of the form
\be
A=\sum_{a\in \CC} d_a ~a~.
\ee
In fact, $A$ admits the structure of an algebra (is non-anomalous) if and only if $\CC$ admits a trivially gapped phase. To see this, suppose $\CC$ admits a $\CC$-symmetric TQFT that is trivially gapped. Then this TQFT only has the identity local operator. Correspondingly, there is only one simple 1D gapped boundary, say $\cm$. 
Therefore the module category $\CM$ has only one simple object, $\cm$. Consider the algebra 
\be
A_{\cm}:=\sum_{a\in \CC} N_{\cm}^a ~ a~,
\ee
where $N_{\cm}^a$ is the dimension of Hilbert space of operators at the junction of $a$ with the gapped boundary $\cm$. Since $\cm$ is the only gapped boundary, all line operators in $\CC$ must be able to form a junction on it. Moreover, $N_{\cm}^a=d_a$ to preserve quantum dimensions. Therefore, we have 
\be
A_{\cm}= \sum_{a\in \CC} ~ d_a ~a~.
\ee
Using \cite{ostrik2003module} (see section \ref{sec:Csym TQFTs}), we know that $A_{\cm}$ is non-anomalous. 

Conversely, suppose 
\be
A=\sum_{a\in \CC} d_a ~a~,
\ee
is non-anomalous. Consider the module category $\CM$ corresponding to this algebra. Let $\cm \in \CM$ and $A_{\cm}$ be the line operator as defined above. We have the relation \cite[Proposition 2.2]{etingof2010fusion}
\be
\label{eq:module dimension constraint}
\sum_{\cm} d_{\cm}^2= \text{dim}(\CC)~,
\ee
where
\be
d_{\cm}^2:=d_{A_{\cm}}~.
\ee
Now, since $\CM$ is the module category defined using the algebra $A$, there is some $\cm'\in \CM$ such that
\be
A_{\cm'}=A~.
\ee
Then, we have $d_{A_{\cm'}}=d_{A}=\text{dim}(\CC)$. To be compatible with the equation \eqref{eq:module dimension constraint}, $\cm'$ must be the only simple object in $\CM$. In other words, the $\CC$-symmetric TQFT defined using $\CM$ has only one simple gapped boundary. Consequently, it is trivially gapped and the whole symmetry $\CC$ is preserved. It follows from this argument that $A$ is not Morita equivalent to any other haploid algebra. 

From Theorem \ref{th: F(L)}, we know that 
\be
F(L)=\sum_{\cm\in \CM_L} A_{\cm}~.
\ee
The number of haploid algebras in the sum above is equal to the number of identity line operators in $F(L)$. Since $F$ preserves quantum dimensions and  $d_{L}=\text{dim}(\CC)$ by definition, the largest haploid algebra object that we can get on the R.H.S of the above equation is precisely $A$. Therefore, we find 
\be
F(L)=A
\ee
if and only if $F(L)$ contains a single identity operator. Since $F(L)$ is obtained from the perpendicular fusion of $L$ on the canonical gapped boundary $\CB_{\CC}$, this implies that $L$ does not contain any non-trivial line operators which can end on $\CB_{\CC}$. Therefore, $L$ is a magnetic Lagrangian algebra with respect to $\CB_{\CC}$.

\subsection{Transporting non-anomalous line operators between fusion categories}

\label{sec:transporting algebras}

Consider two fusion categories $\CC$ and $\CD$ which share the same SymTFT. In many cases, $\CD$ might be a simpler fusion category in which the non-anomalous line operators can be explicitly classified, while it could be hard to do the same in $\CC$. For example, $\CC$ could be a fusion category without multiplicity in its fusion coefficients while $\CD$ is not multiplicity-free. 

Let $[B_1],...[B_n]$ be the $n$ Morita equivalence classes of algebra objects/non-anomalous line operators in $\CD$ for some integer $n$. Using the data of this algebra, we can determine all the Lagrangian algebras $L_1,\dots,L_n$ of the SymTFT $\CZ(\CD)$. As we will see in explicit examples below, in many cases, even without knowledge of the multiplication in the algebras in $\CD$, we can determine the objects $L_1,...,L_n$ which admit the structure of a Lagrangian algebra. Now, since the fusion category $\CC$ share the same SymTFT, there must be a boundary condition $\CB_{\CC}$ of $\CZ(\CD)$ on which the line operators form the fusion category $\CC$. The non-anomalous line operators in $\CC$ can then be determined by fusing $L_1,\dots,L_n$ with the gapped boundary $\CB_{\CC}$ to get $F_{\CB_{\CC}}(L_1),F_{\CB_{\CC}}(L_2),\dots,F_{\CB_{\CC}}(L_n)$. In this way, the non-anomalous line operators in $\CD$ can be `transported' through the SymTFTs to non-anomalous line operators in $\CC$.  

In the following subsections, we will go through three examples in detail to illustrate the idea described above. First we will consider the fusion categories 
\be
\CC=\text{Rep}(D_8), ~~~ \CD=\text{Vec}_{D_8}~,
\ee
where $D_8$ is the dihedral group with $8$ elements. We will use the results from section \ref{sec:D8 example} to determine all algebra objects in Rep$(D_8)$ recovering the classification of algebra objects of Rep$(D_8)$ in \cite{Perez-Lona:2023djo,Diatlyk:2023fwf}. In Appendix \ref{ap:A4}, we use a similar method to classify the algebra objects in $A_4$, where $A_4$ is the alternating group of order $12$. This group is the smallest for which the category Rep$(A_4)$ has multiplicities in its fusion rules. We will show how the non-anomalous line operators in Vec$_{A_4}$ can be used to determine the non-anomalous line operators in Rep$(A_4)$. The final example that we consider is the three Morita equivalent Haagerup fusion categories.
\be
\CC=\SH_1, ~~~~ \CD=\SH_2, ~~~ \CE=\SH_3~.
\ee

\subsubsection{Rep$(D_{8})$ and Vec$_{D_8}$}

Rep$(D_8)$ is a non-invertible symmetry of various CFTs \cite{Choi:2023vgk,Diatlyk:2023fwf} and lattice models \cite{Seifnashri:2024dsd}. In this section, we will identify the non-anomalous line operators (algebra objects) in Rep$(D_8)$ using the Lagrangian algebras in $\CZ(\text{Rep}(D_8))$ and Theorem \ref{th: F(L)}.

Note that the SymTFT $\CZ(\text{Vec}_{D_8})$ is the same as $\CZ(\text{Rep}(D_8))$ since the symmetries Vec$_{D_8}$ and Rep$(D_8)$ are dual to each other under gauging. Therefore, the Lagrangian algebras in $\CZ(\text{Rep}(D_8))$ are the same as in $\CZ(\text{Vec}_{D_8})$ which were constructed explicitly in section \ref{sec:D8 example}. Recall that the bosonic line operators in $\CZ(\text{Vec}_{D_8})$ are
\bea
&& ([e],\mathds{1})~,~([e],\pi_1)~,~([e],\pi_2)~,~([e],\pi_3)~,~([e],\pi_4)~,~\\
&& ([r^2],\mathds{1})~,~([r^2],\pi_1)~,~([r^2],\pi_2)~,~([r^2],\pi_3)~,~([s],\mathds{1}_s)~,~\\
&& ([s],\omega_2)~,~([rs],\mathds{1}_{rs})~,~([rs],\tilde \omega_2)~,~([r],\mathds{1}_r)~.
\eea
All Lagrangian algebras in $\CZ(\text{Vec}_{D_8})$ are certain linear combinations of these bosons. The Lagrangian algebra
\be
L_{\text{Rep}(D_8)}= ([e],\mathds{1})+([r^2],\mathds{1}) + ([s],\mathds{1}_s) + ([r],\mathds{1}_r) + ([rs],\mathds{1}_{rs})~.
\ee
specifies a gapped boundary on which the line operators form the fusion category Rep$(D_8)$. 
Therefore, in order to determine the algebra objects in Rep$(D_8)$ we can use the map 
\be
F_{\CB_{\text{Rep}(D_8)}}: \CZ(\text{Rep}(D_8)) \to \text{Rep}(D_8)~,
\ee
and using Theorem \ref{th: F(L)}. In order to determine $F_{\CB_{\text{Rep}(D_8)}}$ it is useful to write the bosonic line operators in the form $(a,e_a)$ for some $a \in \text{Rep}(D_8)$ and half-braiding $e_a$. Then the map $F_{\CB_{\text{Rep}(D_8)}}$ is just given by 
\be
F_{\CB_{\text{Rep}(D_8)}}((a,e_a))=a~.
\ee
Using the construction of $\CZ(\text{Rep}(D_8))$ as the Drinfeld center of the Tambara-Yamagami category Rep$(D_8)$ we have \cite{gelaki2009centers,Zhang:2023wlu}
\bea
&& ([e],\mathds{1}) \to (\mathds{1},e_{\mathds{1}})~,~([e],\pi_1) \to (\pi_1,e_{\pi_1})~,~([e],\pi_2)\to (\pi_2,e_{\pi_2})~,~([e],\pi_3)\to (\pi_3,e_{\pi_3})~,~\cr
&& ([e],\pi_4)\to (\pi_4,e^{(1)}_{\pi_4})~,~([r^2],\mathds{1})\to (\mathds{1},e'_{\mathds{1}})~,~([r^2],\pi_1)\to (\pi_1,e'_{\pi_1})~,~\cr
&&([r^2],\pi_2)\to (\pi_2,e'_{\pi_2})~,~([r^2],\pi_3)\to (\pi_3,e'_{\pi_3})~,~ 
([s],\mathds{1}_s)\to (\mathds{1}+ \pi_2,e_{\mathds{1}+ \pi_2})~,~ \nonumber\\
&& ([s],\omega_2)\to (\pi_4,e^{(2)}_{\pi_4})~,~([rs],\mathds{1}_{rs})\to (1+\pi_3,e_{1+\pi_3})~,~([rs],\tilde \omega_2)\to (\pi_4,e^{(3)}_{\pi_4})~,~\cr 
&&([r],\mathds{1}_r)\to (\mathds{1}+\pi_1,e_{\mathds{1}+\pi_1})~.
\eea
where the explicit form of the half-braidings is not specified as they will not play a role in the following discussion. 
In general, the line operators in $\CZ(\text{Vec}_{D_8})$ and $\CZ(\text{Rep}(D_8))$ are related by  
\be
([g], \pi_g) \to (\text{Ind}_{C_g}^{D_8}(\pi_g), e)~,
\ee
for some half-braiding $e$ and $\text{Ind}_{C_g}^{D_8}(\pi_g)$ is the induction of the representation $\pi_g$ of $C_g$ to the full group $D_8$. The Lagrangian algebra objects in $\CZ(\text{Rep}(D_8))$ and the corresponding non-anomalous line operators in Rep$(D_8)$ are given in the following table. 
\bea
\arraycolsep=1.4pt\def\arraystretch{1.5}
\begin{array}{|c|c|}
\hline
\text{Lagrangian algebra objects in } \CZ(\text{Rep}(D_8)) & \text{Non-anomalous lines in Rep}(D_8) \\ \hline

(\mathds{1},e_{\mathds{1}}) + (\pi_1, e_{\pi_1}) + (\pi_2, e_{\pi_2}) + (\pi_3, e_{\pi_3}) + 2(\pi_4, e_{\pi_4}) & \mathds{1} + \pi_1 + \pi_2 + \pi_3 + 2 \pi_4 \\ \hline

\begin{tabular}{@{}c@{}}
$(\mathds{1},e_{\mathds{1}}) + (\pi_1, e_{\pi_1}) + (\pi_2, e_{\pi_2}) + (\pi_3, e_{\pi_3})  $ \\ + $(\mathds{1},e_{\mathds{1}}) + (\pi_1,e'_{\pi_1}) + (\pi_2,e'_{\pi_2}) + (\pi_3,e'_{\pi_3})$
\end{tabular}
& 2\mathds{1} + 2\pi_1 + 2\pi_2 + 2\pi_3 \\ \hline

(\mathds{1},e_{1})+ (\pi_1,e_{\pi_1}) + 2(\mathds{1}+\pi_1,e_{\mathds{1}+\pi_1}) + (\mathds{1},e'_{\mathds{1}})+ (\pi_1,e'_{\pi_1})

& 4\mathds{1} + 4 \pi_1 \\ \hline

(\mathds{1},e_{1}) +  (\pi_2, e_{\pi_2}) + 2(\mathds{1}+\pi_2, e_{\mathds{1}+\pi_2}) + (\mathds{1}, e'_{\mathds{1}})+(\pi_2,e'_{\pi_2}) 

& 4 \mathds{1} + 4 \pi_2 \\ 
\hline

(\mathds{1},e_{1}) +  (\pi_2, e_{\pi_2}) + 2(\pi_4, e^{(2)}_{\pi_4}) + (\pi_1, e'_{\pi_1})+(\pi_3,e'_{\pi_3}) 

& \mathds{1} + \pi_1 + \pi_2 + \pi_3 + 2\pi_4\\ \hline

(\mathds{1},e_{1}) +  (\pi_3, e_{\pi_3}) + 2(\mathds{1}+\pi_3, e_{\mathds{1}+\pi_3}) + (\mathds{1}, e'_{\mathds{1}})+(\pi_3,e_{\pi_3}) 

& 4\mathds{1} + 4\pi_3\\ \hline

(\mathds{1},e_{1}) +  (\pi_3, e_{\pi_3}) + 2(\pi_4, e^{(3)}_{\pi_4}) + (\pi_1, e'_{\pi_1})+(\pi_2,e'_{\pi_2}) 

& \mathds{1} + \pi_1 + \pi_2 + \pi_3 + 2\pi_4\\ \hline

\begin{tabular}{@{}c@{}} $(\mathds{1},e_{\mathds{1}})+  (\mathds{1},e'_{\mathds{1}}) + (\mathds{1}+\pi_2,e_{\mathds{1}+\pi_2})+ $ \\ $(\mathds{1}+\pi_1,e_{\mathds{1}+\pi_1})+(\mathds{1}+\pi_3,e_{\mathds{1}+\pi_3})$ \end{tabular}

& 5 \mathds{1} + \pi_1 + \pi_2 + \pi_3 \\  \hline

\begin{tabular}{@{}c@{}} $(\mathds{1},e_{\mathds{1}})+  (\pi_1,e'_{\pi_1}) + (\pi_4,e^{(2)}_{\pi_4})+ $ \\ $(\mathds{1}+\pi_1,e_{\mathds{1}+\pi_1})+(\pi_4,e^{(3)}_{\pi_4})$ \end{tabular}
 
 & 2 \mathds{1} + 2 \pi_1 + 2 \pi_4\\ \hline

(\mathds{1},e_{\mathds{1}}) + (\pi_2, e_{\pi_2}) + (\pi_4, e_{\pi_4}) + (\mathds{1}+\pi_2, e_{\mathds{1}+\pi_2}) + (\pi_4,e^{(2)}_{\pi_4}) 

& 2 \mathds{1} + 2\pi_2 + 2\pi_4 \\ \hline

(\mathds{1},e_{\mathds{1}}) + (\pi_3, e_{\pi_3}) + (\pi_4, e_{\pi_4}) + (\mathds{1}+\pi_3, e_{\mathds{1}+\pi_3}) + (\pi_4,e^{(3)}_{\pi_4}) 

& 2 \mathds{1} + 2\pi_3 + 2\pi_4 \\ \hline

\end{array}
\eea
We find that there are three different Lagrangian algebra objects in the SymTFT which gives the algebra object
\be
\mathds{1}+\pi_1+\pi_2+\pi_3+2\pi_4~.
\ee
Therefore, there are three distinct multiplications on this algebra object. This agrees with the fact that Rep$(D_8)$ admits three fibre functors. The algebra object
\be
2\mathds{1}+2\pi_1+2\pi_2+2\pi_3~,
\ee
decomposes into two copies of the algebra object $\mathds{1}+\pi_1+\pi_2+\pi_3$. This corresponds to gauging the $\DZ_2\times \DZ_2$ invertible  line operators in Rep$(D_8)$. The algebra object 
\be
4\mathds{1}+4\pi_1~,
\ee
decomposes into four copies of the algebra object $\mathds{1}+\pi_1$. The same is true for the algebra objects $4\mathds{1}+4\pi_2$ and $4\mathds{1}+4\pi_3$. These again correspond to gauging invertible line operators in Rep$(D_8)$. 

We also have the algebra object
\be
5\mathds{1}+\pi_1+\pi_2+\pi_3~,
\ee
which decomposes into $4A_{1}+A_2$ where
$A_1:=\mathds{1}, A_2:=\mathds{1}+\pi_1+\pi_2+\pi_3$. This shows that $\mathds{1}+\pi_1+\pi_2+\pi_3$ is a Morita trivial algebra which agrees with the fusion rules of Rep$(D_8)$. Finally, we have the algebra object
\be
2\mathds{1}+2\pi_1+2\pi_4
\ee
which decomposes into two copies of the algebra $\mathds{1}+\pi_1+\pi_4$.\footnote{At the level of the object we can also consider the decomposition of $2\mathds{1}+2\pi_2+2\pi_4$ into two copies of $\mathds{1}+\pi_1$ and $\mathds{1}+\pi_4$. $\mathds{1}+\pi_1$ admits a unique multiplication on it. Therefore, this decomposition is not consistent with the fact that $\mathds{1}+\pi_1$ corresponds to the Lagrangian algebra object $(\mathds{1},e_{1})+ (\pi_1,e_{\pi_1}) + 2(\mathds{1}+\pi_1,e_{\mathds{1}+\pi_1}) + (\mathds{1},e'_{\mathds{1}})+ (\pi_1,e'_{\pi_1})$.} The same is true for the algebra objects $2\mathds{1}+2\pi_2+2\pi_4$ and $2\mathds{1}+2\pi_3+2\pi_4$. The haploid algebra objects obtained above from Lagrangian algebra objects of the bulk SymTFT agree with \cite{Perez-Lona:2023djo,Diatlyk:2023fwf}.

\subsubsection{Haagerup fusion categories}

The Haagerup fusion categories $\SH_1$, $\SH_2$ and $\SH_3$ is a Morita equivalence class of three fusion categories \cite{Izumi:2001mi,grossman2012quantum}. 
The fusion category $\SH_1$ has line operators $\trl,\nu,\eta,\mu$ with fusion rules given by
\be
\begin{array}{|c||c|c|c|c|}
 \hline 
     & \trl & \nu & \eta & \mu \\ \hline \hline
    \nu & \nu & \trl+2\nu+2\eta+\mu & 2\nu+\eta+\mu & \mu \\ \hline
   \eta  & \eta & 2\nu+\eta+\mu & \trl+\nu+\eta+\mu & \nu + \eta \\ \hline
   \mu & \mu & \nu +\eta+mu & \nu +\eta & \trl+\nu \\ \hline
\end{array}
\ee
These line operators have quantum dimensions given by 
\be
1, d+1,d,d-1~,
\ee
respectively where $d=\frac{3+\sqrt{13}}{2}$. The fusion rules are commutative and have multiplicity. 

Both $\SH_2$ and $\SH_3$ have line operators $\trl,\alpha,\alpha^2,\rho,\alpha \rho, \alpha^2 \rho$ and the same fusion rules given by  
\be
\begin{array}{|c||c|c|c|c|c|c|c|}
\hline
     & \trl & \alpha & \alpha^2 & \rho & \alpha \rho & \alpha^2 \rho \\ \hline \hline
   \trl  & \trl & \alpha & \alpha^2 & \rho & \alpha \rho & \alpha^2 \rho \\ \hline
   \al & \al & \al^2 & \trl & \al \rho & \al^2 \rho & \rho \\ \hline
   \al^2 & \al^2 & \trl & \al & \al^2 \rho & \rho & \al \rho \\ \hline
   \rho & \rho & \al^2\rho & \al \rho & \trl+H & \al^2 +H & \al+H\\ \hline
   \al \rho & \al\rho & \rho & \al^2\rho & \al + H & \trl+H & \al^2 +H\\ \hline
   \al^2 \rho & \al^2 \rho & \al \rho & \rho & \al^2+H & \al+H & \trl+H\\ \hline
\end{array}
\ee
where $H=\rho+\al\rho+\al^2\rho$. The quantum dimensions of the lines are
\be
1,1,1,d,d,d~,
\ee
respectively. Note that the fusion rules are non-commutative and without multiplicity.

The fusion category $\SH_2$ has three module categories $\CM_1,\CM_2$ and $\CM_3$ and we have  \cite{grossman2012quantum} (see also \cite[Section 7]{Huang:2021ytb}, Appendix \ref{ap:more on Haagerup})

\bea
&& \bigoplus_{\cm\in\CM_{1}} A_{\cm}= 4\trl+\al+\al^2+3H~.  \nonumber \\
\label{eq:sum of algebras in H2}
&&\bigoplus_{\cm\in\CM_{2}}  A_{\cm}= 6\trl+3H~. \\
&&\bigoplus_{m\in\CM_{3}} A_{\cm}= 2 \trl+2\al+2\al^2+3H~. \nonumber
\eea
These three algebras correspond to three Lagrangian algebra objects in the SymTFT $\CZ(\SH_2)$.

$\CZ(\SH_2)$ has line operators \cite{Izumi:2001mi}
\be
\trl, \pi_1, \pi_2, \sigma_1, \sigma_2,\sigma_3,\mu_1,\mu_2,\dots \mu_6~.
\ee
These have quantum dimensions 
\be
1,3d+1,3d+2,3d+2,3d+2,3d+2, 3d,3d,3d,3d,3d,3d~,
\ee
respectively. The fusion rules and modular data of these line operators can be found, for example, in \cite{hong2008exotic}. The only bosonic line operators are $\trl,\pi_1,\pi_2,\sigma_1$. All Lagrangian algebras in this SymTFT must be supported on a linear combination of these bosonic line operators. In order to identity the Lagrangian algebra corresponding to the three classes of algebras in $\SH_2$, it is useful to write the bosonic line operators as $(a,e_a)$ for some $a\in \SH_2$ and half-braidings $e_a$. This is given explicitly in \cite[Section 2.2]{Evans:2010yr} (see also \cite[Section 4.4.3]{Lin:2022dhv}) and is given by
\be
\label{eq:rewriting bosons centre of H_2}
\trl \to (\trl,e_\trl),~ \pi_1 \to (\trl+\rho,e_{\trl+\rho}),~ \pi_2\to (\al+\al^2+H,e_{\al+\al^2+H}),~ \sigma_1 \to (\trl+\trl+H,e_{\trl+\trl+H})
\ee
where $H=\rho+\al\rho+\al^2\rho$. Consider the Lagrangian algebra 
\be
L=N_L^\trl \trl + N_L^{\pi_1} \pi_1 + N_L^{\pi_2}\pi_2 +N_L^{\sigma_1}\sigma_1. 
\ee
Moving $L$ to the canonical gapped boundary $\CB_{\SH_2}$ corresponding to the fusion category $\SH_2$, we get
\be
F_{\CB_{\SH_2}}(L)= N_L^{\trl} \trl + N_L^{\pi_1} (\trl+\rho)+ N_L^{\pi_2} (\al+\al^2+H)+N_L^{\sigma_1}(\trl+\trl+H)~.
\ee
where $N_L^{\pi_1},N_L^{\pi_2}$ and $N_L^{\sigma_1}$ are non-negative integers. Using Theorem \ref{th: F(L)} we know that $F_{\CB_{\SH_2}}(L)$ must agree with one of the algebra objects in \eqref{eq:sum of algebras in H2}. Using this, we get the three Lagrangian algebras
\bea
&& \CM_1 \to L_1:=\trl+\pi_1+\pi_2+\sigma_1~,\\
&& \CM_2 \to L_2:= \trl+\pi_1+2\sigma_1~,\\
&& \CM_3 \to L_3:=\trl+\pi_1+2\pi_2~.
\eea
The canonical gapped boundary corresponding to $\SH_2$ is given by the Lagrangian algebra $L_2$. $L_3$ corresponds to the fusion category $\SH_3$ and $L_1$ corresponds to $\SH_1$. 

The algebra objects/non-anomalous line operators in $\SH_3$ can be found by computing 
\be
F_{\CB_{L_3}}(L_1),~ F_{\CB_{L_3}}(L_2) \text{ and } F_{\CB_{L_3}}(L_3).
\ee
Using \cite[Section 4.4.3]{Lin:2022dhv}, we find
\be
F_{\CB_{L_3}}(\pi_1)=\trl+H,~ F_{\CB_{L_3}}(\pi_2)=\trl+\trl+H \text{ and } F_{\CB_{L_3}}(\sigma_1)=\al+\al^2+H~.
\ee
Using this, we get
\bea
&& F_{\CB_{L_3}}(L_1)=\trl+\trl+H+\trl+\trl+H+\al+\al^2+H=4\trl+\al+\al^2+3H~. \nonumber \\ 
&& F_{\CB_{L_3}}(L_2)=\trl+\trl+H+2(\al^2+\al^2+H)=2\trl+2\al+2\al^2+3H~,\\
&& F_{\CB_{L_3}}(L_3)=\trl+\trl+H+2(\trl+\trl+H)=6\trl+3H~,\nonumber
\eea
These three algebra objects are consistent with the three module categories in $\SH_3$ \cite{grossman2012quantum} (see Appendix \ref{ap:more on Haagerup}).

Finally, by computing 
\be
F_{\CB_{L_1}}(\pi_1),~ F_{\CB_{L_1}}(\pi_2) \text{ and } F_{\CB_{L_1}}(\sigma_1)~,
\ee
we can find the non-anomalous line operators in $\SH_1$. This requires finding the bulk-to-boundary map for the boundary condition $\CB_{L_1}$. This can be determined by taking the quotient of $\CZ(\SH_2)$ by the Lagrangian algebra $L_1$ as described in \cite{Cong:2017hcl}. The details of this calculation are given in Appendix \ref{ap:quotient functor}. We get,
\be
\label{eq:Haageup bulk to boundary}
F_{\CB_{L_1}}(\pi_1)=\trl+\nu+\eta+\mu,~ F_{\CB_{L_1}}(\pi_2)=\trl+2\eta+\nu \text{ and } F_{\CB_{L_1}}(\sigma_1)=\trl+2\nu+\mu~,
\ee
Using this, we find
\bea
&& F_{\CB_{L_1}}(L_1)=\trl+\trl+\nu+\eta+\mu+\trl+2\eta+\nu+\trl+2\nu+\mu=4\trl+4\nu+3\eta+2\mu~. \nonumber \\
&& F_{\CB_{L_1}}(L_2)=\trl+\trl+\nu+\eta+\mu+2(\trl+2\nu+\mu)=4\trl+5\nu+\eta+3\mu~,\\
&& F_{\CB_{L_1}}(L_3)=\trl+\trl+\nu+\eta+\mu+2(\trl+2\eta+\nu)=4\trl+3\nu+5\eta+\mu~. \nonumber 
\eea
These three algebra objects are consistent with the three module categories of $\SH_{1}$ \cite{grossman2012quantum} (see Appendix \ref{ap:more on Haagerup}).

\section{Conclusion}

In this paper, we studied the non-anomalous line operators in a 1+1D QFT with fusion category symmetry $\CC$ using gapped boundaries of the SymTFT $\CZ(\CC)$. Physically equivalent gaugings in 1+1D correspond to the same gapped boundary of the SymTFT. Our analysis determines the explicit map between them. 
\begin{itemize}
    \item Given a Lagrangian algebra object $L$ in $\CZ(\CC)$ the corresponding physically equivalent gaugings in $\CC$ can be found from $F(L)$, where $F$ is the bulk-to-boundary map. (Theorem \ref{th: F(L)}).
    \item In fact, $F$ maps the algebra structure of $L$ to the line operator $F(L)$ through equation \eqref{eq:relating multiplications}. 
    \item The explicit examples considered in this paper show that in some cases Theorems \ref{th: F(L)} and \ref{th: non-anomalous} can be used to determine all non-anomalous line operators in $\CC$ even without the knowledge of the algebra structure on the Lagrangian algebras in $\CZ(\CC)$.
    \item Conversely, knowledge of non-anomalous line operators in $\CC$ puts strong constraints on which line operators in the bulk SymTFT can form a Lagrangian algebra. We used this to find a necessary condition for a line operator $A \in \CC$ to be non-anomalous. This was then used to prove that $\CC$-symmetric trivially gapped  phases exist if and only if the bulk SymTFT admits a magnetic Lagrangian algebra.
    \item We discussed the notion of transporting non-anomalous line operators between fusion categories which share the same SymTFT. This method can be used to determine non-anomalous line operators in a fusion category from those in a ``simpler" fusion category. This provides an alternative to using NIM-reps to determine algebra objects in a fusion category $\CC$. 
\end{itemize}

\noindent The following are some interesting future directions:

\begin{itemize}
    \item It will be interesting tp generalize the analysis in this paper to SymTFTs for fermionic symmetries \cite{Wen:2024udn}. 
    \item Generalization to higher dimensions. It will be interesting to apply the ideas presented here to the setting in \cite{Putrov:2023jqi,Antinucci:2023ezl,Cordova:2023bja,Argurio:2024oym}. 
    \item General relation between anomalies and gapped boundaries for SymTFTs of non-finite and/or continuous symmetries \cite{kong2017boundary,Putrov:2022pua,Brennan:2024fgj,Antinucci:2024zjp,Apruzzi:2024htg,Bonetti:2024cjk}.
    \item For an invertible non-anomalous symmetry $G$, the discrete torsion takes values in $H^2(G,U(1))$. It would be interesting to explore the structure of generalized discrete torsions for a fixed non-anomalous line operator using gapped boundaries of the SymTFT. 
    \item In \cite{Cordova:2024vsq}, the authors show that massive 1+1D QFTs with spontaneously broken non-invertible symmetries often have particle degeneracies. A crucial part of finding the degeneracies involves determining the kernel and cokernel of certain maps relating different 1D gapped boundaries of the QFT. It will be interesting to use the map from  gapped interfaces between 1+1D gapped boundaries of $\CZ(\CC)$ and 1D gapped boundaries of 1+1D QFTs/TQFTs studied in this work to give a bulk perspective on the results in \cite{Cordova:2024vsq}. 
\end{itemize}

\acknowledgments

We thank Mahesh Balasubramanian, Matthew Buican, Arkya Chatterjee, Christian Copetti, Clement Delcamp, Nicholas Holfester, Adrian Padellaro, Brandon Rayhaun, Ingo Runkel, Sakura Schafer-Nameki,  Yifan Wang for discussions and comments related to this work. R.R. thanks Perimeter Institute for Theoretical Physics for hospitality during the workshop `Higher Categorical Tools for Quantum Phases of Matter' during which part of this work was completed. P.P. thanks the Galileo Galilei Institute for Theoretical Physics for the hospitality and the INFN for partial support during the completion of this work. We thank ULB Brussels and International Solvay Institutes for hospitality during the workshop  `Symmetries, Anomalies and Dynamics of Quantum Field theory' where some of the results in this paper were announced. 

\appendix

\section{Constraints on the action of line operators on gapped interfaces}

\label{ap:action on gapped interfaces}

In Theorem \ref{th: F(L)}, we learned that 
\be
F(L)= \sum_{\cm \in \CM_L} A_{\cm}~.
\ee
From Fig. \ref{fig:lines fixing gapped interface} and surrounding discussion we know that finding the haploid algebra objects in $F(L)$ directly from the SymTFT requires determining how the line operators on the gapped boundary $\CB_{\CC}$ act on the gapped interfaces between $\CB_{\CC}$ and $\CB_{L}$. When $\CZ(\CC)$ is a Dijkgraaf-Witten theory, this can be determined using the results in \cite{Cong:2017hcl}. 

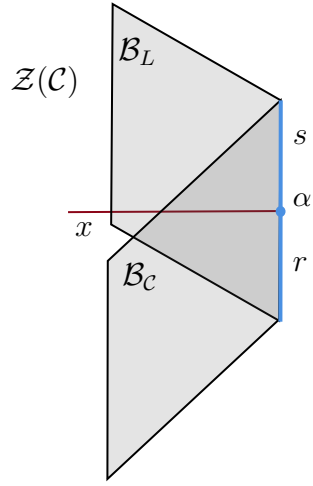
\begin{figure}[h!]
    \centering

\tikzset{every picture/.style={line width=0.75pt}} %set default line width to 0.75pt        

\begin{tikzpicture}[x=0.75pt,y=0.75pt,yscale=-1,xscale=1]

%Straight Lines [id:da7786415188429874] 
\draw [color={rgb, 255:red, 139; green, 6; blue, 24 }  ,draw opacity=1 ]   (285.3,130.11) -- (391.39,129.56) ;
%Shape: Parallelogram [id:dp9196071144712918] 
\draw  [fill={rgb, 255:red, 74; green, 74; blue, 74 }  ,fill opacity=0.15 ] (391.4,73.83) -- (390.65,184.73) -- (306.81,136.32) -- (307.56,25.42) -- cycle ;
%Shape: Parallelogram [id:dp8411268989150091] 
\draw  [fill={rgb, 255:red, 74; green, 74; blue, 74 }  ,fill opacity=0.15 ] (391.4,73.83) -- (391.21,183.56) -- (304.94,264.36) -- (305.13,154.63) -- cycle ;
%Straight Lines [id:da8277615712798359] 
\draw [color={rgb, 255:red, 74; green, 144; blue, 226 }  ,draw opacity=1 ][line width=1.5]    (391.3,185.22) -- (391.47,73.89) ;
%Shape: Ellipse [id:dp9274696764852814] 
\draw  [color={rgb, 255:red, 74; green, 144; blue, 226 }  ,draw opacity=1 ][fill={rgb, 255:red, 74; green, 144; blue, 226 }  ,fill opacity=1 ] (389.43,129.78) .. controls (389.43,128.55) and (390.31,127.56) .. (391.39,127.56) .. controls (392.47,127.56) and (393.35,128.55) .. (393.35,129.78) .. controls (393.35,131.02) and (392.47,132.01) .. (391.39,132.01) .. controls (390.31,132.01) and (389.43,131.02) .. (389.43,129.78) -- cycle ;

% Text Node
\draw (311,154.4) node [anchor=north west][inner sep=0.75pt]    {$\mathcal{B}_{\CC}$};
% Text Node
\draw (309,41.4) node [anchor=north west][inner sep=0.75pt]    {$\mathcal{B}_{L}$};
% Text Node
\draw (396,151.4) node [anchor=north west][inner sep=0.75pt]    {$r$};
% Text Node
\draw (255,55.4) node [anchor=north west][inner sep=0.75pt]    {$\CZ(\CC)$};
% Text Node
\draw (396,87.4) node [anchor=north west][inner sep=0.75pt]    {$s$};
% Text Node
\draw (287.3,133.51) node [anchor=north west][inner sep=0.75pt]    {$x$};
% Text Node
\draw (396,119.4) node [anchor=north west][inner sep=0.75pt]  [font=\small]  {$\alpha $};

\end{tikzpicture}
    \caption{$N_{xr}^{s}\neq 0$ implies a non-trivial junction between the lines $x$ and the gapped interfaces $r$ and $s$ where $\alpha$ labels a basis of point operators at the junction.}
    \label{fig:action of L}
\end{figure}

\begin{figure}[h!]
    \centering

\tikzset{every picture/.style={line width=0.75pt}} %set default line width to 0.75pt        

\begin{tikzpicture}[x=0.75pt,y=0.75pt,yscale=-1,xscale=1]
%uncomment if require: \path (0,302); %set diagram left start at 0, and has height of 302

%Shape: Parallelogram [id:dp4971322019632848] 
\draw  [fill={rgb, 255:red, 74; green, 74; blue, 74 }  ,fill opacity=0.15 ] (232.4,76.83) -- (231.65,187.73) -- (147.81,139.32) -- (148.56,28.42) -- cycle ;
%Shape: Parallelogram [id:dp5055780901923934] 
\draw  [fill={rgb, 255:red, 74; green, 74; blue, 74 }  ,fill opacity=0.15 ] (232.4,76.83) -- (232.21,186.56) -- (145.94,267.36) -- (146.13,157.63) -- cycle ;
%Shape: Parallelogram [id:dp6475162910796773] 
\draw  [fill={rgb, 255:red, 74; green, 74; blue, 74 }  ,fill opacity=0.15 ] (531.4,81.83) -- (530.65,192.73) -- (446.81,144.32) -- (447.56,33.42) -- cycle ;
%Shape: Parallelogram [id:dp010251932199833114] 
\draw  [fill={rgb, 255:red, 74; green, 74; blue, 74 }  ,fill opacity=0.15 ] (531.4,81.83) -- (531.21,191.56) -- (444.94,272.36) -- (445.13,162.63) -- cycle ;
%Straight Lines [id:da9324179520197682] 
\draw    (507.3,159.58) -- (531.39,137.02) ;
%Curve Lines [id:da8227111315043494] 
\draw [color={rgb, 255:red, 139; green, 6; blue, 24 }  ,draw opacity=1 ]   (398,137) .. controls (416,135) and (479,135) .. (507.3,158.58) ;
%Straight Lines [id:da7308274905052713] 
\draw    (204.34,123.84) -- (232.39,133.01) ;
%Straight Lines [id:da008395135573828871] 
\draw [color={rgb, 255:red, 74; green, 144; blue, 226 }  ,draw opacity=1 ][line width=1.5]    (531.3,193.22) -- (531.47,81.89) ;
%Shape: Ellipse [id:dp18020667801312995] 
\draw  [color={rgb, 255:red, 139; green, 6; blue, 24 }  ,draw opacity=1 ][fill={rgb, 255:red, 139; green, 6; blue, 24 }  ,fill opacity=1 ] (506.89,158.85) .. controls (506.89,157.89) and (507.52,157.11) .. (508.3,157.11) .. controls (509.07,157.11) and (509.7,157.89) .. (509.7,158.85) .. controls (509.7,159.8) and (509.07,160.58) .. (508.3,160.58) .. controls (507.52,160.58) and (506.89,159.8) .. (506.89,158.85) -- cycle ;
%Shape: Ellipse [id:dp4164387321001376] 
\draw  [color={rgb, 255:red, 74; green, 144; blue, 226 }  ,draw opacity=1 ][fill={rgb, 255:red, 74; green, 144; blue, 226 }  ,fill opacity=1 ] (529.43,137.78) .. controls (529.43,136.55) and (530.31,135.56) .. (531.39,135.56) .. controls (532.47,135.56) and (533.35,136.55) .. (533.35,137.78) .. controls (533.35,139.02) and (532.47,140.01) .. (531.39,140.01) .. controls (530.31,140.01) and (529.43,139.02) .. (529.43,137.78) -- cycle ;
%Curve Lines [id:da4340793739058266] 
\draw [color={rgb, 255:red, 139; green, 6; blue, 24 }  ,draw opacity=1 ]   (85,138) .. controls (107,137) and (187.99,149.48) .. (202.94,124.84) ;
%Straight Lines [id:da009432880916373065] 
\draw [color={rgb, 255:red, 74; green, 144; blue, 226 }  ,draw opacity=1 ][line width=1.5]    (232.3,188.22) -- (232.47,76.89) ;
%Shape: Ellipse [id:dp5301373081390559] 
\draw  [color={rgb, 255:red, 139; green, 6; blue, 24 }  ,draw opacity=1 ][fill={rgb, 255:red, 139; green, 6; blue, 24 }  ,fill opacity=1 ] (202.94,123.84) .. controls (202.94,122.88) and (203.57,122.11) .. (204.34,122.11) .. controls (205.12,122.11) and (205.75,122.88) .. (205.75,123.84) .. controls (205.75,124.8) and (205.12,125.57) .. (204.34,125.57) .. controls (203.57,125.57) and (202.94,124.8) .. (202.94,123.84) -- cycle ;
%Shape: Ellipse [id:dp7757515167044661] 
\draw  [color={rgb, 255:red, 74; green, 144; blue, 226 }  ,draw opacity=1 ][fill={rgb, 255:red, 74; green, 144; blue, 226 }  ,fill opacity=1 ] (230.43,132.78) .. controls (230.43,131.55) and (231.31,130.56) .. (232.39,130.56) .. controls (233.47,130.56) and (234.35,131.55) .. (234.35,132.78) .. controls (234.35,134.02) and (233.47,135.01) .. (232.39,135.01) .. controls (231.31,135.01) and (230.43,134.02) .. (230.43,132.78) -- cycle ;

% Text Node
\draw (518.34,147.7) node [anchor=north west][inner sep=0.75pt]    {$a$};
% Text Node
\draw (211,131.4) node [anchor=north west][inner sep=0.75pt]    {$b$};
% Text Node
\draw (451,162.4) node [anchor=north west][inner sep=0.75pt]    {$\mathcal{B}_{\CC}$};
% Text Node
\draw (449,49.4) node [anchor=north west][inner sep=0.75pt]    {$\mathcal{B}_{L}$};
% Text Node
\draw (535,164.4) node [anchor=north west][inner sep=0.75pt]    {$r$};
% Text Node
\draw (395,63.4) node [anchor=north west][inner sep=0.75pt]    {$\CZ(\CC)$};
% Text Node
\draw (536,91.4) node [anchor=north west][inner sep=0.75pt]    {$s$};
% Text Node
\draw (408.3,141.51) node [anchor=north west][inner sep=0.75pt]    {$x$};
% Text Node
\draw (536,129.4) node [anchor=north west][inner sep=0.75pt]  [font=\small]  {$\beta $};
% Text Node
\draw (152,157.4) node [anchor=north west][inner sep=0.75pt]    {$\mathcal{B}_{\CC}$};
% Text Node
\draw (150,44.4) node [anchor=north west][inner sep=0.75pt]    {$\mathcal{B}_{L}$};
% Text Node
\draw (236,160.4) node [anchor=north west][inner sep=0.75pt]    {$r$};
% Text Node
\draw (96,58.4) node [anchor=north west][inner sep=0.75pt]    {$\CZ(\CC)$};
% Text Node
\draw (237,87.4) node [anchor=north west][inner sep=0.75pt]    {$s$};
% Text Node
\draw (102.3,141.51) node [anchor=north west][inner sep=0.75pt]    {$x$};
% Text Node
\draw (237,124.4) node [anchor=north west][inner sep=0.75pt]  [font=\small]  {$\gamma $};
% Text Node
\draw (201,104.4) node [anchor=north west][inner sep=0.75pt]  [font=\small]  {$\psi $};
% Text Node
\draw (503.3,162.98) node [anchor=north west][inner sep=0.75pt]  [font=\small]  {$\chi $};

\end{tikzpicture}
    \caption{If $N_{xr}^{s} \neq 0$, then there must be some line operator $b\in F_{\CB_{L}}(L)$ such that the diagram on the left is non-trivial. Similarly, there must be some line operator $a \in F_{\CB_{\CC}}(L)$ such that the diagram on the right is non-trivial.}
    \label{fig:two ways of acting with l}
\end{figure}
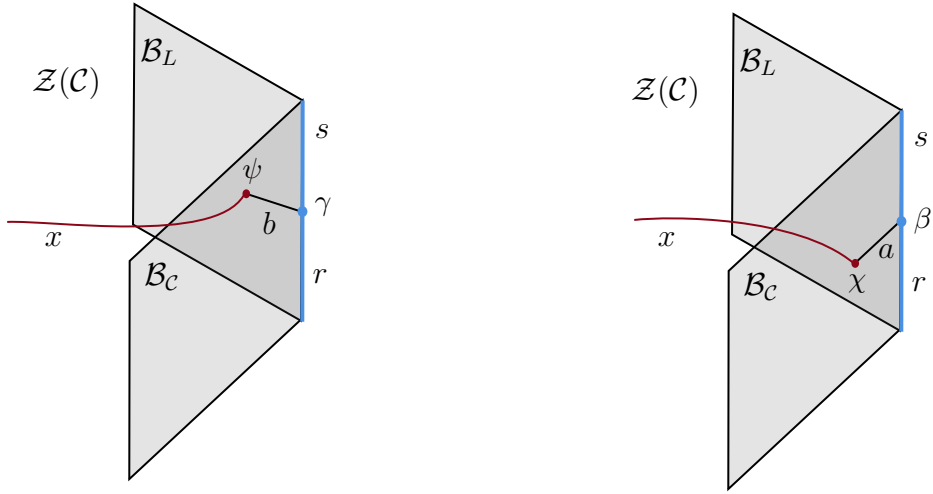

In this section, we will determine some general constraints on the action of the line operators in $F(L)$ on gapped interfaces which may be used in explicit computations to determine haploid sub-objects of $F(L)$. To this end, let us consider the action of $x \in L$ on the gapped interface $r$. We get
\be
x \times r = \sum_{s} N_{xr}^{s} ~ s~.
\ee
where $N_{xr}^{s}$ are non-negative integers and $s$ labels all the simple gapped interfaces between $\CB_{\CC}$ and $\CB_{L}$. If $N_{xr}^{s}\neq 0$, then we get the configuration in Fig. \ref{fig:action of L}.  
The condition $N_{xr}^s\neq 0$ also implies the non-trivial configurations in Fig. \ref{fig:two ways of acting with l}.
Therefore, we have
\be
\label{eq:two ways of L action}
N_{xr}^{s}= \sum_{b \in \CC_{L}} N_{x}^{b} N_{b r}^{s}=  \sum_{a\in \CC} N_{x}^{a} N_{a r}^{s}~.
\ee
where $\CC_{L}$ is the category of line operators on the gapped boundary $\CB_{L}$. Since $x \in L$, the fusion of $x$ with the gapped boundary $\CB_{L}$ contains the identity line in $\CC_{L}$. In other words,
\be
F_{\CB_{L}}(x)= 1 + \dots~,
\ee
Therefore, we have
\be
N_{xr}^{r} \neq 0 ~\forall~ x \in L~.
\ee
for any gapped interface $r$. Using \eqref{eq:two ways of L action}, we find that 
\be
\sum_{a\in \CC} N_{x}^{a} N_{a r}^{r} \neq 0~.
\ee
Therefore, for a fixed simple gapped interface $r$, there is some $a \in F_{\CB_{\CC}}(x)$ which acts trivially on it.

\section{Proof of Theorem \ref{th: non-anomalous}}

\label{ap:proof of theorem}

In this section, we will review some properties of characters of simple line operators in $\CZ(\text{Vec}_{G}^{\omega})$ and give a proof of Theorem \ref{th: non-anomalous}.

Recall that Vec$_{G}^{\omega}$ is the category of $G$-graded vector spaces with associator given by the 3-cocycle $\omega$.  An $\omega$-projective $G$-action on a $G$-graded vector space $V=\bigoplus_{g\in G}V_g$ is a set of linear maps
\be
f_{g} : V \to V~,~ g\in G~,
\ee
such that 
\be
f_g(V_h)=V_{ghg^{-1}}~, \text{and } (f_gf_h)(v)= \omega(g,h|k)~f_g(f_h(v))~,~ \forall ~ v\in V_k ~,
\ee
where 
\be
\omega(g,h|k):= \omega(g,h,k)^{-1}\omega(g,hgh^{-1},h)\omega(gh k (gh)^{-1},g,h)^{-1}~.
\ee
It is known that any line operator in the SymTFT $\CZ(\text{Vec}_{G}^{\omega})$ can be labeled by a $G$-graded vector space $V$ with $\alpha$-projective $G$-action (see \cite[Section 2.1, 3.1]{davydov2010centre}). 
Its character is defined as
\be
\chi_{V}(g,h):= \text{Tr}_{V_g}(h)~,
\ee
where $g,h$ is a pair of commuting elements in $G$. For any two $x, y \in \CZ(\text{Vec}_{G}^{\omega})$, it can be shown that the characters satisfy \cite{davydov2017lagrangian}
\bea
\label{eq: 2-character properties}
&& \chi_x(kgk^{-1},khk^{-1})= \frac{\omega(k,h|g)}{\omega(khk^{-1},k|g)} \chi_x(g,h)~, \\
&& \chi_{x \times y}(g,h)= \chi_{x}(g,h) \times \chi_{y}(g,h)~, \\
&& \chi_{x + y}(g,h)= \chi_{x}(g,h) + \chi_{y}(g,h) ~.
\eea 
Consider the following inner product on the characters
\be
\langle \chi_x, \chi_y \rangle = \frac{1}{|G|} \sum_{g,h\in G, ~ gh=hg}  \omega(h^{-1},h|g) \chi_x(g,h^{-1})\chi_y(g,h)~.
\ee
We will use the following important property of this inner product \cite{davydov2017lagrangian}
\be
\label{eq:innerprod property}
\langle \chi_x, \chi_y \rangle= |\text{Hom}_{\CZ(\text{Vec}_G^{\omega})}(x,y)|~.
\ee
It will also be useful to recall some properties of characters of projective representations. 
\bea
\label{eq: proj char properties 1}
&& \overline{\chi_{\pi_g}(h)}= \omega(h^{-1},h|g) ~ \chi_{\pi_g}(h^{-1})~, \\
\label{eq: proj char properties 2}
&& \chi_{\pi_g}(khk^{-1})= \frac{\omega(k,h|g)}{\omega(khk^{-1},k|f)}~  \chi_{\pi_g}(g)~,
\eea
where $\overline{\chi_{\pi_g}(h)}$ is the complex conjugate of $\chi_{\pi_g}(h)$ \cite[Proposition 2.2]{karpilovsky1985projective}. Now, we are ready to prove Theorem \ref{th: non-anomalous}.
\vspace{0.2cm}

\noindent \textbf{Theorem 4.2:}\textit{
$H\subseteq G$ is non-anomalous if and only if there exists a Lagrangian algebra $L$ in $\CZ(\text{Vec}_{G}^{\omega})$ such that
\be
([h],\pi_h)\in L ~~~ \forall ~ [h] \in C(H)~,
\ee
for some representation $\pi_h$ of the centralizer of $h$. }

\noindent {\bf Proof:} Let $H$ is a non-anomalous group. Then $A(H,1)$ is an algebra object. Using \eqref{eq:character of L}, we can compute the character of $\CZ(A(H,1))$ to get
\be
\chi_{\CZ(A(H,1))}(k,l):= \sum_{y \in Y} \frac{\omega(y^{-1}ly,y^{-1}|k)}{\omega(y^{-1},l|k)}~.
\ee
Let $([l],\pi_l)$ be a simple object in $\CZ(\text{Vec}_{G}^{\omega})$ such that $[l] \in C(H)$. Consider the inner product
\be
\langle \chi_{([l],\pi_l)}, \chi_{\CZ(A(H,1))} \rangle = \frac{1}{|G|} \sum_{g,h\in G, ~ gh=hg}  \omega(h^{-1},h|g) \chi_{([l],\pi_l)}(g,h^{-1})\chi_{\CZ(A(H,1))}(g,h)~.
\ee
Using \eqref{eq:character of simple object}, we get
\bea
\langle \chi_{([l],\pi_l)}, \chi_{\CZ(A(H,1))} \rangle &=& \frac{1}{|G|} \sum_{g\in [l], ~ gh=hg}  \omega(h^{-1},h|g) \chi_{\pi_g}(h^{-1}) \chi_{\CZ(A(H,1))}(g,h)~,\\
&=& \frac{1}{|[l]|} \sum_{g\in [l]} \bigg [ \frac{1}{|C_g|} \sum_{h \in C_g}  \omega(h^{-1},h|g) \chi_{\pi_g}(h^{-1}) \chi_{\CZ(A(H,1))}(g,h) \bigg ] ~,\nonumber \\
\eea
where we used the fact that $|G|=|[l]||C_{g}|$ for any $l \in G$ and $g \in [l]$, $|[l]|$ is the size of the conjugacy class $[l]$ and $|C_l|$ is the order of the group $C_l$. Using \eqref{eq: proj char properties 1}, we can write the above expression as
\be
\frac{1}{|[l]|} \sum_{g\in [l]} \bigg [ \frac{1}{|C_g|} \sum_{h \in C_g} \overline{\chi_{\pi_g}(h)} \chi_{\CZ(A(H,1))}(g,h) \bigg ]~.
\ee
Now, note that for a fixed $g$, $\chi_{\CZ(A(H,1))}(g,h)$ is a function on the group $C_g$. Moreover, if $k,h \in C_g$, we find that this function satisfies
\be
\chi_{\CZ(A(H,1))}(g,khk^{-1})= \chi_{\CZ(A(H,1))}(kgk^{-1},khk^{-1})= \frac{\omega(k,h|g)}{\omega(khk^{-1},k|g)} \chi_{\CZ(A(H,1))}(g,h)~.
\ee
where in the first equality we used the fact that $k \in C_g$ commutes with $g$ and in the second equality we used \eqref{eq: 2-character properties}. Moreover, since the conjugacy class $[l] \in C(H)$,  $\chi_{\CZ(A(H,1))}(g,h)$ is non-zero for some $g$ and $h$. (For example, $\chi_{\CZ(A(H,1))}(g,1)$ is non-zero.) Therefore, for a fixed $g$, $\chi_{\CZ(A(H,1))}(g,h)$ is a non-zero projective character on the group $C_g$ and 
\be
\label{eq:innerproduct}
\frac{1}{|C_g|} \sum_{h \in C_g} \overline{\chi_{\pi_g}(h)} ~ \chi_{\CZ(A(H,1))}(g,h)~,
\ee
is the inner-product of the characters $\chi_{\pi_g}(h)$ and $\chi_{\CZ(A(H,1))}(g,h)$. Since the characters of irreducible projective representations form a basis of projective class functions \cite[Theorem 3.1]{karpilovsky1985projective}, we can always choose a $\pi_l$ in $([l],\pi_l)$ such that the inner-product \eqref{eq:innerproduct} is non-zero. Therefore, there is always some $\pi_l$ for which $\langle \chi_{([l],\pi_l)}, \chi_{\CZ(A(H,1))} \rangle \neq 0$. 

Conversely, consider some line operator $([l],\pi_l)$ such that $\langle \chi_{([l],\pi_l)}, \chi_{\CZ(A(H,\sigma))} \rangle \neq 0$ is non-zero. We have
\bea
&& \hspace{-1.0cm} \langle \chi_{([l],\pi_l)}, \chi_{\CZ(A(H,\sigma))} \rangle \\
&=& \frac{1}{|G|} \sum_{g,h\in G, ~ gh=hg}  \omega(h^{-1},h|g) \chi_{([l],\pi_l)}(g,h^{-1})\chi_{\CZ(A(H,\sigma))}(g,h)~,\\
&=& \frac{1}{|G|} \sum_{g\in [l], ~ gh=hg}  \omega(h^{-1},h|g) \chi_{\pi_g}(h^{-1}) \chi_{\CZ(A(H,\sigma))}(g,h)~, \\
&=& \frac{1}{|G|} \sum_{g\in [l], ~ gh=hg}  \omega(h^{-1},h|g) \chi_{\pi_g}(h^{-1}) \chi_{\CZ(A(H,\sigma))}(g,h)~, \\
&=& \frac{1}{|G|} \sum_{g\in [l], ~ gh=hg}  \omega(h^{-1},h|g) \chi_{\pi_l}(h^{-1}) \sum_{y \in Y} \frac{\omega(y^{-1}hy,y^{-1}|g)\sigma(y^{-1}gy,y^{-1}hy)}{\omega(y^{-1},h|g)\sigma(y^{-1}hy,y^{-1}gy)}~, \nonumber \\
\eea
where 
\bea
Y:= \{y\in G, y^{-1}gy\in H, y^{-1}hy\in H\}/H \subset G/H~.
\eea
Since $\langle \chi_{([l],\pi_l)}, \chi_{\CZ(A(H,\sigma))} \rangle \neq 0$ is non-zero, the set $Y$ must be non-zero, which implies that $y^{-1}gy$ must be in $H$ for some $y$ where $g \in [l]$. In other words, $[l] \in C(H)$. 
$\hfill \square$

\section{Algebras in Rep$(A_{4})$ from algebras in Vec$_{A_4}$}

\label{ap:A4}

Consider the alternating group $A_4$ of even permutations of four objects. 
\be
A_4 = \langle (123), (234) \rangle~.
\ee
The category Vec$_{A_4}$ has 12 line operators labeled by even permutations. The Morita equivalence classes of algebras in Vec$_{A_4}$ are determined by conjugacy classes of subgroups of $A_4$ and associated 2-coycles. The conjugacy classes of subgroups of $A_4$ are given by
\bea
&& \DZ_1, ~ \{\langle (12)(34)\rangle, \langle (13)(24) \rangle, \langle (14)(23) \rangle\}\cr
&& \{\langle (243)\rangle, \langle (123) \rangle, \langle (142) \rangle,\langle (134) \rangle\}, \cr
&& \{\langle (13)(24), (12)(34) \rangle\}, A_4~.
\eea
The groups in the second conjugacy class are all isomorphic to $\DZ_2$, those in the third conjugacy class are isomorphic to $\DZ_3$ and those in the fourth are isomorphic to $\DZ_2 \times \DZ_2$. Therefore, the Morita equivalence classes of algebras in Vec$_{A_4}$ can be labeled as $(G,\sigma)$ where $H$ is the isomorphism class of the groups in a conjugacy class of subgroups and $\sigma \in H^2(G,U(1))$. We get
\be
\label{eq:A12 algebra objects}
(\DZ_1,1), (\DZ_2,1), (\DZ_3,1), (\DZ_2 \times \DZ_2,1), (\DZ_2 \times \DZ_2,\alpha), (A_4,1), (A_4,\beta)
\ee
where $\alpha, \beta$ are the non-trivial 2-cocycles corresponding to the non-trivial elements in $H^2(\DZ_2\times \DZ_2,U(1))\cong \DZ_2$ and $H^2(A_4,U(1))\cong \DZ_2$, respectively. These seven Morita equivalence classes of algebras correspond to seven Lagrangian algebras in the SymTFT $\CZ(\text{Vec}_{A_4})$. 

The SymTFT $\CZ(\text{Vec}_{A_4})$ is the $A_4$ Dijkgraaf-Witten theory. The line operators in this theory are labeled by 
\be
([g],\pi_g)
\ee
where $[g]$ is a conjugacy class of $A_4$ with representative $g$ and $\pi_g$ is an irreducible representation of the centralizer $C_g$ of $g$. The conjugacy classes of $A_4$ are 
\bea
&&[()], [(12)(34)]=\{(12)(34),(13)(24),(14)(23)\}, 
 [(123)]=\{(123),(142),(243),(134)\},\nonumber \\ 
 && [(124)]=\{(124),(234),(132),(143)\}~.
\eea
where $()$ denotes the trivial permutation. The centralizers of the representatives of these conjugacy classes are
\be
A_4, \DZ_2 \times \DZ_2 =\langle (13)(24), (12)(34) \rangle, \DZ_3=\langle (123)\rangle,\DZ_3=\langle (124)\rangle~.
\ee
The line operators in $\CZ(\text{Vec}_{A_4})$, their quantum dimensions and topological spins are given by 
\begin{center}
\begin{tabular}{ |c|c|c|c|c|c| } 
\hline
Line operator & $([()],\mathds{1})$ &$([()],\pi_1)$& $([()],\pi_2)$, & $([()],\pi_3)$& $([(12)(34)],\mathds{1}_1)$ \\ \hline
$d_x$ & 1 & 1 & 1 & 3 & 3 \\ \hline
$\theta_x$ & 1 & 1 & 1 & 1 & 1\\ \hline
\end{tabular}
\end{center}
\begin{center}
\begin{tabular}{ |c|c|c|c|c| } 
\hline
Line operator & $([(12)(34)],\omega_1)$ & $([(12)(34)],\omega_2)$  & $([(12)(34)],\omega_1\omega_2)$ & $([(123),\mathds{1}_2])$ \\ \hline
$d_x$ & 3 & 3 & 3 & 4 \\ \hline
$\theta_x$  & 1 & -1 & -1 & 1\\ \hline
\end{tabular}
\end{center}
\begin{center}
\begin{tabular}{ |c|c|c|c|c|c| } 
\hline
Line operator & $([(123),\omega])$ & $([(123)],\omega^2)$ & $([(124)],\mathds{1}_3)$ & $([(124),\tilde \omega])$ & $([(124),\tilde \omega^2 ])$\\ \hline
$d_x$ & 4 & 4 & 4 & 4 & 4 \\ \hline
$\theta_x$ & $e^{\frac{2\pi i}{3}}$ & $e^{-\frac{2 \pi i}{3}}$ & 1 & $e^{\frac{2\pi i}{3}}$ & $e^{-\frac{2\pi i}{3}}$ \\ \hline
\end{tabular}
\end{center}
where $\mathds{1},\pi_1,\pi_2$ are the 1-dimensional representations of $A_4$ and $\pi_3$ is the 3-dimensional irreducible representation. $\mathds{1}_1,\omega_1,\omega_2,\omega_1\omega_2$ are the four 1-dimensional representations of $\DZ_2 \times \DZ_2$. 
$\mathds{1}_2,\omega,\omega^2$ are the three 1-dimensional representations of $\DZ_3=\langle (123) \rangle $. 
$\mathds{1}_3,\tilde \omega,\tilde \omega^2$ are the three 1-dimensional representations of $\DZ_3=\langle (124) \rangle $. 

Now, using the character theory developed in section \ref{sec:invertible symmetry}, we can find the Lagrangian algebras in this theory from the algebra objects \eqref{eq:A12 algebra objects}. In fact, we can completely identify all algebra objects by using just Theorem \ref{th: non-anomalous}. Since $\langle (13)(24), (12)(34) \rangle$ is a $\DZ_2 \times \DZ_2$ normal subgroup of $A_{4}$, the discussion in section \ref{sec:invertible symmetry} and \cite{naidu2008lagrangian} implies that it must correspond to a Lagrangian subcategory of $\CZ(\text{Vec}_{A_4})$. The algebra objects and corresponding Lagrangian algebras as summarised in the following table. 
\bea
\arraycolsep=1.4pt\def\arraystretch{1.5}
\begin{array}{|c|c|}
\hline
\text{Algebra objects in Vec}_{A_4} & \text{Lagrangian algebra objects in } \CZ(\text{Vec}_{A_4}) \\ \hline
(\DZ_1,1) & ([()],\mathds{1})+([()],\pi_1) + ([()],\pi_2) + 3 ([()],\pi_3) \\
\hline
(\DZ_2,1) & 
\begin{tabular}{@{}c@{}}
$([()],\mathds{1})+([(12)(34)],\mathds{1}_1) + ([(12)(34)],\omega_1)$ \\ + $([()],\pi_1) + ([()],\pi_2) + ([()],\pi_3)$
\end{tabular}
\\
\hline
(\DZ_3,1) & ([()],\mathds{1})+([(123)],\mathds{1}_2) + ([(124)],\mathds{1}_3) + ([()],\pi_3) \\
\hline
(\DZ_2 \times \DZ_2,1) & ([()],\mathds{1})+([()],\pi_1) + ([()],\pi_2) + 3([(12)(34)],\mathds{1}_1) \\
\hline
(\DZ_2 \times \DZ_2,\alpha) & ([()],\mathds{1})+([()],\pi_1) + ([()],\pi_2) + 3([(12)(34)],\omega_1) \\
\hline
(A_4,1) & ([()],\mathds{1})+([(12)(34)],\mathds{1}_1) + ([(123)],\mathds{1}_2) + ([(124)],\mathds{1}_3) \\
\hline
(A_4,\beta) & ([()],\mathds{1}) + ([(12)(34)],\omega_1)+([(123)],\mathds{1}_2) + ([(124)],\mathds{1}_3) \\
\hline
\end{array}
\eea

The line operators of the gapped boundary of the SymTFT $\CZ(\text{Vec}_{A_4})$ specified by the Lagrangian algebra 
\be
L_6= ([()],\mathds{1})+([(12)(34)],\mathds{1}_1) + ([(123)],\mathds{1}_2) + ([(124)],\mathds{1}_3)~.
\ee
form the category Rep$(A_4)$. The fusion category Rep$(A_4)$ contains four line operators $\mathds{1},\pi_1,\pi_2,\pi_3$ corresponding to the three 1-dimensional representations and one 3-dimensional representation of $A_4$, respectively. The fusion rules are
\be
\pi_1 \times \pi_1 = \pi_2, ~ \pi_1 \times \pi_2 = \mathds{1}, ~ \pi_3 \times \pi_3 =\mathds{1} + \pi_1 + \pi_2 + 2\pi_3~.  
\ee
The non-anomalous line operators in Rep$(A_4)$ can be identified by applying the map $F_{L_6}$ to all the Lagrangian algebras in the table above. In order to do this, it is convenient to use the isomorphism 
\be
\label{eq:SymTFT iso}
\CZ(\text{Vec}_{A_4}) \cong \CZ(\text{Rep}(A_4))~,
\ee
and write all line operators in the SymTFT in the form $(a,e_a)$ for some $a \in \text{Rep}(A_4)$ and half-braiding $e_a$. Once we write all line operators in this form, we know that 
\be
F_{L_6}(a,e_a)= a~.
\ee
Therefore, in order to find the non-anomalous line operators in Rep$(A_4)$, we don't have to explicitly determine the half-braidings $e_a$. Under the isomorphism \eqref{eq:SymTFT iso}, we get
\be
([g], \pi_g) \to (\text{Ind}_{C_g}^{A_4}(\pi_g), e)~,
\ee
for some half-braiding $e$ and $\text{Ind}_{C_g}^{A_4}(\pi_g)$ is the induction of the representation $\pi_g$ of $C_g$ to the full group $A_4$. The induction of representations can be found using the character table of $A_4$ and Frobenius reciprocity. We get 
\bea
&& ([()],\mathds{1}) \to (\mathds{1},e_1), ~ ([()],\pi_1) \to (\pi_1,e_{\pi_1}), ~ ([()],\pi_2) \to (\pi_2,e_{\pi_2}),  ~ ([()],\pi_3) \to (\pi_3,e_{\pi_3}), \cr
&& ([(12)(34)],\mathds{1}_1) \to (\mathds{1}+\pi_1+\pi_2,e_{\mathds{1}+\pi_1+\pi_2}), ~ ([(12)(34)],\omega_1) \to (\pi_3, e_{\pi_3}^{(1)}),~ \cr
&& ([(12)(34)],\omega_2) \to (\pi_3, e_{\pi_3}^{(2)}), ~ ([(12)(34)],\omega_1\omega_2) \to (\pi_3, e_{\pi_3}^{(3)})~, ~ \cr
&& ([(123),\mathds{1}_2]) \to (\mathds{1}+\pi_3,e_{\mathds{1}+\pi_3})~,([(123)],\omega) \to (\pi_2 + \pi_3, e_{\pi_2+\pi_3}),~ \cr 
&& ([(123)],\omega^2) \to (\pi_1 + \pi_3, e_{\pi_1+\pi_3}),~ ([(124)],\mathds{1}_3) \to (\mathds{1} + \pi_3, \tilde e_{\mathds{1}+\pi_3}),~ \cr
&&([(124)],\tilde \omega) \to (\pi_1 + \pi_3, \tilde e_{\pi_1+\pi_3})~, ([(124)],\tilde \omega^2) \to (\pi_2 + \pi_3, \tilde e_{\pi_1+\pi_3})  ~.
\eea
Using this isomorphism, the $7$ Lagrangian algebras can be written in terms of line operators in $\CZ(\text{Rep}(A_4))$. We can apply the map $F_{L_6}$ to these Lagrangian algebras to find all the non-anomalous line operators in Rep$(A_4)$. 
\bea
\arraycolsep=1.4pt\def\arraystretch{1.5}
\begin{array}{|c|c|}
\hline
\text{Lagrangian algebra objects in } \CZ(\text{Rep}(A_4)) & \text{Non-anomalous lines in Rep}(A_4) \\ \hline
(\mathds{1},e_{1}) + (\pi_1, e_{\pi_1}) + (\pi_2, e_{\pi_2}) + 3(\pi_3, e_{\pi_3}) & \mathds{1} + \pi_1 + \pi_2 + 3 \pi_3 \\ \hline
\begin{tabular}{@{}c@{}}
$(\mathds{1},e_{1}) + (\mathds{1} + \pi_1 + \pi_2, e_{\mathds{1} + \pi_1 + \pi_2}) + (\pi_3, e_{\pi_3}^{(1)}) $ \\ + $(\pi_1,e_{\pi_1}) + (\pi_2,e_{\pi_2}) + (\pi_3,e_{\pi_3})$
\end{tabular}
& 2\mathds{1} + 2\pi_1 + 2\pi_2 + 2\pi_3 \\ \hline
(\mathds{1},e_{1})+ (\mathds{1}+\pi_3,e_{\mathds{1}+\pi_3}) + (\mathds{1}+\pi_3,\tilde e_{1+\pi_3})+  (\pi_3,e_{\pi_3}) & 3\mathds{1} + 3 \pi_3 \\ \hline
(\mathds{1},e_{1}) +  (\pi_1, e_{\pi_1}) + (\pi_2, e_{\pi_2}) +  3 (\mathds{1} + \pi_1 + \pi_2, e_{\mathds{1} + \pi_1 + \pi_2}) & 4 \mathds{1} + 4 \pi_1 + 4 \pi_2 \\ 
\hline
(\mathds{1},e_{1}) + (\pi_1, e_{\pi_1}) + (\pi_2, e_{\pi_2}) +  3 (\pi_3, e_{\pi_3}^{(1)}) & \mathds{1} + \pi_1 + \pi_2 + 3 \pi_3 \\ \hline
 \begin{tabular}{@{}c@{}} $(\mathds{1},e_{1}) + (\mathds{1} + \pi_1 + \pi_2, e_{\mathds{1} + \pi_1 + \pi_2})$ \\ $(\mathds{1}+\pi_3,e_{\mathds{1}+\pi_3}) + (\mathds{1}
+\pi_3,\tilde e_{\mathds{1}+\pi_3})$ \end{tabular} & 4 \mathds{1} + \pi_1 + \pi_2 + 2 \pi_3 \\ \hline
(\mathds{1},e_{1}) + (\pi_3,e_{\pi_3}^{(1)})+(\mathds{1}+\pi_3,e_{\mathds{1}+\pi_3}) + (\mathds{1}
+\pi_3,\tilde e_{\mathds{1}+\pi_3}) & 3 \mathds{1} + 3 \pi_3 \\
\hline
\end{array}
\eea
$A_1:=\mathds{1} + \pi_1 + \pi_2 + 3 \pi_3 $ being an algebra object implies the well-known fact that Rep$(A_4)$ admits a fibre functor. In fact, Rep$(A_4)$ admits two fibre functors, which is reflected in the above table by another Lagrangian algebra which gives the same algebra object. The algebra object $2\mathds{1} + 2\pi_1 + 2\pi_2 + 2\pi_3 $ can be decomposed into two copies of the haploid algebra object
\be
A_2:=\mathds{1} + \pi_1 + \pi_2 + \pi_3 ~.
\ee
The algebra object $3 \mathds{1} + 3 \pi_3$ can be decomposed into three copies of the algebra object
\be
A_3= \mathds{1} + \pi_3~.
\ee
From the table above, it is clear that $A_3$ admits at least two distinct algebra structures. The algebra object $4 \mathds{1} + 4\pi_1 + 4\pi_2$ can be decomposed into four copies of the algebra
\be
A_4= \mathds{1} + \pi_1 + \pi_2~.
\ee
These lines form a non-anomalous $\DZ_3$ group. Finally, from the Lagrangian algebra $L_6$, we get the algebra object
\be
4 \mathds{1} + \pi_1 + \pi_2 + 2 \pi_3 = 2 A_5 + A_6~,
\ee
where $A_6:=\mathds{1}$ and $A_6:=\mathds{1}+\pi_1+\pi_2+2\pi_3$. The algebra $A_6$ is Morita trivial which is consistent with the fusion rules of Rep$(A_4)$.

\section{Bulk-to-boundary map from Lagrangian algebra}
\label{ap:quotient functor}

Given the line operators in the SymTFT $\CZ(\CC)$ in the notation
\be
(a,e_a)~,
\ee
for some $a\in \CC$ and half-braiding $e_a$, recall that the bulk to boundary map $F: \CZ(\CC) \to \CB_{\CC}$ is given by the forgetful functor
\be
F((a,e_a))=a~.
\ee
However, often we may have to study the bulk-to-boundary map to a different gapped boundary, say $\CB_{L}$ determined by the Lagrangian algebra $L$. This is particularly important for transporting non-anomalous line operators between Morita equivalent fusion categories, as described in section \ref{sec:transporting algebras}. In this section, we will describe how to use the Lagrangian algebra $L$ to determine the relationship between bulk and boundary line operators. 

Given a Lagrangian algebra $L$ in $\CZ(\CC)$, the fusion category of line operators on the gapped boundary $\CB_{L}$ is given by the quotient category construction in \cite[Propositions 2.15, 2.16]{muger2004galois} (see also \cite{Cong:2017ffh,Cong:2017hcl}). We first define the quotient pre-category $\CZ(\CC)/L$ as follows: 
\begin{itemize}
    \item Objects of $\CZ(\CC)/L$ are objects of $\CZ(\CC)$.
    \item Morphisms are given by
\be
\text{Hom}_{\CZ(\CC)/L}(x,y):=\text{Hom}_{\CZ(\CC)}(x,L\times y)~.
\ee
\end{itemize}
Note that, in general, the category $\CZ(\CC)/L$ is not semi-simple, since $\text{Hom}_{\CZ(\CC)}(x,x)$ could be higher-dimensional for some simple line operator $x$ while it must be $\mathds{C}$ in a semi-simple category. Given the pre-quotient category $\CZ(\CC)/L$, the fusion category $\CC_{L}$ of line operators on the gapped boundary is given by the quotient category defined as follows:
\begin{itemize}
    \item The objects of $\CC_L$ are $(x,p)$, where $x\in \CZ(\CC)/L$ and $p=p^2\in \text{Hom}_{\CZ(\CC)}(x,x)$. 
    \item Morphisms are given by
    \be
\text{Hom}_{\CC_L}((x,p),(y,q)):=\{f\in \text{Hom}_{\CZ(\CC)/L}(x,y) ~ | ~ f\circ p=p\circ f, f\circ q=q\circ f\}~.
\ee
\end{itemize}
The above definition must be understood as follows. If $x$ is a simple line operator in $\CZ(\CC)$ such that $\text{Hom}_{\CZ(\CC)/L}(x,x)$ is $n$-dimensional, then $x$ is split into $n$ objects using the idempotents $p$. This will then result in a semi-simple category $\CC_L$. See \cite[Section 3.6]{Cong:2017ffh} for an explicit calculation of the bulk to boundary map using the above quotient category construction for gapped boundary of $S_3$ Dijkgraaf-Witten theory. 

Let us now explain the bulk to boundary map \eqref{eq:Haageup bulk to boundary} using the quotient category construction described above.
Consider the SymTFT $\CZ(\SH_2)$ and its gapped boundary $\CB_{L_1}$ determined by the Lagrangian algebra object
\be
L_1=\trl+\pi_1 + \pi_2 + \sigma_1~.
\ee
In order to understand how the bulk line operators are related to boundary lines, we look at the Hom space of the pre-quotient category $\CZ(\SH_2)/L_1$ 
\be
\text{Hom}_{\CZ(\SH_2)/L_1}(x,y):=\text{Hom}_{\CZ(\SH_2)}(x,L_1\times y)= \text{Hom}_{\CZ(\SH_2)}(x, (\trl+\pi_1+\pi_2+\sigma_1) \times y)~.
\ee
We will use the fusion rules for $\CZ(\SH_2)$ given in \cite{hong2008exotic} to determine the above Hom spaces. 
\begin{itemize}
    \item $ y=1 ~:~ \text{Hom}_{\CZ(\SH_2)/L_1}(x,\trl):= \text{Hom}_{\CZ(\SH_2)}(x, (\trl+\pi_1+\pi_2+\sigma_1) \times \trl)$. This Hom space is non-zero if and only if $x=\trl,\pi_1,\pi_2$ or $\sigma_1$ and it implies that these line operators get identified with the vacuum in $\CZ(\SH_2)/L_1$. This is expected as the boundary condition $\CB_{L_1}$ is obtained from gauging the line operator $L_1$.
   
    \item $ y=\pi_1 ~:~ \text{Hom}_{\CZ(\SH_2)/L_1}(x,\pi_1):= \text{Hom}_{\CZ(\SH_2)}(x, (\trl+\pi_1+\pi_2+\sigma_1) \times \pi_1)$. Using the fusion rules of $\CZ(\SH_2)$, we get
    \be
    \text{Hom}_{\CZ(\SH_2)/L_1}(x,\pi_1):= \text{Hom}_{\CZ(\SH_2)}\bigg (x, \trl+4\pi_1+4\pi_2+4\sigma_1+3\sigma_2+3\sigma_3+3 \sum_{i=1}^{6}\rho_i \bigg )~.
    \ee
    
    \item $ y=\pi_2 ~:~ \text{Hom}_{\CZ(\SH_2)/L_1}(x,\pi_2):= \text{Hom}_{\CZ(\SH_2)}(x, (\trl+\pi_1+\pi_2+\sigma_1) \times \pi_2)$. Using the fusion rules of $\CZ(\SH_2)$, we get
    \be
    \text{Hom}_{\CZ(\SH_2)/L_1}(x,\pi_2):=\text{Hom}_{\CZ(\SH_2)}\bigg (x, \trl+4\pi_1+6\pi_2+3\sigma_1+4\sigma_2+4\sigma_3+3 \sum_{i=1}^{6}\rho_i \bigg )~.
    \ee

    \item $ y=\sigma_1 ~:~ \text{Hom}_{\CZ(\SH_2)/L_1}(x,\sigma_1):= \text{Hom}_{\CZ(\SH_2)}(x, (\trl+\pi_1+\pi_2+\sigma_1) \times \sigma_1)$. Using the fusion rules of $\CZ(\SH_2)$, we get
    \be
    \text{Hom}_{\CZ(\SH_2)/L_1}(x,\sigma_1):=\text{Hom}_{\CZ(\SH_2)}\bigg (x, \trl+4\pi_1+3\pi_2+6\sigma_1+4\sigma_2+4\sigma_3+3 \sum_{i=1}^{6}\rho_i \bigg )~.
    \ee

    \item $ y=\sigma_2 ~:~ \text{Hom}_{\CZ(\SH_2)/L_1}(x,\sigma_2):= \text{Hom}_{\CZ(\SH_2)}(x, (\trl+\pi_1+\pi_2+\sigma_1) \times \sigma_2)$. Using the fusion rules of $\CZ(\SH_2)$, we get
    \be
    \text{Hom}_{\CZ(\SH_2)/L_1}(x,\sigma_2):=\text{Hom}_{\CZ(\SH_2)}\bigg (x, 3\pi_1+4\pi_2+4\sigma_1+5\sigma_2+5\sigma_3+3 \sum_{i=1}^{6}\rho_i \bigg )~.
    \ee

    \item $ y=\sigma_3 ~:~ \text{Hom}_{\CZ(\SH_2)/L_1}(x,\sigma_3):= \text{Hom}_{\CZ(\SH_2)}(x, (\trl+\pi_1+\pi_2+\sigma_1) \times \sigma_3)$. Using the fusion rules of $\CZ(\SH_2)$, we get
    \be
    \text{Hom}_{\CZ(\SH_2)/L_1}(x,\sigma_3):=\text{Hom}_{\CZ(\SH_2)}\bigg (x, 3\pi_1+4\pi_2+4\sigma_1+5\sigma_2+5\sigma_3+3 \sum_{i=1}^{6}\rho_i \bigg )~.
    \ee

    \item $ y=\rho_1 ~:~ \text{Hom}_{\CZ(\SH_2)/L_1}(x,\rho_1):= \text{Hom}_{\CZ(\SH_2)}(x, (\trl+\pi_1+\pi_2+\sigma_1) \times \rho_1)$. Using the fusion rules of $\CZ(\SH_2)$, we get
    \be
    \text{Hom}_{\CZ(\SH_2)/L_1}(x,\rho_1):=\text{Hom}_{\CZ(\SH_2)}\bigg (x, 3\bigg (\pi_1+\pi_2+\sigma_1+\sigma_2+\sigma_3+ \sum_{i=1}^{6}\rho_i\bigg ) \bigg )~.
    \ee
    We get the same Hom space for all the other $\rho_i$, $2\leq i\leq 6$.
\end{itemize}
Let $\mathds{1},$\footnote{Unlike in other sections, here we denote the identity object of $\SH_1$ as $\mathds{1}$ to avoid confusion with the identity object $\trl$ of $\SH_2$.}$\mu, \nu, \eta$ be the simple line operators on the gapped boundary $\CB_{L_1}$. These line operators can be written as combinations of the bulk line operators as follows:
\bea
&&\mathds{1} \to L_3 = \trl + \pi_1 + \pi_2 + \sigma_1~.\\
&&\mu \to \pi_1 + \pi_2 + \sum_{i=1}^6 \rho_i ~.\\
&& \nu \to \pi_1 + 2\pi_2 + \sigma_1 + 2\sigma_2 + 2 \sigma_3 + \sum_{i=1}^6 \rho_i ~.\\
&& \eta \to \pi_1 + 2\sigma_1 + \sigma_2  + \sigma_3 + \sum_{i=1}^6 \rho_i ~.
\eea
The above is precisely the $K$ map introduced in section \ref{sec:A to L}. This map satisfies the consistency condition relating quantum dimensions. 
\be
d_{a}= \frac{d_{K(a)}}{d_{K(\mathds{1})}}~,~~ a\in \{\mathds{1},\mu,\nu,\eta\}~.
\ee
and is chosen such that the linear combination of bulk lines in the Hom spaces computed above can be written as linear combinations of the boundary line operators $\mathds{1},\mu,\nu,\eta$. The bulk-to-boundary map is given by
\bea
&& F_{\CB_{L_1}}(\trl)= \mathds{1}~,\\
&& F_{\CB_{L_1}}(\pi_1)= \mathds{1} + \mu + \eta + \nu~, \\
&& F_{\CB_{L_1}}(\pi_2)= \mathds{1} + 2\nu + \mu~,\\
&& F_{\CB_{L_1}}(\sigma_1)= \mathds{1} + 2\eta + \nu~,\\
&& F_{\CB_{L_1}}(\sigma_2)= 2\nu + \eta~,\\
&& F_{\CB_{L_1}}(\sigma_3)= 2\nu + \eta~,\\
&& F_{\CB_{L_1}}(\rho_i)= \mu + \eta + \nu~,~ \forall ~ 1\leq i\leq 6 ~.
\eea
Note that the $F_{\CB_{L_1}}$ above preserves quantum dimensions. 

\section{Morita equivalence classes of algebras in Haagerup fusion categories}

\label{ap:more on Haagerup}

The algebras in three Morita equivalent Haagerup fusion categories have been classified in \cite{grossman2012quantum}. In this section, we give a brief review of the results in this paper.

\subsection{Module categories over $\SH_1$}

Consider the Haagerup fusion category $\SH_1$ with simple line operators
\be
\trl, \mu, \nu, \eta~.
\ee
It has three module categories $\CM_1,\CM_2,\CM_3$ each with four simple objects, say $\cm_1,\dots,\cm_4$. The Morita equivalent classes of algebras corresponding to these module categories can be read off from the graphs in \cite[Theorem 3.25]{grossman2012quantum}.
Recall the definition of the algebra object
\be
A_{\cm}:=\sum_{a \in \SH_1} N_{\cm}^a ~ a~,
\ee
where $N_{\cm}^a$ is the dimension of Hilbert space of operators at the junction of $a$ with the gapped boundary $\cm \in \CM$. 

$\CM_1$ is the regular module and we get the algebras
\bea
&&A_{\cm_1}=\trl,\\
&&A_{\cm_2}= \trl+\nu,\\
&&A_{\cm_3}= \trl+ \mu + 2\nu + 2\eta,\\
&&A_{\cm_4}=\trl+ \mu + \nu + \eta~.
\eea
Therefore, we get
\be
\sum_{\cm \in \CM_1} A_{\cm}= 4\trl+4\nu + 3\eta+2\mu~.  
\ee

For the module category $\CM_2$, we have
\bea
&&A_{\cm_1}=\trl+\mu + \nu,\\
&&A_{\cm_2}= \trl+2\nu + \eta,\\
&&A_{\cm_3}= \trl+ \mu +\nu ,\\
&&A_{\cm_4}=\trl+ \mu + \nu~.
\eea
Therefore, we get
\be
\sum_{\cm \in \CM_2} A_{\cm}= 4\trl+5\nu + \eta+3\mu~.  
\ee

Finally, for the module category $\CM_3$, we have
\bea
&&A_{\cm_1}=\trl+\eta,\\
&&A_{\cm_2}= \trl+\eta,\\
&&A_{\cm_3}= \trl+ \mu + 3\nu + 2\eta,\\
&&A_{\cm_4}=\trl+ \eta~.
\eea
Therefore, we get
\be
\sum_{\cm \in \CM_3} A_{\cm}= 4\trl+3\nu + 5\eta+\mu~.  
\ee

\subsection{Module categories over $\SH_2$}
Consider the Haagerup fusion category $\SH_2$ with simple line operators
\be
\trl, \alpha, \alpha^2, \rho, \alpha \rho, \alpha^2 \rho~.
\ee
There are three module categories $\CM_1$, $\CM_2$ and $\CM_3$ over $\SH_2$ given by the category of modules over the algebras $\trl$, $\trl+\rho$ and $\trl+\alpha + \alpha^2$, respectively . 

The module category $\CM_1$ contains 4 objects, say $\cm_1,\dots,\cm_4$. We have
\bea
&&A_{\cm_1}=\trl+\rho,\\
&&A_{\cm_2}= \trl + \alpha \rho,\\
&&A_{\cm_3}= \trl+ \alpha^2 \rho,\\
&&A_{\cm_4}=\trl+ \alpha + \alpha^2 + 2 H ~.
\eea
Therefore, we get
\be
\label{eq:H2 M2 sum}
\sum_{\cm \in \CM_1} A_{\cm}= 4 \trl+\alpha + \alpha^2 + 3H~.  
\ee

The module category $\CM_2$ is the regular module with $6$ objects corresponding to the simple objects in $\SH_2$. 
Using the graphs in \cite[Corollary 3.16]{grossman2012quantum} (see also \cite[Section 7]{Huang:2021ytb}), we can read off the following 
\bea
&&A_{1}=A_{\alpha}=A_{\alpha^2}=\trl~,\\
&&A_{\rho}= \trl+ H,\\
&&A_{\alpha\rho}= \trl+ H,\\
&&A_{\alpha^2\rho}= \trl+ H~.
\eea
where $H=\rho+\alpha \rho + \alpha^2 \rho$. Therefore, we get
\be
\sum_{\cm \in \CM_2} A_{\cm}= 6\trl+3H~.  
\ee

Finally, the module category $\CM_3$ contains two simple objects, say $\cm_1$ and $\cm_2$. 
We have
\bea
&&A_{\cm_1}=\trl+\alpha + \alpha^2,\\
&&A_{\cm_2}= \trl+\alpha + \alpha^2 + 3H~.
\eea
Therefore, we get
\be
\sum_{\cm \in \CM_3} A_{\cm}= 2\trl +2\alpha + 2\alpha^2 + 3H~.  
\ee

\subsection{Module categories over $\SH_3$}

There are again three module categories $\CM_1,\CM_2$ and $\CM_3$ over $\SH_3$. The Morita equivalent classes of algebras corresponding to these module categories can be read off from the graphs in \cite[Theorem 3.25]{grossman2012quantum}.

$\CM_1$ is a module category with four objects say $\cm_1,\dots,\cm_4$. We have
\bea
&&A_{\cm_1}=\trl+\alpha\rho+\alpha^2 \rho,\\
&&A_{\cm_2}= \trl+\alpha + \alpha^2 + H,\\
&&A_{\cm_3}= \trl+ \rho+ \alpha^2 \rho,\\
&&A_{\cm_4}=\trl+ \rho + \alpha\rho~.
\eea
Therefore, we get
\be
\sum_{\cm \in \CM_1} A_{\cm}= 4\trl +\alpha + \alpha^2 + 3H~.  
\ee
Note that even though this sum is the same as \eqref{eq:H2 M2 sum}, the individual algebras $A_{\cm_i}$ that constitute this sum is different in $\SH_2$ and $\SH_3$. 

The structure of the module category $\CM_2$ is the same as in the module category $\CM_3$ in $\SH_2$. Therefore, we get
\be
\sum_{\cm \in \CM_2} A_{\cm}= 2\trl  +2\alpha + 2\alpha^2 + 3H~.  
\ee

Finally, $\CM_3$ is the regular module and the Morita equivalent algebras are the same as in the module category $\CM_2$ in $\CH_2$. We get
\be
\sum_{\cm \in \CM_3} A_{\cm}= 6\trl+3H~.  
\ee

\newpage
\bibliography{refs}

%\bibitem{}
 %       \emph{""}, \emph{} {\bf} ()  

\end{document}